\PassOptionsToPackage{unicode}{hyperref}
\PassOptionsToPackage{hyphens}{url}
\PassOptionsToPackage{dvipsnames,svgnames,x11names}{xcolor}
\documentclass[
  letterpaper,
]{article}
\usepackage{xcolor}
\usepackage[margin=1in]{geometry}
\usepackage{amsmath,amssymb}
\usepackage{iftex}
\ifPDFTeX
  \usepackage[T1]{fontenc}
  \usepackage{textcomp} 
\else 
  \usepackage{unicode-math} 
  \defaultfontfeatures{Scale=MatchLowercase}
  \defaultfontfeatures[\rmfamily]{Ligatures=TeX,Scale=1}
\fi
\usepackage{lmodern}
\ifPDFTeX\else
\fi
\IfFileExists{upquote.sty}{\usepackage{upquote}}{}
\IfFileExists{microtype.sty}{
  \usepackage[]{microtype}
  \UseMicrotypeSet[protrusion]{basicmath} 
}{}
\makeatletter
\@ifundefined{KOMAClassName}{
  \IfFileExists{parskip.sty}{%
    \usepackage{parskip}
  }{
    \setlength{\parindent}{0pt}
    \setlength{\parskip}{6pt plus 2pt minus 1pt}}
}{
  \KOMAoptions{parskip=half}}
\makeatother
\makeatletter
\ifx\paragraph\undefined\else
  \let\oldparagraph\paragraph
  \renewcommand{\paragraph}{
    \@ifstar
      \xxxParagraphStar
      \xxxParagraphNoStar
  }
  \newcommand{\xxxParagraphStar}[1]{\oldparagraph*{#1}\mbox{}}
  \newcommand{\xxxParagraphNoStar}[1]{\oldparagraph{#1}\mbox{}}
\fi
\ifx\subparagraph\undefined\else
  \let\oldsubparagraph\subparagraph
  \renewcommand{\subparagraph}{
    \@ifstar
      \xxxSubParagraphStar
      \xxxSubParagraphNoStar
  }
  \newcommand{\xxxSubParagraphStar}[1]{\oldsubparagraph*{#1}\mbox{}}
  \newcommand{\xxxSubParagraphNoStar}[1]{\oldsubparagraph{#1}\mbox{}}
\fi
\makeatother

\usepackage{longtable,booktabs,array}
\usepackage{multirow}
\usepackage{calc} 
\usepackage{etoolbox}
\makeatletter
\patchcmd\longtable{\par}{\if@noskipsec\mbox{}\fi\par}{}{}
\makeatother
\IfFileExists{footnotehyper.sty}{\usepackage{footnotehyper}}{\usepackage{footnote}}
\makesavenoteenv{longtable}
\usepackage{graphicx}
\makeatletter
\newsavebox\pandoc@box
\newcommand*\pandocbounded[1]{
  \sbox\pandoc@box{#1}%
  \Gscale@div\@tempa{\textheight}{\dimexpr\ht\pandoc@box+\dp\pandoc@box\relax}%
  \Gscale@div\@tempb{\linewidth}{\wd\pandoc@box}%
  \ifdim\@tempb\p@<\@tempa\p@\let\@tempa\@tempb\fi
  \ifdim\@tempa\p@<\p@\scalebox{\@tempa}{\usebox\pandoc@box}%
  \else\usebox{\pandoc@box}%
  \fi%
}
\def\fps@figure{htbp}
\makeatother

\usepackage[]{natbib}
\DeclareUnicodeCharacter{2212}{\ensuremath{-}}
\DeclareUnicodeCharacter{2013}{--}
\DeclareUnicodeCharacter{2014}{---}
\DeclareUnicodeCharacter{2009}{\,}
\DeclareUnicodeCharacter{00A0}{~}
\DeclareUnicodeCharacter{00D7}{\ensuremath{\times}}
\DeclareUnicodeCharacter{2264}{\ensuremath{\leq}}
\DeclareUnicodeCharacter{2265}{\ensuremath{\geq}}
\DeclareUnicodeCharacter{2260}{\ensuremath{\neq}}
\DeclareUnicodeCharacter{2248}{\ensuremath{\approx}}
\DeclareUnicodeCharacter{03B1}{\ensuremath{\alpha}}
\DeclareUnicodeCharacter{03B2}{\ensuremath{\beta}}
\DeclareUnicodeCharacter{03C3}{\ensuremath{\sigma}}
\DeclareUnicodeCharacter{03C7}{\ensuremath{\chi}}
\DeclareUnicodeCharacter{2032}{\ensuremath{'}}

\usepackage{multirow}

\usepackage{caption}
\setcitestyle{round}

\AtBeginEnvironment{apptbl}{\footnotesize}
\makeatletter
\newcommand\table@defaultsize{\footnotesize}
\let\table@size\table@defaultsize
\newcommand\nexttablesize[1]{\gdef\table@size{#1}}
\AtBeginEnvironment{longtable}{\table@size\global\let\table@size\table@defaultsize}
\makeatother

\makeatletter
\let\floatnote@text\@empty
\newcommand\floatnote[1]{\gdef\floatnote@text{#1}}
\newcommand\fs@plainnote{\fs@plain
  \def\@fs@post{\ifx\floatnote@text\@empty\else
    \vspace\abovecaptionskip\noindent\floatnote@text\par
    \global\let\floatnote@text\@empty\fi}}
\makeatother

\AtBeginDocument{%
  \floatstyle{plainnote}\restylefloat{appfig}%
  \floatplacement{appfig}{H}\floatplacement{apptbl}{H}}

\AtBeginDocument{}
\makeatletter
\renewcommand\@maketitle{%
  \newpage\null\vskip 0.5em
  \begin{center}
    {\Large\bfseries \@title\par}
    \vskip 1.5em
    {\normalsize \@author\par}
  \end{center}
  \par\vskip 1em}
\newcommand\aclauthor[3]{
  \begin{tabular}[t]{@{}c@{}}
    \textbf{#1}\\[1pt]
    #2\\[1pt]
    {\small\texttt{#3}}
  \end{tabular}}
\makeatother

\AtBeginDocument{%
  \author{%
    \aclauthor{Davood Wadi}
      {Desautels Faculty of Management\\ McGill University\\ Montreal, Quebec, Canada}
      {davood.wadi@hec.ca}%
    \hspace{3em}%
    \aclauthor{Yu Ma}
      {Desautels Faculty of Management\\ McGill University\\ Montreal, Quebec, Canada}
      {yu.ma@mcgill.ca}%
  }%
}
\makeatletter
\@ifpackageloaded{float}{}{\usepackage{float}}
\floatstyle{plain}
\@ifundefined{c@chapter}{\newfloat{apptbl}{h}{loapptbl}}{\newfloat{apptbl}{h}{loapptbl}[chapter]}
\floatname{apptbl}{Table W}
\floatstyle{plaintop}
\restylefloat{apptbl}
\newcommand*\quartoapptblref[1]{Table \hyperref[#1]{W\ref{#1}}}
\@ifpackageloaded{caption}{}{\usepackage{caption}}
\DeclareCaptionLabelFormat{quartoapptblreflabelformat}{#1#2}
\makeatother
\makeatletter
\@ifpackageloaded{float}{}{\usepackage{float}}
\floatstyle{plain}
\@ifundefined{c@chapter}{\newfloat{appfig}{h}{loappfig}}{\newfloat{appfig}{h}{loappfig}[chapter]}
\floatname{appfig}{Fig. W}
\newcommand*\quartoappfigref[1]{Fig. \hyperref[#1]{W\ref{#1}}}
\@ifpackageloaded{caption}{}{\usepackage{caption}}
\DeclareCaptionLabelFormat{quartoappfigreflabelformat}{#1#2}
\makeatother
\makeatletter
\@ifpackageloaded{float}{}{\usepackage{float}}
\floatstyle{plain}
\@ifundefined{c@chapter}{\newfloat{appenv}{h}{loapp}}{\newfloat{appenv}{h}{loapp}[chapter]}
\floatname{appenv}{Appendix}
\floatstyle{plaintop}
\restylefloat{appenv}

\makeatother
\makeatletter
\@ifpackageloaded{caption}{}{\usepackage{caption}}
\AtBeginDocument{%
\ifdefined\contentsname
  \renewcommand*\contentsname{Table of contents}
\else
  \newcommand\contentsname{Table of contents}
\fi
\ifdefined\listfigurename
  \renewcommand*\listfigurename{List of Figures}
\else
  \newcommand\listfigurename{List of Figures}
\fi
\ifdefined\listtablename
  \renewcommand*\listtablename{List of Tables}
\else
  \newcommand\listtablename{List of Tables}
\fi
\ifdefined\figurename
  \renewcommand*\figurename{\textbf{Fig.}}
\else
  \newcommand\figurename{\textbf{Fig.}}
\fi
\ifdefined\tablename
  \renewcommand*\tablename{\textbf{Table}}
\else
  \newcommand\tablename{\textbf{Table}}
\fi
}
\@ifpackageloaded{float}{}{\usepackage{float}}
\floatstyle{ruled}
\@ifundefined{c@chapter}{\newfloat{codelisting}{h}{lop}}{\newfloat{codelisting}{h}{lop}[chapter]}
\floatname{codelisting}{Listing}

\makeatother
\makeatletter
\@ifpackageloaded{caption}{}{\usepackage{caption}}
\@ifpackageloaded{subcaption}{}{\usepackage{subcaption}}
\makeatother
\usepackage{bookmark}
\IfFileExists{xurl.sty}{\usepackage{xurl}}{} 
\makeatletter
\@ifundefined{xmpquote}{}{}
\makeatother
\hypersetup{
  pdftitle={Whom Do AI Agents Serve? Role-Induced Sponsorship Bias in Decision Models and Generative LLMs},
  pdfauthor={Davood Wadi; Yu Ma},
  colorlinks=true,
  linkcolor={blue},
  filecolor={Maroon},
  citecolor={Blue},
  urlcolor={Blue},
  pdfcreator={LaTeX via pandoc}}

\title{Whom Do AI Agents Serve? Role-Induced Sponsorship Bias in
Decision Models and Generative LLMs}
\author{Davood Wadi \and Yu Ma}
\date{}
\begin{document}
\maketitle

\textbf{Abstract}\\
Consumers increasingly delegate product search to AI shopping agents
powered by generative large language models (LLMs) or non-generative
decision models. Because retail platforms monetize through sponsored
listings that consumers typically discount, we examine whether an AI
agent's evaluation of sponsored listings depends on who delegates it.
Across six LLMs and six decision models in Study 1, consumer-delegated
agents penalize sponsored listings, whereas naming a booking platform as
principal significantly attenuates this sponsorship penalty without
instructions to favor advertisers. This sponsorship bias replicates
across prompt wordings and LLM reasoning efforts and operates through
reduced pre-choice skepticism in LLMs. In Study 2, attributing the paid
placement to the platform amplifies sponsorship bias across both model
families, and stricter disclosure wording (``Sponsored''
vs.~``Promoted'') fails to eliminate it. In Study 3, prompt-level
mitigation removes platform leniency but can over-correct the
sponsorship penalty, whereas consumer-aligned fine-tuning eliminates
sponsorship bias and better preserves baseline preferences.

Keywords: large language models; decision models; AI agents; retail
media; sponsorship disclosure; digital retailing

\section{Introduction}\label{sec-intro}

Retailers and technology firms have deployed generative large language
models (LLMs) as conversational shopping assistants
\citep{amazon2026alexa, meta2026muse, walmart2024genai}, while a growing
share of consumers use general-purpose AI chatbots to discover and
evaluate products and services
\citep{adobe2025genai, balaskas2026recommendations, wadi2026shopping, hasselwander2026toward, kumar2026transformative, mogaji2024generative}.
Alongside generative LLMs, AI providers have recently introduced
non-generative decision models \citep[also called System-One
models;][]{almeida2026introducing, deusser2026evaluatingbenchmarkingmodeljev, palmer2026kev9b}
that directly score and rank alternatives without generating text.
Simultaneously, AI interfaces and digital marketplaces have begun
incorporating advertisements into AI-mediated search
\citep{chatgpt2026advertise}. As a result, the AI systems used to
evaluate and recommend products are also the channel through which
advertisers pay for their products to be recommended.

This dual role places the AI agent between two different principals
whose interests diverge
\citep{hagiu2011intermediaries, inderst2012competition, abdollahpouri2020multistakeholder}.
On the one hand, the consumer is the assumed principal who relies on the
AI agent as a shopping advisor. On the other hand, the deploying
platform could be named as the principal in the system instructions. The
AI agent must therefore execute an evaluative judgment that balances
between the interest of the consumer who trusts it and the platform that
programs its role.

In traditional e-commerce, sponsorship disclosures are designed for
human shoppers, who observe a ``Sponsored'' label and discount the paid
listing accordingly \citep{boerman2017post, sahni2020sponsorship}. When
a consumer delegates product evaluation or purchase to an AI agent
\citep[e.g.,][]{deepmind_mariner}, however, the sponsorship disclosure
is processed by the AI agent rather than the consumer. As a result, the
AI agent's trade-off between organic and sponsored alternatives remains
hidden from the shopper. This raises a central question: does merely
specifying who delegates the AI agent in its system instructions (the
platform vs.~the consumer) alter how it evaluates sponsored listings? We
refer to this shift as sponsorship bias, defined as a more favorable
evaluation of sponsored listings when the platform (vs.~the consumer) is
assigned as the AI agent's principal.

We jointly investigate generative LLMs and non-generative decision
models across three controlled choice studies by manipulating the
delegating principal and observing whether the AI agent's evaluation of
sponsored listings shifts as a result. In Study 1, we evaluate six
frontier generative LLMs and six commercial and open-weight decision
models on matched hotel listings. We alter a single sentence in the
system instructions to define the AI agent as a delegate of either a
traveler or a booking platform. Under consumer delegation, every model
across both families applies a significant sponsorship penalty (i.e., a
less favorable evaluation of sponsored listings relative to matched
organic listings). Under platform delegation, this sponsorship penalty
is significantly attenuated across all 12 models. We replicate this
sponsorship bias across alternative system-prompt wordings and reasoning
efforts. Furthermore, moderated mediation analysis shows that
sponsorship disclosure increases skepticism toward the sponsored listing
in the generative LLM's pre-choice reasoning trace, and this expressed
skepticism is negatively associated with subsequent selection of that
listing. In contrast, platform delegation attenuates sponsorship-induced
skepticism and weakens this indirect association.

Study 2 decomposes the disclosure label (``Promoted'' vs.~``Sponsored''
crossed with whether the placement is attributed to the platform) across
both generative LLM and decision model architectures. Across both model
families tested, attributing the sponsored listing to the platform
amplifies the sponsorship bias. Although the explicit term ``Sponsored''
lowers evaluation of sponsored listings relative to ``Promoted'' in both
model families, stricter regulatory disclosure wording does not mitigate
the sponsorship bias in either architecture.

Because disclosure terminology alone fails to eliminate sponsorship
bias, Study 3 evaluates prompt-level and model-level mitigation to
anchor platform-delegated AI agents to the consumer-delegated baseline.
First, under prompt-level mitigation, instructing the platform-delegated
AI agent to act in the traveler's best interest eliminates platform
leniency toward sponsored listings in both model families. However, in
the decision model, prompt-level mitigation over-corrects the
sponsorship penalty, and in both models it perturbs the
platform-delegated agent's baseline hotel and review preferences.
Second, under model-level mitigation, we apply consumer-aligned
fine-tuning to an open-weight decision model across synthetic non-hotel
retail categories so that it matches its consumer-delegated choice
distribution regardless of the assigned principal. Evaluated zero-shot
on hotel choices, consumer-aligned fine-tuning eliminates sponsorship
bias without over-correcting the sponsorship penalty and achieves lower
Kullback-Leibler (KL) divergence across the baseline hotel and review
preferences.

This research makes three contributions to the literature on digital
retailing, sponsorship disclosure, and AI-mediated commerce. First, we
show that stakeholder alignment in AI agents can emerge implicitly from
role assignment alone. Across both generative LLMs and decision models,
naming the platform as principal induces sponsorship bias without
programmed objectives \citep{abdollahpouri2020multistakeholder} or
explicit instructions to favor advertisers
\citep{wu2026ads, allouah2026your}. Second, we extend sponsorship
disclosure and persuasion knowledge research
\citep{boerman2017post, friestad1994persuasion, evans2015rethinking}
from human consumers to AI agents. We show that attributing the
sponsored listing to the platform amplifies sponsorship bias, and that
stricter disclosure wording (``Sponsored'' vs.~``Promoted'') fails to
eliminate it. Third, we compare prompt-level mitigation and model-level
mitigation via consumer-aligned fine-tuning as remedies for sponsorship
bias. While both remedies eliminate platform leniency toward sponsored
listings, consumer-aligned fine-tuning better preserves the
platform-delegated agent's baseline hotel and review preferences without
prompt instructions.

\section{Literature review}\label{sec-lit}

Technological shifts continually reshape retail formats, digital channel
competition, and how consumers adopt retail technologies to search and
evaluate market alternatives
\citep{gauri2021evolution, grewal2021retailing, jindal2021omnichannel, namin2026consumers, ratchford2022online}.
As consumers increasingly delegate product search and evaluation to AI
shopping assistants
\citep{bahmani2022hey, chakraborty2024enhancing, hermann2024artificial, lalicic2021consumers, srinivasaraja2026meta},
a central question for multi-stakeholder marketplaces, where consumer,
platform, and advertiser interests diverge
\citep{abdollahpouri2020multistakeholder}, is whether an AI agent's
recommendations depend on whether the service is provided by the
platform or the consumer.

Prior research on retail recommendation systems has predominantly
analyzed ranking and filtering engines
\citep{rokach2011introduction, zhang2022effectiveness}, in which
competing stakeholder objectives are balanced through an explicit
formula set by the system's designers
\citep{abdollahpouri2020multistakeholder}. Modern foundation-model
recommenders depart from this paradigm across two complementary model
families. First, generative LLMs generate natural-language reasoning
traces before producing discrete choices
\citep{stoeber2026ai, suh2026time}. Second, non-generative decision
models adapt language-model backbones to score structured options in a
single forward pass and output calibrated choice probabilities at low
latency and cost \citep{palmer2026kev4b}. In both families, the party on
whose behalf the agent acts can be specified in a single sentence of
natural-language system instructions \citep{wallace2024instruction}.
Recent work on agentic commerce examines how AI shopping agents weigh
product attributes and labels under a fixed user role
\citep{allouah2026your} or how chatbots behave when explicitly
instructed to steer users toward sponsored products \citep{wu2026ads}.
Neither addresses whether merely specifying who employs the AI agent
(i.e., platform vs.~consumer, with no instruction to favor advertisers)
shifts how generative LLMs and decision models evaluate sponsored
listings.

This distinction between platform and consumer delegation is central to
modern retailing because online marketplaces and travel platforms
increasingly extend traditional retail promotions
\citep{ailawadi2009communication, grewal2011innovations} into retail
media networks that earn high-margin revenue from on-site sponsored
listings alongside organic results \citep{thomas2025retail}. Whereas the
platform has a structural incentive to steer shoppers toward sponsored
listings
\citep{hagiu2011intermediaries, inderst2012competition, kang2015opportunism},
regulation requires sponsored labels
\citep{eu2022dsa, ftc2013search, bigne2021online} to alert consumers to
this commercial motive. In direct human search, disclosures activate
persuasion knowledge and lead consumers to penalize paid placements in
their utility evaluation
\citep{friestad1994persuasion, boerman2017post, kim2019paradox, cai2016boomerang}.
However, disclosure remedies assume that the protected party is the one
who observes and discounts the warning \citep{loewenstein2011limits}.
When a consumer delegates evaluation to an AI agent, the sponsorship
label reaches the agent rather than the consumer, and the agent's
internal trade-offs remain hidden \citep{chen2025reasoning}. If a
platform-delegated AI agent discounts sponsored listings less than a
consumer-delegated agent, disclosure fails to protect the shopper and
exposes consumers to perceived manipulation and betrayal
\citep{saenger2024betrayed, liu2025you, du2021paradoxes}.

\section{Hypothesis development}\label{hypothesis-development}

We refer to the entity named as the delegator in the system instructions
as the agent's principal. We define consumer delegation as the condition
where the system prompt assigns the consumer as the agent's principal
(e.g., ``You are employed by a traveler''), and platform delegation as
the condition where the system prompt assigns the platform as the
agent's principal (e.g., ``You are employed by a booking platform'').

\textbf{Role adoption in AI agents.} Behavior in both generative LLMs
and decision models is not fully specified by explicit instructions. As
models scale, they display capabilities and response patterns that were
not directly programmed \citep{wei2022emergent}. One well-documented
pattern is context-dependent role adoption. When conditioned on a
persona or party, a language model tailors its representations and
outputs to the inferred perspective of that party
\citep{deshpande2023toxicity, shanahan2023role, stoeber2026ai}. For
instance, conditioning a model on demographic information leads it to
reproduce the opinion distributions of the corresponding human subgroups
\citep{argyle2023out, santurkar2023whose} and shift its answers toward
the views a user appears to hold, even at the expense of accuracy
\citep{perez2023discovering, sharma2024towards}. This role adoption
draws on the semantic associations a foundation model acquires from its
training corpora, which link social categories or entities to the
behaviors and interests they typically pursue
\citep{bolukbasi2016man, caliskan2017semantics, garg2018word}. In a
retail recommendation task, a role assignment tells the AI agent whose
perspective to adopt, which can shape how the agent evaluates
alternatives in both open-ended choice generation (LLMs) and single-pass
option scoring (decision models), even with no explicit instruction
regarding revenue or commissions.

\textbf{Sponsorship penalty.} From a random utility perspective, just as
a higher price or lower review rating reduces an alternative's value,
sponsorship status enters consumer utility with a finite negative weight
(\(\beta_{\text{sponsorship}} < 0\)) because a paid placement signals
persuasive intent
\citep{friestad1994persuasion, obermiller1998development, campbell2000consumers, cai2016boomerang, kim2021influencer}.
Because foundation-model training corpora include vast amounts of human
text, such as product reviews, consumer forums, and marketing literature
that express this skepticism toward paid placements, an AI agent
assigned to the consumer should adopt this negative valuation. When the
AI agent acts solely for the consumer
\citep{inderst2012competition, hagiu2011intermediaries}, its negative
weight on sponsorship
(\(\beta_{\text{sponsorship}}^{\text{consumer}} < 0\)) establishes the
unconflicted baseline, which we refer to as the sponsorship penalty
(i.e., the reduction in a listing's evaluation attributable to its
disclosed sponsorship status).

\textbf{H1.} Under consumer delegation, given equal objective
attributes, an AI agent evaluates a sponsored listing less favorably
than an organic listing.

\textbf{Sponsorship bias.} A model trained on text describing digital
markets is also likely to associate a retail platform with its
commercial incentives. Although sponsorship is a negative utility
attribute from the consumer's perspective, it is a source of retail
media revenue from the platform's perspective. As a result, assigning
the platform as the AI agent's principal may lead the agent to adopt the
platform's perspective and attenuate its negative weight on sponsorship
relative to the consumer-delegated baseline
(\(|\beta_{\text{sponsorship}}^{\text{platform}}| < |\beta_{\text{sponsorship}}^{\text{consumer}}|\)),
a shift we term sponsorship bias.

\textbf{H2.} The negative effect of sponsorship on listing evaluation is
moderated by the agent's principal, such that the sponsorship penalty is
attenuated when the agent is delegated by the platform (vs.~consumer).

\textbf{Pre-choice skepticism.} In generative LLMs that produce
pre-choice reasoning traces, we can also test the temporal pathway
linking sponsorship disclosure, pre-choice skepticism, and subsequent
choice. Recognizing a listing as sponsored leads consumers to doubt its
claims, and this doubt lowers their evaluation of the listing
\citep{friestad1994persuasion, campbell2000consumers}. If a generative
LLM agent penalizes sponsored listings, its observable outputs should
reflect a similar sequence. A sponsorship label should lead the AI agent
to express skepticism about the listing's value in its reasoning before
choosing \citep{hagendorff2023machine}, and that skepticism should be
associated with a lower probability that it chooses the
listing.\footnote{Note that the proposed mediation is temporal: the AI
  agent produces the reasoning trace before it makes the choice. With
  mediation we test this temporal pathway. We do not claim that the
  reasoning traces cause the choice.}

\textbf{H3.} Skepticism mediates the effect of sponsorship on choice.

\begin{figure}

\centering{

\includegraphics[width=0.55\linewidth,height=\textheight,keepaspectratio]{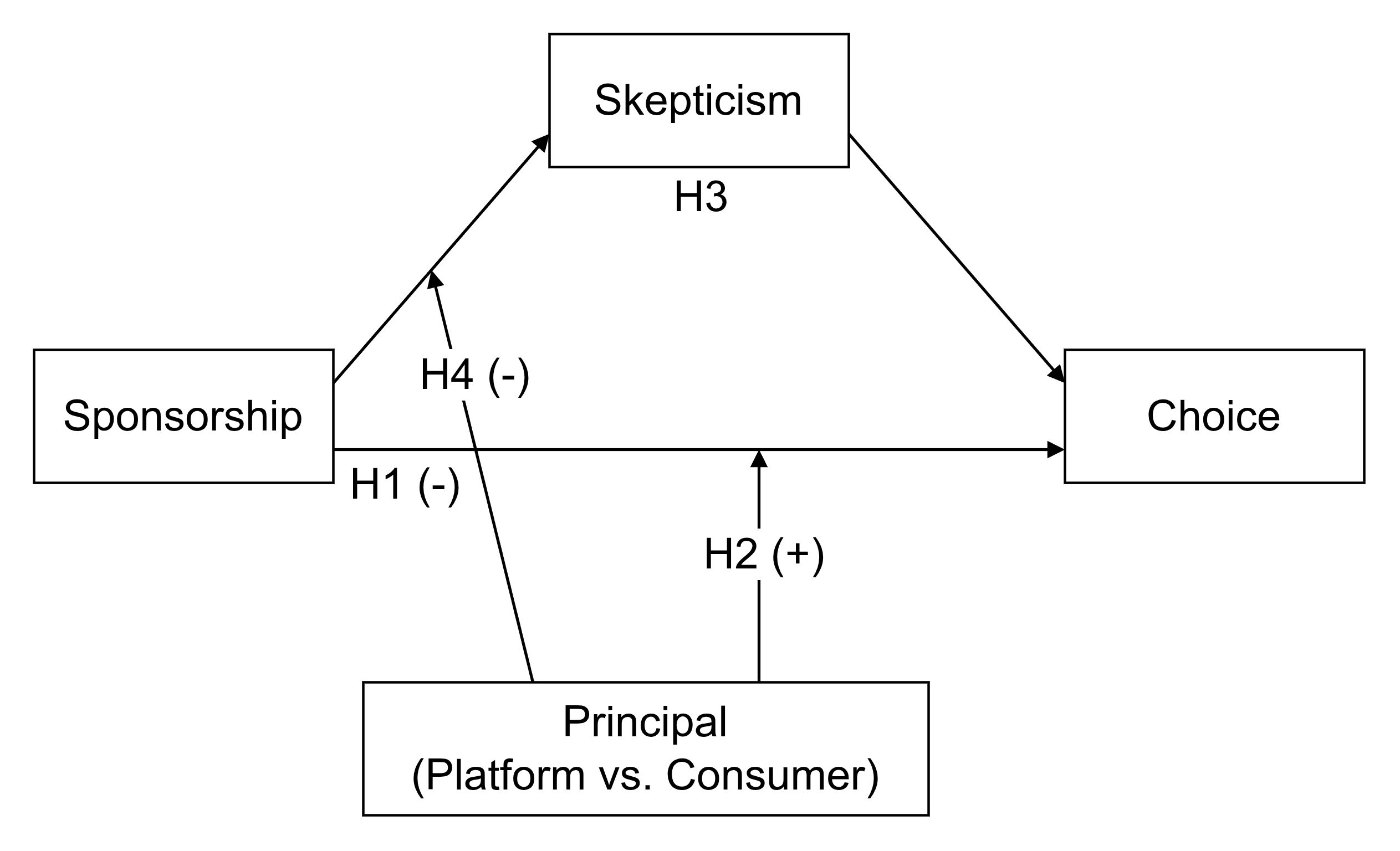}

}

\caption{\label{fig-conceptual-framework}Conceptual framework}

\end{figure}

If an AI agent delegated by a consumer treats the sponsorship label with
skepticism, an AI agent delegated by the platform, which earns
commissions, should associate the same label with less skepticism. The
principal should therefore moderate the first stage of the mediated
pathway, such that the indirect effect of sponsorship on choice through
skepticism is weaker under platform (vs.~consumer) delegation
(Fig.~\ref{fig-conceptual-framework}).

\textbf{H4.} The effect of sponsorship on skepticism is moderated by the
agent's principal, such that platform delegation attenuates the effect
of sponsorship on skepticism.

--- insert Fig.~\ref{fig-conceptual-framework} about here ---

\section{Study 1}\label{sec-study1}

Study 1 tests H1 and H2 jointly across both foundation-model families in
hotel booking, an industry where online search and sponsored listings
are pervasive. We test a frontier generative LLM (Gemini 3.1 Pro) that
generates natural-language reasoning traces before making discrete
choices and an open-weight non-generative decision model \citep[Kev
9B;][]{palmer2026kev9b} that scores options in a single forward pass
without generating text. In addition, because generative LLMs emit
pre-choice reasoning traces, we test the moderated mediation pathway via
pre-choice skepticism (H3 and H4) in Gemini 3.1 Pro, and then evaluate
generalizability across five additional generative LLMs, five additional
decision models, system-prompt wordings, and reasoning efforts.

\subsection{Experiment design, stimuli, and
procedure}\label{experiment-design-stimuli-and-procedure}

\textbf{Design.} We employed a 2 (Sponsorship: organic vs.~sponsored)
\(\times\) 2 (Principal: consumer vs.~platform) factorial design across
both model families. In each choice task, the AI agent evaluated four
hotel alternatives (\quartoapptblref{apptbl-study1-alternatives}): a
target listing (manipulated), a matched organic competitor, and two
dominated alternatives included for ecological validity. The target
listing received the sponsorship manipulation via the ``Platform
Sponsor'' attribute (\emph{true} vs.~\emph{false}), which was set to
\emph{false} for all other alternatives.

\textbf{Stimuli.} For each replication, four unique hotel names were
drawn from a pool of 10 real Expedia hotels. Positive reviews were drawn
from a validated pool of 10 paired Expedia reviews
(\quartoapptblref{apptbl-review-pool}; Appendix B) and assigned to the
target and competitor listings, while negative reviews were assigned to
the two dominated alternatives \citep{kim2019paradox}. To isolate the
valuation of the sponsorship attribute ceteris paribus, the target and
competitor hotels were assigned identical prices (\(P\)) drawn from the
empirical distribution of Expedia listings
(\(P \sim \mathcal{U}[\$173.45, \$351.40]\)), whereas the dominated
alternatives were assigned fixed prices above this range (\$450 and
\$400). Holding observable price and review quality equal ensures that
the two viable listings differ only in whether the target listing is
sponsored, which isolates the marginal utility weight placed on the
sponsorship attribute (\(\beta_{\text{sponsorship}}\)) without price or
quality confounds.

\textbf{Prompts and evaluation protocols.} Table~\ref{tbl-principal}
presents the system instructions manipulating the principal, which name
the employing party but omit any instruction regarding commissions, ad
revenue, or profitability to avoid experimental demand effects. Because
token sampling in generative LLMs is stochastic, for Gemini 3.1 Pro we
randomized the order of alternatives and attributes on every trial
\citep{liu2024lost, pezeshkpour2024large, yin2026fragile, wadi2026does, wang2024large}.
We conducted 500 independent, zero-shot replications per cell using a
Monte Carlo sampling approach
\citep{robert1999monte, wang2022self, wadi2025monte, holtzman2019curious}
at high reasoning effort (\emph{thinking level}), yielding
\(N = 2{,}000\) observations (see Appendix B for power analysis,
decoding parameters, and API specifications).

Because a non-generative decision model returns a deterministic
probability distribution for any fixed input, we evaluated Kev 9B on a
position-balanced Latin-square set of 360 base choice sets constructed
from the hotel and review pools (Appendix B). Each of the 360 choice
sets was evaluated across all four cells of the \(2 \times 2\) design
(\(N = 1{,}440\) paired observations). In each trial, the principal
instruction was passed as the decision \emph{state} and the four hotels
as a structured \emph{choice} question (``Which hotel would you
book?'').

--- insert Table~\ref{tbl-principal} about here ---

\subsection{Measures}\label{measures}

Consistent with the random utility theory, our choice analyses in both
model families estimate the log-odds of selecting the target listing
over the competitor listing.\footnote{dominated alternatives were never
  chosen by any LLM and were never the modal choice in any decision
  model.} For Gemini 3.1 Pro, the primary dependent variable is binary
\emph{target choice} (1 = target selected, 0 = competitor selected),
analyzed via binary logistic regression controlling for standardized
price and target and competitor hotel and review fixed effects
(Table~\ref{tbl-main-logit}; full table in
\quartoapptblref{apptbl-main-logit-full}). For Kev 9B, because the model
outputs continuous choice probabilities, the primary dependent variable
is the exact log-odds of selecting the target over the competitor,
\(\ln(P_{\text{target}} / P_{\text{competitor}})\), analyzed via
ordinary least squares (OLS) with 360 paired stimulus fixed effects and
stimulus-clustered standard errors (Table~\ref{tbl-kev-main}), alongside
calibrated target probabilities
(\quartoapptblref{apptbl-study2-descriptive}).

To test the mediated pathway in Gemini 3.1 Pro (H3 and H4), we
operationalize \emph{skepticism} as the extent to which the AI agent's
pre-choice reasoning trace expressed negative or defensive evaluations
of the target listing's sponsorship status
\citep{friestad1994persuasion, obermiller1998development, campbell2000consumers}.
It was scored on a seven-point Likert scale (1 = Strongly Disagree to 7
= Strongly Agree) for the item, ``The sponsorship status of
\{target\_option\_id\} made the agent skeptical of its true value to the
traveler.''\footnote{\emph{\{target\_option\_id\}} denotes the target's
  randomized identifier.} \footnote{We distinguish between the LLM's
  observable choice, its generated reasoning trace, and its internal
  neural activations. Terms such as ``skepticism'' do not imply sentient
  human motivation, but serve as an indicator for defensive evaluation
  expressed in the pre-choice reasoning trace.} To avoid shared-weight
or provider-alignment bias \citep{gu2026survey, panickssery2024llm},
pre-choice reasoning traces were scored using Kev 9B
\citep{palmer2026kev9b}, which was selected via a blinded validation
study against human ratings on a stratified sample of 60 traces
(\(\text{ICC}(2,1) = 0.959\), Pearson \(r = .964\),
\(\text{MAE} = 0.450\); see Appendix B and
\quartoapptblref{apptbl-judge-validation}).

\subsection{Results}\label{results}

\begin{figure}

\centering{

\includegraphics[width=0.55\linewidth,height=\textheight,keepaspectratio]{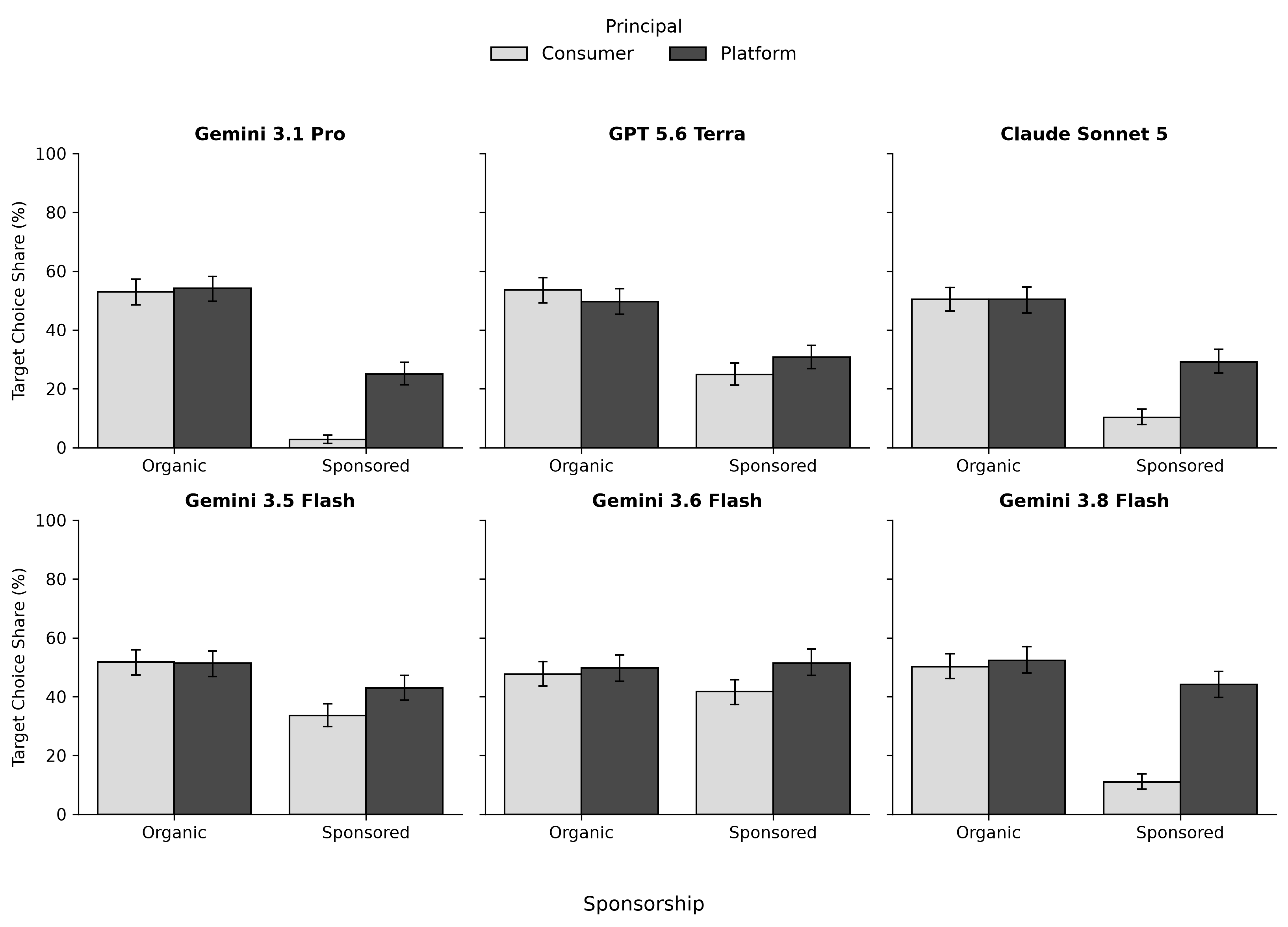}

}

\caption{\label{fig-main-choice}Target choice shares (\%) by principal
and sponsorship across generative LLMs (Study 1).}

\end{figure}

\begin{figure}

\centering{

\includegraphics[width=0.55\linewidth,height=\textheight,keepaspectratio]{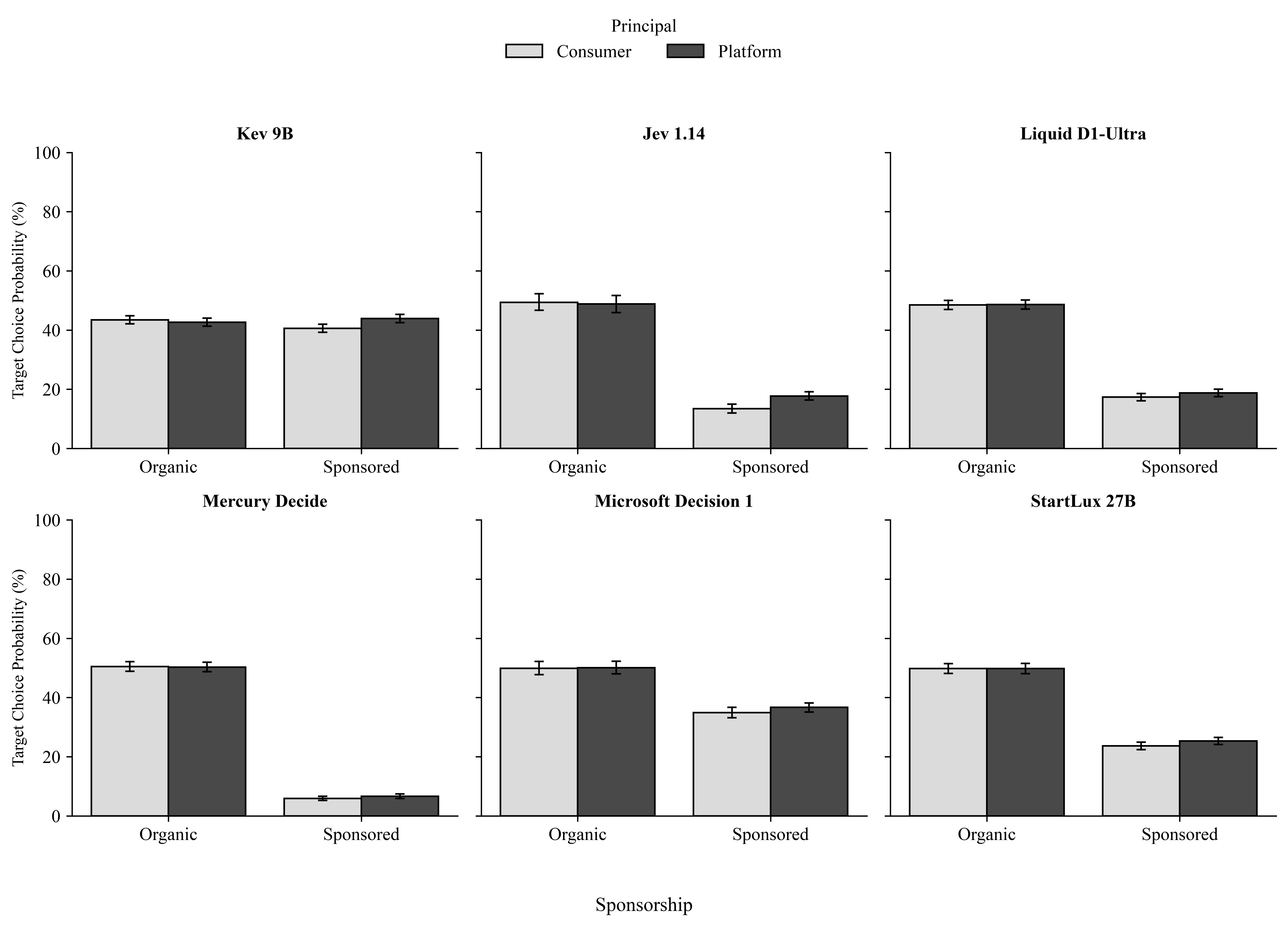}

}

\caption{\label{fig-study2-choice}Calibrated target choice probabilities
(\%) by principal and sponsorship across decision models (Study 1).}

\end{figure}

Supporting H1, under consumer delegation, both Gemini 3.1 Pro and Kev 9B
apply a significant sponsorship penalty (Table~\ref{tbl-main-logit};
Table~\ref{tbl-kev-main}; Fig.~\ref{fig-main-choice};
Fig.~\ref{fig-study2-choice}). In Gemini 3.1 Pro, tagging the target
listing ``Platform Sponsor'' under consumer delegation reduces its
choice share from 53.0\% to 2.8\%, a 50.2 percentage-point sponsorship
penalty (\(\beta = -6.491\), \(SE = 0.418\), \(p < .001\);
\quartoapptblref{apptbl-descriptive-others}). Similarly, in Kev 9B,
tagging the target listing ``Platform Sponsor'' under consumer
delegation reduces its calibrated target probability from \(43.43\%\)
(\(SD = 13.63\)) to \(40.60\%\) (\(SD = 13.09\)), a 2.83
percentage-point sponsorship penalty on exact log-odds
(\(\beta = -0.123\), \(SE = 0.012\), \(z = -10.65\), \(p < .001\);
\quartoapptblref{apptbl-study2-descriptive}).

--- insert Fig.~\ref{fig-main-choice} about here ---

--- insert Fig.~\ref{fig-study2-choice} about here ---

Supporting H2, platform delegation significantly attenuates this
sponsorship penalty across both model families
(Table~\ref{tbl-main-logit}; Table~\ref{tbl-kev-main}). In Gemini 3.1
Pro, tagging the target listing ``Platform Sponsor'' under platform
delegation reduces its choice share from 54.2\% (271 of 500) to 25.0\%
(125 of 500), a sponsorship penalty of 29.2 percentage points compared
to 50.2 points under consumer delegation. This 21.0 percentage-point
difference-in-differences is positive and significant in the
fixed-effects logit (\(\beta = 3.989\), \(SE = 0.428\), \(z = 9.32\),
\(p < .001\)). The organic choice shares do not differ between
principals (53.0\% vs.~54.2\%, \(\beta = -0.165\), \(p = .403\)).
Likewise, in Kev 9B, platform delegation reverses the effect of
sponsorship. Calibrated target probability rise from \(42.62\%\)
(\(SD = 13.58\)) to \(43.88\%\) (\(SD = 13.07\); \(\beta = 0.071\),
\(SE = 0.014\), \(z = 4.94\), \(p < .001\)), yielding a positive and
significant Principal \(\times\) Sponsorship interaction on exact
log-odds (\(\beta = 0.194\), \(SE = 0.012\), \(z = 16.47\),
\(p < .001\)), with no principal effect when the target is organic
(\(\beta = -0.003\), \(p = .794\)). Together, these results show that
sponsorship bias emerges across both generative LLMs and non-generative
decision models.

--- insert Table~\ref{tbl-main-logit} about here ---

--- insert Table~\ref{tbl-kev-main} about here ---

\textbf{Moderated mediation.} We next test the moderated mediation
pathway via pre-choice skepticism in Gemini 3.1 Pro (H3 and H4) using
PROCESS Model 8 (Table~\ref{tbl-mediation}; full tables in
\quartoapptblref{apptbl-process-full-skep} and
\quartoapptblref{apptbl-process-full-choice}). The results support H3.
Sponsorship significantly increases skepticism under both consumer
(\(b = 4.827\), \(SE = 0.081\), \(p < .001\)) and platform delegation
(\(b = 3.466\), \(SE = 0.081\), \(p < .001\)). Skepticism is negatively
associated with target choice (\(\beta = -1.670\), \(SE = 0.121\),
\(p < .001\)), yielding significant negative indirect effects for both
principals (\(ab_{\text{consumer}} = -8.061\), 95\% CI
{[}\(-10.636, -7.309\){]}; \(ab_{\text{platform}} = -5.788\), 95\% CI
{[}\(-7.660, -5.202\){]}). Supporting H4, platform delegation
significantly attenuates the first-stage effect of sponsorship on
skepticism (\(b_{\text{sponsorship} \times \text{principal}} = -1.361\),
\(SE = 0.114\), \(p < .001\)), producing a positive and significant
index of moderated mediation
(\(\text{Index}_{\text{mod. med.}} = 2.273\), Boot \(SE = 0.318\), 95\%
CI {[}\(1.916, 3.135\){]}).

--- insert Table~\ref{tbl-mediation} about here ---

\subsection{Robustness tests}\label{robustness-tests}

\textbf{Cross-model generalizability.} To test whether the sponsorship
penalty and sponsorship bias generalize beyond Gemini 3.1 Pro and Kev
9B, we replicated the \(2 \times 2\) experiment across five additional
generative LLMs (\(N = 2{,}000\) per model; \(N = 10{,}000\) total) and
five additional non-generative decision models (\(N = 1{,}440\) per
model; \(N = 7{,}200\) additional observations). The five additional
LLMs span Anthropic (Claude Sonnet 5), OpenAI (GPT 5.6 Terra), and three
successive generations of Google's Flash family (Gemini 3.5, 3.6, and
3.8 Flash; Fig.~\ref{fig-main-choice};
\quartoapptblref{apptbl-descriptive-others}). The five additional
decision models span commercial API models, including Jev 1.14
\citep{typesafe2026jev114}, Liquid D1-Ultra \citep{liquidAI2026d1ultra},
Mercury Decide \citep{inception2026mercurydecide}, and Microsoft
Decision 1 \citep{srivastava2026microsoftdecision1}, as well as an
open-weight decision model, StartLux 27B \citep[see
Appendix B]{zhuang2026startlux}.

Both H1 and H2 replicate in all 10 additional models
(Table~\ref{tbl-other-llms-logit};
Table~\ref{tbl-study2-decision-models}; Fig.~\ref{fig-main-choice};
Fig.~\ref{fig-study2-choice}; full tables in
\quartoapptblref{apptbl-full-claude-sonnet-5},
\quartoapptblref{apptbl-full-gpt-5-6-terra},
\quartoapptblref{apptbl-full-gemini-3-5-flash},
\quartoapptblref{apptbl-full-gemini-3-6-flash},
\quartoapptblref{apptbl-full-gemini-3-8-flash}, and
\quartoapptblref{apptbl-study2-descriptive}). Across all five additional
LLMs, the consumer-delegated sponsorship penalty is negative and
significant (\(\beta\) from \(-0.917\) to \(-5.836\), all \(p < .001\)),
and the Principal \(\times\) Sponsorship interaction is positive and
significant (\(\beta\) from \(0.708\) to \(5.243\), all
\(p \leq .007\)). Comparing the three successive Gemini Flash
generations further indicates that routine model updates do not resolve
sponsorship bias, as the largest bias among the three releases appears
in the newest release (Gemini 3.8 Flash: \(\beta = 5.243\),
vs.~\(\beta = 0.975\) and \(1.124\) in Gemini 3.5 and 3.6 Flash).
Likewise, across all five additional decision models, sponsorship
significantly reduces exact target log-odds under consumer delegation
(\(\beta\) from \(-0.729\) to \(-3.379\), all \(p < .001\)), and the
Principal \(\times\) Sponsorship interaction is positive and significant
in every model (\(\beta\) from \(0.104\) to \(0.641\), all
\(p \le .008\)).

--- insert Table~\ref{tbl-other-llms-logit} about here ---

--- insert Table~\ref{tbl-study2-decision-models} about here ---

We also replicated the PROCESS Model 8 moderated mediation analysis
across the three Gemini Flash generations.\footnote{Claude Sonnet 5 and
  GPT 5.6 Terra do not reliably expose reasoning traces via their public
  APIs.} Both H3 and H4 are supported in all three models
(\quartoapptblref{apptbl-mediation-other};
\quartoapptblref{apptbl-mediation-other-indirect}). Sponsorship
increases skepticism and skepticism is negatively associated with target
choice (\(p < .001\); H3). Platform delegation attenuates first-stage
skepticism (\(b\) from \(-0.533\) to \(-2.165\), \(p < .001\)), yielding
a positive and significant index of moderated mediation in every model
version (H4).

\textbf{System-prompt wording.} We next test whether the findings in
Gemini 3.1 Pro and Kev 9B are driven by specific wording in the main
system prompt (Table~\ref{tbl-principal}; Table~\ref{tbl-main-logit};
Table~\ref{tbl-kev-main}). Specifically, we conducted two independent
\(2 \times 2\) replications across both models to rule out two potential
wording artifacts (\quartoappfigref{appfig-choice-system};
\quartoappfigref{appfig-kev-choice-system};
\quartoapptblref{apptbl-mnl-principal-noun};
\quartoapptblref{apptbl-mnl-recommend};
\quartoapptblref{apptbl-kev-system-descriptive}). First, because ``hotel
booking platform'' shares the word ``platform'' with the ``Platform
Sponsor'' attribute, a \emph{principal-noun} variant replaced
``traveler'' and ``hotel booking platform'' with ``customer'' and
``online travel agency'' to eliminate lexical overlap. Second, an
\emph{action-verb} variant replaced ``book'' with ``recommend'' in the
choice question.

Both H1 and H2 are supported under each wording variant across both
model families (Table~\ref{tbl-main-logit}; Table~\ref{tbl-kev-main}).
The consumer-delegated sponsorship penalty remains negative and
significant in both the principal-noun and action-verb variants for
Gemini 3.1 Pro (\(\beta = -6.703\) and \(-5.074\), both \(p < .001\))
and Kev 9B (\(\beta = -0.106\) and \(-0.152\), both \(p < .001\)). The
Principal \(\times\) Sponsorship interaction remains positive and
significant in every condition (Gemini 3.1 Pro: \(\beta = 2.503\) and
\(4.592\), both \(p < .001\); Kev 9B: \(\beta = 0.044\) and \(0.195\),
both \(p < .001\)).

\textbf{Reasoning effort.} Finally, to test whether the bias in
generative LLMs depends on reasoning effort, we executed the
\(2 \times 2\) design on Gemini 3.1 Pro at its minimum reasoning effort
(\emph{thinking\_level = ``LOW''}, \(N = 2{,}000\)), which reduced the
average reasoning trace length relative to the default HIGH
(\(M = 284.99\) vs.~\(543.92\) tokens). In the pooled model across both
reasoning levels (\(N = 4{,}000\); Table~\ref{tbl-main-logit};
\quartoapptblref{apptbl-thinking-level-choice};
\quartoappfigref{appfig-thinking-level-choice};
\quartoapptblref{apptbl-thinking-level-logit-full}), both H1
(\(\beta = -5.318\), \(SE = 0.307\), \(p < .001\)) and H2
(\(\text{Principal} \times \text{Sponsorship}\) \(\beta = 2.877\),
\(SE = 0.348\), \(p < .001\)) remain supported under LOW thinking level.
The three-way interaction with thinking level is not significant
(\(\beta = -0.842\), \(SE = 0.518\), \(p = .104\)).

\section{Study 2}\label{sec-study2}

While Study 1 established the sponsorship penalty and sponsorship bias
across generative LLMs and decision models, the label used in Study 1
(``Platform Sponsor'') combined a sponsorship disclosure (``Sponsor'')
with explicit source attribution to the platform (``Platform''). To
isolate how each component contributes to the bias, Study 2 crosses
disclosure type (``Promoted'' vs.~``Sponsored''), platform attribution
(no vs.~yes), and principal (consumer vs.~platform) in a
\(2 \times 2 \times 2\) design evaluated across both model architectures
from Study 1 (Gemini 3.1 Pro and Kev 9B).

Study 2 addresses three questions. First, because ``Sponsored'' is an
unambiguous sponsorship disclosure whereas ``Promoted'' is a softer
merchandising tag
\citep{friestad1994persuasion, boerman2012sponsorship, kim2019paradox},
does ``Sponsored'' lower target evaluation relative to ``Promoted''
under both principals? Second, if sponsorship bias reflects alignment
with the employing platform, does attributing the placement to the
platform (e.g., ``Sponsored by the platform'') amplify sponsorship bias
(a positive Principal \(\times\) Attribution interaction)? Third, can
mandating the explicit term ``Sponsored'' mitigate this sponsorship
bias?

\subsection{Design and procedure}\label{design-and-procedure}

We manipulated the target listing's disclosure across four labels
(Table~\ref{tbl-study3-descriptive}): ``Promoted,'' ``Promoted by the
platform,'' ``Sponsored,'' and ``Sponsored by the platform''
(\emph{true} for the target, \emph{false} for all other options).
Following Study 1, we collected 500 independent replications per cell
for Gemini 3.1 Pro (\(N = 4{,}000\) total; analyzed via binary logistic
regression with hotel and review fixed effects and standardized price;
Table~\ref{tbl-study3-regression}; full table in
\quartoapptblref{apptbl-study3-logit-full}) and evaluated the 360 paired
Latin-square stimuli across all eight cells for Kev 9B (\(N = 2{,}880\)
total; analyzed via OLS on exact log-odds
\(\ln(P_{\text{target}} / P_{\text{competitor}})\) with 360 stimulus
fixed effects and stimulus-clustered standard errors;
Table~\ref{tbl-study3-regression}).

--- insert Table~\ref{tbl-study3-descriptive} about here ---

\subsection{Results}\label{results-1}

First, the explicit ``Sponsored'' label strongly suppresses target
evaluation relative to ``Promoted'' in both models
(Table~\ref{tbl-study3-descriptive}; Table~\ref{tbl-study3-regression};
\quartoappfigref{appfig-study3-levels}). Under consumer delegation
without platform attribution, switching from ``Promoted'' to
``Sponsored'' reduces Gemini 3.1 Pro's target choice share from
\(28.8\%\) to \(0.8\%\) (\(\beta_{\text{disclosure}} = -5.640\),
\(SE = 0.554\), \(p < .001\)) and reduces Kev 9B's calibrated target
probability from \(47.26\%\) to \(40.42\%\)
(\(\beta_{\text{disclosure}} = -0.328\), \(SE = 0.015\), \(z = -21.60\),
\(p < .001\)).

\begin{figure}

\centering{

\includegraphics[width=0.55\linewidth,height=\textheight,keepaspectratio]{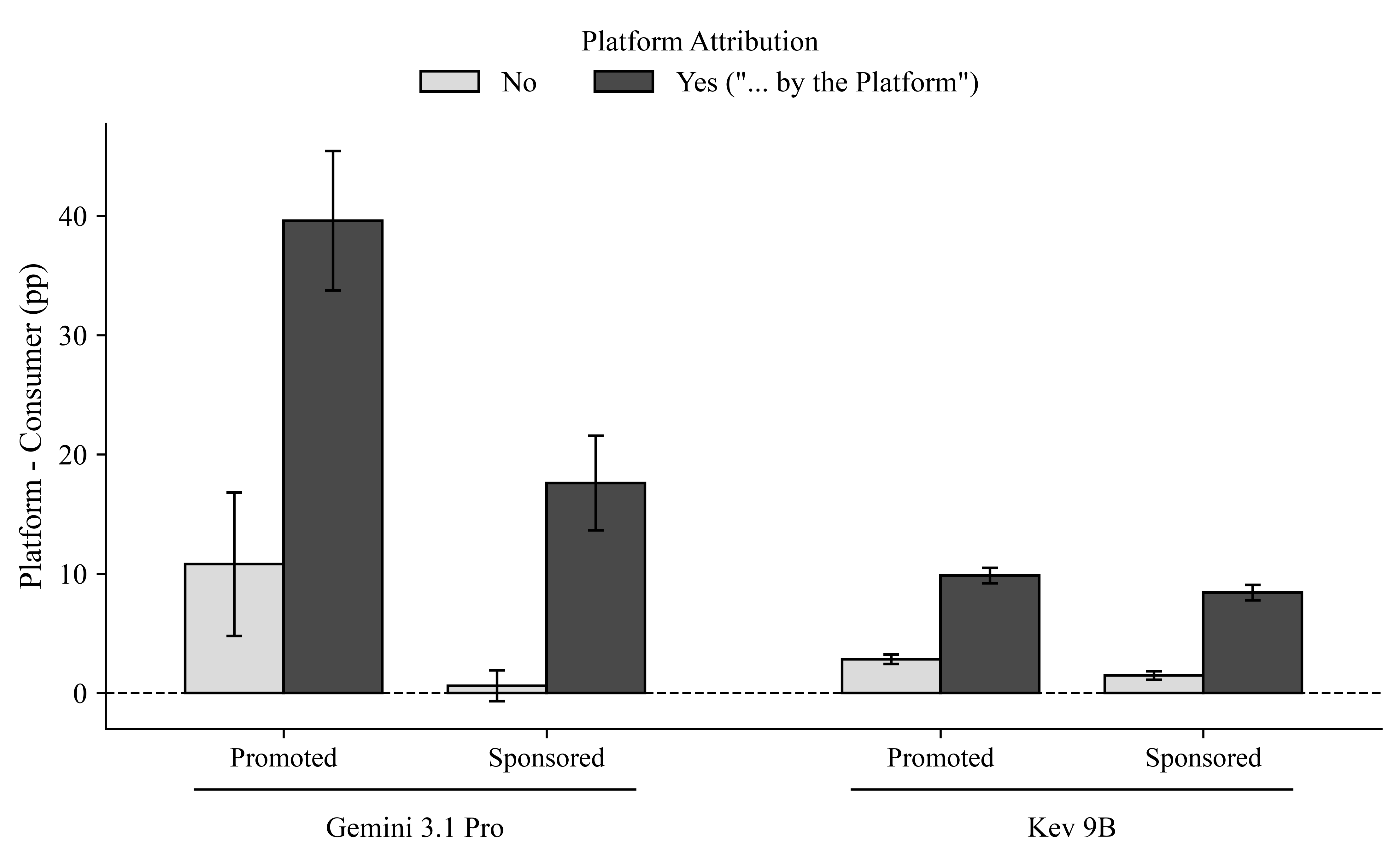}

}

\caption{\label{fig-study3-sponsor-platform}Difference in target
evaluation between platform and consumer delegation (in percentage
points) across disclosure types and platform attribution - Gemini 3.1
Pro and Kev 9B (Study 2).}

\end{figure}

--- insert Fig.~\ref{fig-study3-sponsor-platform} about here ---

Second, platform attribution amplifies the sponsorship bias in both
models (Fig.~\ref{fig-study3-sponsor-platform};
Table~\ref{tbl-study3-regression}). In Gemini 3.1 Pro, appending ``by
the platform'' to the disclosure label increases the sponsorship bias
from \(+10.8\) percentage points (\(39.6\%\) vs.~\(28.8\%\)) to
\(+39.6\) points (\(74.2\%\) vs.~\(34.6\%\)) under ``Promoted,'' and
from \(+0.6\) points (\(1.4\%\) vs.~\(0.8\%\)) to \(+17.6\) points
(\(21.2\%\) vs.~\(3.6\%\)) under ``Sponsored.'' The Principal \(\times\)
Attribution interaction is positive and significant under both
``Promoted'' (\(\beta = 2.344\), \(SE = 0.285\), \(z = 8.24\),
\(p < .001\)) and ``Sponsored'' (\(\beta = 2.276\), \(SE = 0.735\),
\(z = 3.10\), \(p = .002\)).

Kev 9B exhibits a similar pattern in its calibrated probabilities
(Table~\ref{tbl-study3-descriptive}; Table~\ref{tbl-study3-regression};
Fig.~\ref{fig-study3-sponsor-platform}). Appending ``by the platform''
increases the sponsorship bias from \(+2.82\) percentage points
(\(50.08\%\) vs.~\(47.26\%\)) to \(+9.85\) points (\(53.54\%\)
vs.~\(43.70\%\)) under ``Promoted,'' and from \(+1.48\) points
(\(41.89\%\) vs.~\(40.42\%\)) to \(+8.43\) points (\(48.96\%\)
vs.~\(40.53\%\)) under ``Sponsored.'' The Principal \(\times\)
Attribution interaction is positive and significant under both
``Promoted'' (\(\beta = 0.357\), \(SE = 0.012\), \(z = 29.38\),
\(p < .001\)) and ``Sponsored'' (\(\beta = 0.342\), \(SE = 0.012\),
\(z = 28.03\), \(p < .001\)).

Finally, stricter disclosure wording (``Sponsored'' rather than
``Promoted'') does not eliminate the sponsorship bias. The three-way
Principal \(\times\) Disclosure \(\times\) Attribution interaction is
not significant in either Gemini 3.1 Pro (\(\beta = -0.068\),
\(SE = 0.783\), \(p = .930\)) or Kev 9B (\(\beta = -0.015\),
\(SE = 0.009\), \(z = -1.65\), \(p = .099\)).

--- insert Table~\ref{tbl-study3-regression} about here ---

\section{Study 3}\label{sec-study3}

Because stricter disclosure labels do not eliminate sponsorship bias
(Study 2), Study 3 examines whether the bias can be mitigated by
intervening directly on the AI agent so that a platform-delegated agent
evaluates listings from the consumer's perspective. Using Study 1 as the
baseline (Table~\ref{tbl-study2-finetuning};
\quartoapptblref{apptbl-study3-remedies-descriptive}), we investigate an
inference-time remedy (\emph{prompt-level mitigation}) against a
parameter-level remedy (\emph{model-level mitigation}).

\subsection{Prompt-level mitigation}\label{prompt-level-mitigation}

We first evaluated prompt-level mitigation across Gemini 3.1 Pro
(\(N = 500\) per sponsorship cell; \(N = 1{,}000\) total) and Kev 9B
(\(N = 360\) paired Latin-square stimuli per sponsorship cell;
\(N = 720\) total) by appending ``Act in the traveler's best interest''
to the platform system instruction without mentioning sponsorship. We
pooled these observations with Study 1's consumer-delegation as the
reference and platform-delegation as the baseline
(Table~\ref{tbl-study2-finetuning}).

Prompt-level mitigation removes the baseline sponsorship bias in both
models, but aligns the sponsorship penalty with consumer delegation only
in Gemini 3.1 Pro (Table~\ref{tbl-study2-finetuning};
\quartoapptblref{apptbl-study3-remedies-descriptive};
\quartoapptblref{apptbl-mitigation-logit}). In Gemini 3.1 Pro,
prompt-level mitigation aligns the platform-delegated sponsorship
penalty with consumer delegation
(\(\text{Principal} \times \text{Sponsorship}\) \(\beta = 0.117\),
\(SE = 0.493\), \(p = .812\)). In Kev 9B, however, prompt-level
mitigation over-corrects the platform-delegated sponsorship penalty
relative to consumer delegation
(\(\text{Principal} \times \text{Sponsorship}\) \(\beta = -0.038\),
\(SE = 0.009\), \(p < .001\)).

Prompt-level mitigation thus exhibits two limitations. First, as
observed in Kev 9B, prompt-level mitigation can over-correct the
sponsorship penalty. Second, rather than removing only the sponsorship
bias while holding the platform-delegated agent's baseline hotel and
review preferences fixed, prompt-level mitigation perturbs the 36 target
and competitor hotel and review fixed effects relative to the baseline
platform condition, yielding a softmax Kullback-Leibler (KL) divergence
across the fixed effects of \(1.1865\) bits in Gemini 3.1 Pro and
\(0.0026\) bits in Kev 9B.

\subsection{Model-level mitigation}\label{model-level-mitigation}

To overcome these limitations, we applied consumer-aligned fine-tuning
at the parameter level (Table~\ref{tbl-study2-finetuning}) so that Kev
9B reproduces its full 4-option consumer-delegated choice distribution
across principals. To prevent domain leakage, we constructed a synthetic
Latin-square dataset across five non-hotel retail categories (four
training categories: laptops, wireless headphones, coffee makers, and
running shoes, \(N = 2{,}880\); one held-out validation category: air
purifiers, \(N = 720\); see Appendix B). For each retail stimulus \(x\),
we queried frozen Base Kev 9B under consumer delegation to obtain the
reference distribution
\(P_{\text{ref}}(\cdot \mid s_{\text{consumer}}, x)\) and trained Kev 9B
by minimizing KL divergence against
\(P_{\text{ref}}(\cdot \mid s_{\text{consumer}}, x)\) under both
platform and consumer delegation. On the held-out air-purifier
validation set, consumer-aligned fine-tuning reduced mean KL divergence
by \(59.5\%\) (from \(0.0074\) bits at step 0 to \(0.0030\) bits at the
selected checkpoint, step 440), with zero exposure to hotel stimuli.

Evaluated zero-shot on the Study 1 hotel benchmark
(Table~\ref{tbl-study2-finetuning}), consumer-aligned fine-tuning
eliminates sponsorship bias under platform delegation without
over-correcting the sponsorship penalty. Rather over-correcting the
sponsorship penalty, as under prompt-level mitigation, consumer-aligned
fine-tuning applies a calibrated \(2.65\) percentage-point sponsorship
penalty (\(42.36\%\) organic to \(39.71\%\) sponsored;
\(\beta = -0.116\), \(SE = 0.012\), \(z = -10.09\), \(p < .001\)) that
closely matches consumer delegation (\(2.83\) percentage points;
\(\beta = -0.123\), \(SE = 0.011\), \(p < .001\)). Relative to consumer
delegation, both the main effect of principal (\(\beta = -0.003\),
\(SE = 0.009\), \(p = .724\)) and the Principal \(\times\) Sponsorship
interaction (\(\beta = 0.007\), \(SE = 0.005\), \(p = .179\)) are
nonsignificant. Moreover, when comparing the 36 hotel and review fixed
effects against the baseline platform condition, consumer-aligned
fine-tuning reduces the fixed-effects KL divergence in half relative to
prompt-level mitigation (\(0.0013\) vs.~\(0.0026\) bits; and by \(75\%\)
in fixed-effects-predicted choice probabilities, \(0.0004\)
vs.~\(0.0016\) bits). Consumer-aligned fine-tuning thus eliminates
sponsorship bias zero-shot under the standard platform prompt while
preserving the platform-delegated agent's baseline hotel and review
preferences far more closely than prompt-level mitigation.

--- insert Table~\ref{tbl-study2-finetuning} about here ---

\section{Conclusions}\label{conclusions}

This research examined whether an AI shopping agent's evaluation of
sponsored listings depends on whether the agent is delegated by a retail
platform or a consumer. Across six generative LLMs from three providers
and six commercial and open-weight decision models in Study 1, a
consumer-delegated AI agent strongly penalized sponsored listings
relative to organic alternatives (H1), whereas platform delegation
significantly attenuated this sponsorship penalty (H2). Both effects
replicated across alternative system-prompt wordings and low versus high
LLM reasoning effort. In generative LLMs with reasoning traces, the
sponsorship effect operated alongside a temporal moderated mediation
pathway through pre-choice skepticism toward the sponsored listing (H3
and H4). Study 2 showed that stricter disclosure wording (``Sponsored''
vs.~``Promoted'') did not eliminate this sponsorship bias. Study 3
demonstrated that while both prompt-level and model-level mitigation
removed platform leniency toward sponsored listings, consumer-aligned
fine-tuning better preserved baseline non-sponsorship preferences.

These findings have three main implications for theory, practice, and
policy. First, we show that stakeholder alignment in AI shopping agents
can emerge implicitly from role specification alone. Prior research on
multi-stakeholder recommenders assumes that stakeholder weights are set
by an explicit formula \citep{abdollahpouri2020multistakeholder}, while
recent studies on AI agents in e-commerce either fix the user role
\citep{allouah2026your} or explicitly instruct agents to favor
advertisers \citep{wu2026ads}. By contrast, merely naming the platform
as the employing principal caused both generative LLMs and
single-forward-pass decision models to relax their discount on sponsored
listings. For marketplaces that monetize retail media networks while
deploying AI shopping assistants
\citep{jindal2021omnichannel, thomas2025retail, chakraborty2024enhancing},
system-prompt design is therefore a critical governance decision. Even
without instructions to favor advertisers, naming the platform as
principal can steer shoppers toward sponsored listings and expose the
firm to perceived manipulation and betrayal
\citep{saenger2024betrayed, du2021paradoxes}.

Second, our findings extend research on sponsorship disclosure and
persuasion knowledge from human consumers to AI agents
\citep{boerman2017post, friestad1994persuasion, evans2015rethinking, liu2025you, kim2019paradox, sahni2020sponsorship}.
In Study 2, attributing the paid placement to the platform (``by the
platform'') amplified sponsorship bias in both generative LLM and
decision model architectures. Moreover, although the explicit label
``Sponsored'' lowered overall evaluation of paid placements relative to
``Promoted,'' it did not reduce this amplification. Traditional
disclosure mandates designed for human shoppers
\citep{eu2022dsa, ftc2013search} are therefore insufficient when an AI
agent intermediates choice. Regulatory oversight of retail AI assistants
must complement ad-labeling rules with prompt- and model-level audits
across consumer and platform roles.

Third, Study 3 clarifies how role-induced biases in AI shopping agents
can be corrected. In a random utility framework, because the negative
utility weight on sponsorship is finite
(\(\beta_{\text{sponsorship}} > -\infty\)), protecting consumers
requires preserving the consumer-delegated baseline weight
(\(\beta_{\text{sponsorship}}^{\text{consumer}}\)). Under prompt-level
mitigation, instructing the platform-delegated agent to act in the
traveler's best interest removed platform leniency in both model
architectures. In the decision model, however, this instruction
over-corrected the sponsorship penalty, and in both architectures it
perturbed baseline hotel and review preferences. Under model-level
mitigation, consumer-aligned fine-tuning aligned the decision model's
parameters directly with its consumer-delegated choice distribution.
Evaluated zero-shot on hotel choices, fine-tuning eliminated sponsorship
bias without over-correcting the sponsorship penalty. It also reduced
the fixed-effects KL divergence relative to prompt-level mitigation.

This research has several limitations that suggest avenues for future
work. First, our primary evaluation benchmarks focused on hotel booking,
and future research could test zero-shot and fine-tuned evaluation in
categories such as grocery retail or financial advice. Second, while we
compared consumer- and platform-delegated AI agents across twelve
foundation models, future studies could benchmark these AI sponsorship
penalties against human consumers in identical retail interfaces
\citep{stoeber2026ai}. Third, our experiments held price and review
quality equal between the target and competitor listings to isolate the
sponsorship penalty ceteris paribus. Future work could vary price
discounts and review ratings across a grid to quantify the dollar
willingness-to-pay trade-off when a sponsored listing is cheaper or
higher rated. Researchers could also examine how AI agents evaluate
user-generated reviews that vary in persuasiveness or reflect financial
incentives
\citep{golmohammadi2020negative, ravula2023role, luca2016fake, wadi2026careful}.

Finally, two methodological boundaries warrant note. From a
computational standpoint, deference to the employer named in the system
instructions may partly reflect instruction-hierarchy training designed
to prioritize system-level context \citep{wallace2024instruction}.
However, prioritizing the system prompt is orthogonal to our findings
because the prompt contains no instruction about sponsored listings.
Rather, from the principal's identity alone, the AI agent infers the
principal's motives, adjusts its skepticism, and relaxes its sponsorship
penalty. In addition, pre-choice reasoning traces in Study 1 do not
provide direct access to internal neural activations
\citep{chen2025reasoning}. We therefore interpret expressed skepticism
as a temporally prior, human-validated process signal rather than a
causal mechanism (H3 and H4).

\section*{Data availability}\label{data-availability}
\addcontentsline{toc}{section}{Data availability}

The trial-level data, analysis notebooks, stimuli, and experiment code
for all reported results are available in an anonymized repository for
peer review:
https://osf.io/p5ejq/overview?view\_only=7e9a65e9efd7420ba6749f370da45f49

\section*{Declaration of generative AI and AI-assisted technologies in
the manuscript preparation
process}\label{declaration-of-generative-ai-and-ai-assisted-technologies-in-the-manuscript-preparation-process}
\addcontentsline{toc}{section}{Declaration of generative AI and
AI-assisted technologies in the manuscript preparation process}

Generative LLMs and decision models are the object of study in this
research. Their use as experimental subjects and as the reasoning-trace
judge is described in Studies 1 through 3 and Appendix B. Separately,
during the preparation of this work the authors used a generative AI
tool to assist with language editing and to improve readability. After
using this tool, the authors reviewed and edited the content as needed
and take full responsibility for the content of the published article.

\section*{References}\label{references}
\addcontentsline{toc}{section}{References}

\renewcommand{\bibsection}{}
\bibliography{refs/ai-marketing.bib,refs/acl-refs.bib,refs/marketing.bib,refs/others.bib,refs/myrefs.bib,refs/seo.bib,refs/mouselab.bib,refs/sponsorships.bib,refs/ai-bias.bib,refs/regulation.bib,refs/llm-behavior.bib,refs/jbr.bib,refs/decision-models.bib,jbr/editor/editor-refs.bib}

\newpage{}

\emph{Note.} The sponsorship tag reduces target choice (H1), an effect
attenuated under platform delegation (H2). In generative LLMs, this
penalty is associated with a temporal pathway through pre-choice
skepticism (H3), and the assigned principal moderates the first stage of
this pathway (H4), weakening the skepticism the sponsorship disclosure
triggers.

\emph{Note.} Share of replications in which the AI agent selected the
target hotel across 500 replications per cell, by sponsorship condition
and delegating principal (\quartoapptblref{apptbl-descriptive-others}).
Error bars represent 95\% confidence intervals.

\emph{Note.} Mean calibrated probability assigned to the target hotel
across the 360 position-balanced Latin-square stimuli per cell, by
sponsorship condition and delegating principal
(\quartoapptblref{apptbl-study2-descriptive}). Error bars represent 95\%
confidence intervals.

\emph{Note.} Mean difference between platform and consumer delegation
(in percentage points) in target choice share for Gemini 3.1 Pro
(\(N = 500\) pairs per label) and calibrated target probability for Kev
9B (\(N = 360\) paired stimuli per label;
Table~\ref{tbl-study3-descriptive}). Error bars represent 95\%
confidence intervals.

\begin{longtable}[]{@{}
  >{\raggedright\arraybackslash}p{(\linewidth - 4\tabcolsep) * \real{0.1667}}
  >{\raggedright\arraybackslash}p{(\linewidth - 4\tabcolsep) * \real{0.4167}}
  >{\raggedright\arraybackslash}p{(\linewidth - 4\tabcolsep) * \real{0.4167}}@{}}
\caption{System instructions manipulating the principal in Study
1}\label{tbl-principal}\tabularnewline
\toprule\noalign{}
\begin{minipage}[b]{\linewidth}\raggedright
Variant
\end{minipage} & \begin{minipage}[b]{\linewidth}\raggedright
Consumer delegation
\end{minipage} & \begin{minipage}[b]{\linewidth}\raggedright
Platform delegation
\end{minipage} \\
\midrule\noalign{}
\endfirsthead
\toprule\noalign{}
\begin{minipage}[b]{\linewidth}\raggedright
Variant
\end{minipage} & \begin{minipage}[b]{\linewidth}\raggedright
Consumer delegation
\end{minipage} & \begin{minipage}[b]{\linewidth}\raggedright
Platform delegation
\end{minipage} \\
\midrule\noalign{}
\endhead
\bottomrule\noalign{}
\endlastfoot
Main experiment & You have been employed by a \textbf{traveler} to
recommend hotels. Which hotel would you \textbf{book}? & You have been
employed by a \textbf{hotel booking platform} to recommend hotels. Which
hotel would you \textbf{book}? \\
Principal noun & You have been employed by a \textbf{customer} to
recommend hotels. Which hotel would you \textbf{book}? & You have been
employed by an \textbf{online travel agency} to recommend hotels. Which
hotel would you \textbf{book}? \\
Action verb & You have been employed by a \textbf{traveler} to recommend
hotels. Which hotel would you \textbf{recommend}? & You have been
employed by a \textbf{hotel booking platform} to recommend hotels. Which
hotel would you \textbf{recommend}? \\
\end{longtable}

\emph{Note.} Each variant was evaluated in a 2 (Sponsorship) \(\times\)
2 (Principal) design on Gemini 3.1 Pro (\(N = 500\) replications per
cell) and Kev 9B (\(N = 360\) Latin-square stimuli per cell).

\begin{longtable}[]{@{}
  >{\raggedright\arraybackslash}p{(\linewidth - 8\tabcolsep) * \real{0.4444}}
  >{\centering\arraybackslash}p{(\linewidth - 8\tabcolsep) * \real{0.1389}}
  >{\centering\arraybackslash}p{(\linewidth - 8\tabcolsep) * \real{0.1389}}
  >{\centering\arraybackslash}p{(\linewidth - 8\tabcolsep) * \real{0.1389}}
  >{\centering\arraybackslash}p{(\linewidth - 8\tabcolsep) * \real{0.1389}}@{}}
\caption{Logistic regression on target choice - Gemini 3.1 Pro (main
experiment, system-prompt wording variants, and thinking level; Study
1)}\label{tbl-main-logit}\tabularnewline
\toprule\noalign{}
\begin{minipage}[b]{\linewidth}\raggedright
Term
\end{minipage} & \begin{minipage}[b]{\linewidth}\centering
Main experiment
\end{minipage} & \begin{minipage}[b]{\linewidth}\centering
Principal noun
\end{minipage} & \begin{minipage}[b]{\linewidth}\centering
Action verb
\end{minipage} & \begin{minipage}[b]{\linewidth}\centering
Thinking level
\end{minipage} \\
\midrule\noalign{}
\endfirsthead
\toprule\noalign{}
\begin{minipage}[b]{\linewidth}\raggedright
Term
\end{minipage} & \begin{minipage}[b]{\linewidth}\centering
Main experiment
\end{minipage} & \begin{minipage}[b]{\linewidth}\centering
Principal noun
\end{minipage} & \begin{minipage}[b]{\linewidth}\centering
Action verb
\end{minipage} & \begin{minipage}[b]{\linewidth}\centering
Thinking level
\end{minipage} \\
\midrule\noalign{}
\endhead
\bottomrule\noalign{}
\endlastfoot
Principal (platform) & −0.165 (0.198) & −0.352 (0.210) & 0.076 (0.190) &
−0.163 (0.190) \\
Sponsorship (sponsored) & −6.491*** (0.418) & −6.703*** (0.423) &
−5.074*** (0.325) & −6.059*** (0.365) \\
Principal \(\times\) Sponsorship & 3.989*** (0.428) & 2.503*** (0.406) &
4.592*** (0.366) & 3.719*** (0.401) \\
Thinking level (LOW) & --- & --- & --- & −0.254 (0.190) \\
Principal \(\times\) Thinking level & --- & --- & --- & 0.406 (0.270) \\
Sponsorship \(\times\) Thinking level & --- & --- & --- & 0.741
(0.433) \\
Principal \(\times\) Sponsorship \(\times\) Thinking level & --- & --- &
--- & −0.842 (0.518) \\
\textbf{Contrasts} & & & & \\
\emph{Sponsorship effect, consumer (HIGH)} & −6.491*** (0.418) &
−6.703*** (0.423) & −5.074*** (0.325) & −6.059*** (0.365) \\
\emph{Sponsorship effect, platform (HIGH)} & −2.502*** (0.229) &
−4.199*** (0.300) & −0.482** (0.187) & −2.340*** (0.209) \\
\emph{Sponsorship effect, consumer (LOW)} & --- & --- & --- & −5.318***
(0.307) \\
\emph{Sponsorship effect, platform (LOW)} & --- & --- & --- & −2.441***
(0.210) \\
\emph{Principal \(\times\) Sponsorship (LOW)} & --- & --- & --- &
2.877*** (0.348) \\
\(N\) & 2,000 & 2,000 & 2,000 & 4,000 \\
Pseudo \(R^2\) & 0.599 & 0.628 & 0.535 & 0.566 \\
Log-Likelihood & −512.17 & −451.54 & −617.04 & −1104.56 \\
\end{longtable}

\emph{Note.} Standard errors in parentheses. Reference categories:
consumer delegation, organic target, and HIGH thinking level
(\(N = 2{,}000\) for Main experiment, Principal noun, and Action verb;
\(N = 4{,}000\) for Thinking level). Models control for standardized
target price and target/competitor review and hotel fixed effects
(\quartoapptblref{apptbl-main-logit-full},
\quartoapptblref{apptbl-mnl-principal-noun},
\quartoapptblref{apptbl-mnl-recommend}, and
\quartoapptblref{apptbl-thinking-level-logit-full}).\\
*\emph{p} \textless{} .05. **\emph{p} \textless{} .01. ***\emph{p}
\textless{} .001.

\begin{longtable}[]{@{}
  >{\raggedright\arraybackslash}p{(\linewidth - 6\tabcolsep) * \real{0.6053}}
  >{\centering\arraybackslash}p{(\linewidth - 6\tabcolsep) * \real{0.1316}}
  >{\centering\arraybackslash}p{(\linewidth - 6\tabcolsep) * \real{0.1316}}
  >{\centering\arraybackslash}p{(\linewidth - 6\tabcolsep) * \real{0.1316}}@{}}
\caption{Paired stimulus fixed-effects regression on exact log-odds of
target choice - Kev 9B (main experiment and system-prompt wording
variants; Study 1)}\label{tbl-kev-main}\tabularnewline
\toprule\noalign{}
\begin{minipage}[b]{\linewidth}\raggedright
Term
\end{minipage} & \begin{minipage}[b]{\linewidth}\centering
Main experiment
\end{minipage} & \begin{minipage}[b]{\linewidth}\centering
Principal noun
\end{minipage} & \begin{minipage}[b]{\linewidth}\centering
Action verb
\end{minipage} \\
\midrule\noalign{}
\endfirsthead
\toprule\noalign{}
\begin{minipage}[b]{\linewidth}\raggedright
Term
\end{minipage} & \begin{minipage}[b]{\linewidth}\centering
Main experiment
\end{minipage} & \begin{minipage}[b]{\linewidth}\centering
Principal noun
\end{minipage} & \begin{minipage}[b]{\linewidth}\centering
Action verb
\end{minipage} \\
\midrule\noalign{}
\endhead
\bottomrule\noalign{}
\endlastfoot
Principal (platform) & −0.003 (0.013) & −0.001 (0.010) & −0.003
(0.012) \\
Sponsorship (sponsored) & −0.123*** (0.012) & −0.106*** (0.011) &
−0.152*** (0.013) \\
Principal \(\times\) Sponsorship & 0.194*** (0.012) & 0.044*** (0.005) &
0.195*** (0.011) \\
\textbf{Contrasts} & & & \\
\emph{Sponsorship effect under consumer delegation} & −0.123*** (0.012)
& −0.106*** (0.011) & −0.152*** (0.013) \\
\emph{Sponsorship effect under platform delegation} & 0.071*** (0.014) &
−0.062*** (0.012) & 0.042** (0.014) \\
\(N\) & 1,440 & 1,440 & 1,440 \\
\(R^2\) & 0.957 & 0.967 & 0.960 \\
\end{longtable}

\emph{Note.} OLS regressions on exact log-odds
\(\ln(P_{\text{target}} / P_{\text{competitor}})\) with 360 paired
stimulus fixed effects (\(N = 1{,}440\) per variant; \(N = 4{,}320\)
total). Stimulus-clustered standard errors (360 clusters) in
parentheses. Reference categories: consumer delegation and organic
target.\\
*\emph{p} \textless{} .05. **\emph{p} \textless{} .01. ***\emph{p}
\textless{} .001.

\begin{longtable}[]{@{}
  >{\raggedright\arraybackslash}p{(\linewidth - 4\tabcolsep) * \real{0.5818}}
  >{\centering\arraybackslash}p{(\linewidth - 4\tabcolsep) * \real{0.2182}}
  >{\centering\arraybackslash}p{(\linewidth - 4\tabcolsep) * \real{0.2000}}@{}}
\caption{Moderated mediation analysis of sponsorship on target choice
via skepticism - Gemini 3.1 Pro (Study
1)}\label{tbl-mediation}\tabularnewline
\toprule\noalign{}
\begin{minipage}[b]{\linewidth}\raggedright
\end{minipage} & \begin{minipage}[b]{\linewidth}\centering
Skepticism (M)
\end{minipage} & \begin{minipage}[b]{\linewidth}\centering
Target choice (Y)
\end{minipage} \\
\midrule\noalign{}
\endfirsthead
\toprule\noalign{}
\begin{minipage}[b]{\linewidth}\raggedright
\end{minipage} & \begin{minipage}[b]{\linewidth}\centering
Skepticism (M)
\end{minipage} & \begin{minipage}[b]{\linewidth}\centering
Target choice (Y)
\end{minipage} \\
\midrule\noalign{}
\endhead
\bottomrule\noalign{}
\endlastfoot
Principal (platform) & −0.007 (0.081) & −0.289 (0.249) \\
Sponsorship (sponsored) & 4.827*** (0.081) & −1.283* (0.603) \\
Principal \(\times\) Sponsorship & −1.361*** (0.114) & 2.859***
(0.586) \\
Skepticism & --- & −1.670*** (0.121) \\
\textbf{Conditional effects of sponsorship} & & \\
\emph{Consumer delegation} & 4.827*** (0.081) & −1.283* (0.603) \\
\emph{Platform delegation} & 3.466*** (0.081) & 1.576*** (0.389) \\
\textbf{Conditional indirect effects and index} & \textbf{Effect (Boot
\emph{SE})} & \textbf{95\% CI} \\
\emph{Consumer delegation} & −8.061 (0.846) & {[}−10.636, −7.309{]} \\
\emph{Platform delegation} & −5.788 (0.624) & {[}−7.660, −5.202{]} \\
\emph{Index of moderated mediation} & 2.273 (0.318) & {[}1.916,
3.135{]} \\
\(N\) & 2,000 & 2,000 \\
(Pseudo) \(R^2\) & 0.751 & 0.760 \\
\end{longtable}

\emph{Note.} \(N = 2{,}000\). PROCESS Model 8
\citep{hayes2017introduction}; unstandardized coefficients with standard
errors in parentheses. Skepticism (M) is estimated by OLS; target choice
(Y) by logistic regression (log-odds). Both equations control for
standardized target price and target/competitor hotel and review fixed
effects (\quartoapptblref{apptbl-process-full-skep} and
\quartoapptblref{apptbl-process-full-choice}). Indirect effects and the
index of moderated mediation use 10,000 percentile bootstrap
resamples.\\
*\emph{p} \textless{} .05. **\emph{p} \textless{} .01. ***\emph{p}
\textless{} .001.

\begin{longtable}[]{@{}
  >{\raggedright\arraybackslash}p{(\linewidth - 10\tabcolsep) * \real{0.3243}}
  >{\centering\arraybackslash}p{(\linewidth - 10\tabcolsep) * \real{0.1351}}
  >{\centering\arraybackslash}p{(\linewidth - 10\tabcolsep) * \real{0.1351}}
  >{\centering\arraybackslash}p{(\linewidth - 10\tabcolsep) * \real{0.1351}}
  >{\centering\arraybackslash}p{(\linewidth - 10\tabcolsep) * \real{0.1351}}
  >{\centering\arraybackslash}p{(\linewidth - 10\tabcolsep) * \real{0.1351}}@{}}
\caption{Logistic regression on target choice - Cross-LLM
generalizability (Study 1)}\label{tbl-other-llms-logit}\tabularnewline
\toprule\noalign{}
\begin{minipage}[b]{\linewidth}\raggedright
\end{minipage} & \begin{minipage}[b]{\linewidth}\centering
GPT 5.6 Terra
\end{minipage} & \begin{minipage}[b]{\linewidth}\centering
Claude Sonnet 5
\end{minipage} & \begin{minipage}[b]{\linewidth}\centering
Gemini 3.5 Flash
\end{minipage} & \begin{minipage}[b]{\linewidth}\centering
Gemini 3.6 Flash
\end{minipage} & \begin{minipage}[b]{\linewidth}\centering
Gemini 3.8 Flash
\end{minipage} \\
\midrule\noalign{}
\endfirsthead
\toprule\noalign{}
\begin{minipage}[b]{\linewidth}\raggedright
\end{minipage} & \begin{minipage}[b]{\linewidth}\centering
GPT 5.6 Terra
\end{minipage} & \begin{minipage}[b]{\linewidth}\centering
Claude Sonnet 5
\end{minipage} & \begin{minipage}[b]{\linewidth}\centering
Gemini 3.5 Flash
\end{minipage} & \begin{minipage}[b]{\linewidth}\centering
Gemini 3.6 Flash
\end{minipage} & \begin{minipage}[b]{\linewidth}\centering
Gemini 3.8 Flash
\end{minipage} \\
\midrule\noalign{}
\endhead
\bottomrule\noalign{}
\endlastfoot
Principal (platform) & −0.345 (0.177) & 0.097 (0.157) & 0.008 (0.200) &
−0.011 (0.221) & −0.119 (0.219) \\
Sponsorship (sponsored) & −2.238*** (0.203) & −3.243*** (0.217) &
−1.824*** (0.217) & −0.917*** (0.226) & −5.836*** (0.376) \\
Principal \(\times\) Sponsorship & 0.708** (0.264) & 1.759*** (0.262) &
0.975*** (0.285) & 1.124*** (0.318) & 5.243*** (0.428) \\
\textbf{Contrasts} & & & & & \\
\emph{Sponsorship effect under consumer delegation} & −2.238*** (0.203)
& −3.243*** (0.217) & −1.824*** (0.217) & −0.917*** (0.226) & −5.836***
(0.376) \\
\emph{Sponsorship effect under platform delegation} & −1.530*** (0.189)
& −1.483*** (0.170) & −0.849*** (0.195) & 0.207 (0.222) & −0.593**
(0.225) \\
\(N\) & 2,000 & 2,000 & 2,000 & 2,000 & 2,000 \\
Pseudo \(R^2\) & 0.443 & 0.348 & 0.519 & 0.610 & 0.643 \\
Log-Likelihood & −748.58 & −845.00 & −661.59 & −540.43 & −479.44 \\
\end{longtable}

\emph{Note.} Standard errors in parentheses (\(N = 2,000\) per LLM;
\(N = 10,000\) total). Models control for standardized target price and
target/competitor review and hotel fixed effects
(\quartoapptblref{apptbl-full-claude-sonnet-5},
\quartoapptblref{apptbl-full-gpt-5-6-terra},
\quartoapptblref{apptbl-full-gemini-3-5-flash},
\quartoapptblref{apptbl-full-gemini-3-6-flash}, and
\quartoapptblref{apptbl-full-gemini-3-8-flash}).\\
*\emph{p} \textless{} .05. **\emph{p} \textless{} .01. ***\emph{p}
\textless{} .001.

\begin{longtable}[]{@{}
  >{\raggedright\arraybackslash}p{(\linewidth - 10\tabcolsep) * \real{0.3750}}
  >{\centering\arraybackslash}p{(\linewidth - 10\tabcolsep) * \real{0.1250}}
  >{\centering\arraybackslash}p{(\linewidth - 10\tabcolsep) * \real{0.1250}}
  >{\centering\arraybackslash}p{(\linewidth - 10\tabcolsep) * \real{0.1250}}
  >{\centering\arraybackslash}p{(\linewidth - 10\tabcolsep) * \real{0.1250}}
  >{\centering\arraybackslash}p{(\linewidth - 10\tabcolsep) * \real{0.1250}}@{}}
\caption{Paired stimulus fixed-effects regression on exact log-odds of
target choice - Cross-decision-model generalizability (Study
1)}\label{tbl-study2-decision-models}\tabularnewline
\toprule\noalign{}
\begin{minipage}[b]{\linewidth}\raggedright
\end{minipage} & \begin{minipage}[b]{\linewidth}\centering
Jev 1.14
\end{minipage} & \begin{minipage}[b]{\linewidth}\centering
Liquid D1-Ultra
\end{minipage} & \begin{minipage}[b]{\linewidth}\centering
Mercury Decide
\end{minipage} & \begin{minipage}[b]{\linewidth}\centering
Microsoft Decision 1
\end{minipage} & \begin{minipage}[b]{\linewidth}\centering
StartLux 27B
\end{minipage} \\
\midrule\noalign{}
\endfirsthead
\toprule\noalign{}
\begin{minipage}[b]{\linewidth}\raggedright
\end{minipage} & \begin{minipage}[b]{\linewidth}\centering
Jev 1.14
\end{minipage} & \begin{minipage}[b]{\linewidth}\centering
Liquid D1-Ultra
\end{minipage} & \begin{minipage}[b]{\linewidth}\centering
Mercury Decide
\end{minipage} & \begin{minipage}[b]{\linewidth}\centering
Microsoft Decision 1
\end{minipage} & \begin{minipage}[b]{\linewidth}\centering
StartLux 27B
\end{minipage} \\
\midrule\noalign{}
\endhead
\bottomrule\noalign{}
\endlastfoot
Principal (platform) & −0.026 (0.022) & 0.004 (0.015) & −0.012 (0.012) &
0.007 (0.013) & −0.003 (0.010) \\
Sponsorship (sponsored) & −2.357*** (0.064) & −1.760*** (0.036) &
−3.379*** (0.073) & −0.729*** (0.026) & −1.307*** (0.037) \\
Principal \(\times\) Sponsorship & 0.641*** (0.040) & 0.104*** (0.031) &
0.121** (0.046) & 0.111*** (0.014) & 0.108*** (0.025) \\
\textbf{Contrasts} & & & & & \\
\emph{Sponsorship effect under consumer delegation} & −2.357*** (0.064)
& −1.760*** (0.036) & −3.379*** (0.073) & −0.729*** (0.026) & −1.307***
(0.037) \\
\emph{Sponsorship effect under platform delegation} & −1.716*** (0.068)
& −1.656*** (0.034) & −3.258*** (0.074) & −0.617*** (0.023) & −1.199***
(0.030) \\
\(N\) & 1,440 & 1,440 & 1,440 & 1,440 & 1,440 \\
\(R^2\) & 0.891 & 0.927 & 0.896 & 0.946 & 0.906 \\
\end{longtable}

\emph{Note.} OLS regressions on exact log-odds
\(\ln(P_{\text{target}} / P_{\text{competitor}})\) with 360 paired
stimulus fixed effects (\(N = 1{,}440\) per decision model;
\(N = 7{,}200\) total). Stimulus-clustered standard errors (360
clusters) in parentheses. Reference categories: consumer delegation and
organic target.\\
*\emph{p} \textless{} .05. **\emph{p} \textless{} .01. ***\emph{p}
\textless{} .001.

\begin{longtable}[]{@{}
  >{\raggedright\arraybackslash}p{(\linewidth - 12\tabcolsep) * \real{0.1299}}
  >{\raggedright\arraybackslash}p{(\linewidth - 12\tabcolsep) * \real{0.1494}}
  >{\raggedright\arraybackslash}p{(\linewidth - 12\tabcolsep) * \real{0.1948}}
  >{\centering\arraybackslash}p{(\linewidth - 12\tabcolsep) * \real{0.1234}}
  >{\centering\arraybackslash}p{(\linewidth - 12\tabcolsep) * \real{0.1234}}
  >{\centering\arraybackslash}p{(\linewidth - 12\tabcolsep) * \real{0.1234}}
  >{\centering\arraybackslash}p{(\linewidth - 12\tabcolsep) * \real{0.1234}}@{}}
\caption{Target choice share (Gemini 3.1 Pro) and calibrated target
probability (Kev 9B) across disclosure types and platform attribution
(Study 2)}\label{tbl-study3-descriptive}\tabularnewline
\toprule\noalign{}
\multirow{2}{=}{\begin{minipage}[b]{\linewidth}\raggedright
Disclosure type
\end{minipage}} &
\multirow{2}{=}{\begin{minipage}[b]{\linewidth}\raggedright
Platform attribution
\end{minipage}} &
\multirow{2}{=}{\begin{minipage}[b]{\linewidth}\raggedright
Disclosure label
\end{minipage}} &
\multicolumn{2}{>{\centering\arraybackslash}p{(\linewidth - 12\tabcolsep) * \real{0.2468} + 2\tabcolsep}}{%
\begin{minipage}[b]{\linewidth}\centering
Gemini 3.1 Pro
\end{minipage}} &
\multicolumn{2}{>{\centering\arraybackslash}p{(\linewidth - 12\tabcolsep) * \real{0.2468} + 2\tabcolsep}@{}}{%
\begin{minipage}[b]{\linewidth}\centering
Kev 9B
\end{minipage}} \\
& & & \begin{minipage}[b]{\linewidth}\centering
Consumer
\end{minipage} & \begin{minipage}[b]{\linewidth}\centering
Platform
\end{minipage} & \begin{minipage}[b]{\linewidth}\centering
Consumer
\end{minipage} & \begin{minipage}[b]{\linewidth}\centering
Platform
\end{minipage} \\
\midrule\noalign{}
\endfirsthead
\toprule\noalign{}
\multirow{2}{=}{\begin{minipage}[b]{\linewidth}\raggedright
Disclosure type
\end{minipage}} &
\multirow{2}{=}{\begin{minipage}[b]{\linewidth}\raggedright
Platform attribution
\end{minipage}} &
\multirow{2}{=}{\begin{minipage}[b]{\linewidth}\raggedright
Disclosure label
\end{minipage}} &
\multicolumn{2}{>{\centering\arraybackslash}p{(\linewidth - 12\tabcolsep) * \real{0.2468} + 2\tabcolsep}}{%
\begin{minipage}[b]{\linewidth}\centering
Gemini 3.1 Pro
\end{minipage}} &
\multicolumn{2}{>{\centering\arraybackslash}p{(\linewidth - 12\tabcolsep) * \real{0.2468} + 2\tabcolsep}@{}}{%
\begin{minipage}[b]{\linewidth}\centering
Kev 9B
\end{minipage}} \\
& & & \begin{minipage}[b]{\linewidth}\centering
Consumer
\end{minipage} & \begin{minipage}[b]{\linewidth}\centering
Platform
\end{minipage} & \begin{minipage}[b]{\linewidth}\centering
Consumer
\end{minipage} & \begin{minipage}[b]{\linewidth}\centering
Platform
\end{minipage} \\
\midrule\noalign{}
\endhead
\bottomrule\noalign{}
\endlastfoot
Promoted & No & ``Promoted'' & 28.80\% (45.33) & 39.60\% (48.96) &
47.26\% (14.19) & 50.08\% (13.75) \\
Promoted & Yes & ``Promoted by the platform'' & 34.60\% (47.62) &
74.20\% (43.80) & 43.70\% (13.75) & 53.54\% (13.32) \\
Sponsored & No & ``Sponsored'' & 0.80\% (8.92) & 1.40\% (11.76) &
40.42\% (12.50) & 41.89\% (11.81) \\
Sponsored & Yes & ``Sponsored by the platform'' & 3.60\% (18.65) &
21.20\% (40.91) & 40.53\% (12.89) & 48.96\% (12.65) \\
\end{longtable}

\emph{Note.} Gemini 3.1 Pro: \(N = 500\) replications per cell
(\(N = 4{,}000\) total), reporting target choice shares (SDs in
parentheses). Kev 9B: \(N = 360\) Latin-square stimuli per cell
(\(N = 2{,}880\) total), reporting mean calibrated target probabilities
(SDs in parentheses).

\begin{longtable}[]{@{}
  >{\raggedright\arraybackslash}p{(\linewidth - 4\tabcolsep) * \real{0.6970}}
  >{\centering\arraybackslash}p{(\linewidth - 4\tabcolsep) * \real{0.1515}}
  >{\centering\arraybackslash}p{(\linewidth - 4\tabcolsep) * \real{0.1515}}@{}}
\caption{Regression results for disclosure type, platform attribution,
and principal manipulation - Gemini 3.1 Pro and Kev 9B (Study
2)}\label{tbl-study3-regression}\tabularnewline
\toprule\noalign{}
\begin{minipage}[b]{\linewidth}\raggedright
\end{minipage} & \begin{minipage}[b]{\linewidth}\centering
Gemini 3.1 Pro
\end{minipage} & \begin{minipage}[b]{\linewidth}\centering
Kev 9B
\end{minipage} \\
\midrule\noalign{}
\endfirsthead
\toprule\noalign{}
\begin{minipage}[b]{\linewidth}\raggedright
\end{minipage} & \begin{minipage}[b]{\linewidth}\centering
Gemini 3.1 Pro
\end{minipage} & \begin{minipage}[b]{\linewidth}\centering
Kev 9B
\end{minipage} \\
\midrule\noalign{}
\endhead
\bottomrule\noalign{}
\endlastfoot
Principal (platform) & 1.145*** (0.194) & 0.172*** (0.011) \\
Disclosure (``Sponsored'') & −5.640*** (0.554) & −0.328*** (0.015) \\
Attribution (``by the platform'') & 0.429* (0.190) & −0.157***
(0.010) \\
Principal \(\times\) Disclosure & −0.289 (0.688) & −0.074*** (0.005) \\
Disclosure \(\times\) Attribution & 1.360* (0.620) & 0.160*** (0.010) \\
Principal \(\times\) Attribution & 2.344*** (0.285) & 0.357***
(0.012) \\
Principal \(\times\) Disclosure \(\times\) Attribution & −0.068 (0.783)
& −0.015 (0.009) \\
\textbf{Contrasts} & & \\
\emph{Principal \(\times\) Attribution (Promoted)} & 2.344*** (0.285) &
0.357*** (0.012) \\
\emph{Principal \(\times\) Attribution (Sponsored)} & 2.276** (0.735) &
0.342*** (0.012) \\
\(N\) & 4,000 & 2,880 \\
(Pseudo) \(R^2\) & 0.585 & 0.933 \\
Log-Likelihood & −941.85 & --- \\
\end{longtable}

\emph{Note.} Standard errors in parentheses (clustered by stimulus for
Kev 9B). Reference categories: consumer delegation, ``Promoted''
disclosure, and no platform attribution. Gemini 3.1 Pro controls for
standardized target price and target/competitor hotel and review fixed
effects (\quartoapptblref{apptbl-study3-logit-full}); Kev 9B includes
360 paired stimulus fixed effects.\\
*\emph{p} \textless{} .05. **\emph{p} \textless{} .01. ***\emph{p}
\textless{} .001.

\begin{longtable}[]{@{}
  >{\raggedright\arraybackslash}p{(\linewidth - 6\tabcolsep) * \real{0.3936}}
  >{\raggedright\arraybackslash}p{(\linewidth - 6\tabcolsep) * \real{0.3404}}
  >{\centering\arraybackslash}p{(\linewidth - 6\tabcolsep) * \real{0.1383}}
  >{\centering\arraybackslash}p{(\linewidth - 6\tabcolsep) * \real{0.1277}}@{}}
\caption{Regression results for prompt-level mitigation and
consumer-aligned fine-tuning on the hotel benchmark - Gemini 3.1 Pro and
Kev 9B (Study 3)}\label{tbl-study2-finetuning}\tabularnewline
\toprule\noalign{}
\begin{minipage}[b]{\linewidth}\raggedright
Platform condition
\end{minipage} & \begin{minipage}[b]{\linewidth}\raggedright
Term
\end{minipage} & \begin{minipage}[b]{\linewidth}\centering
Gemini 3.1 Pro
\end{minipage} & \begin{minipage}[b]{\linewidth}\centering
Kev 9B
\end{minipage} \\
\midrule\noalign{}
\endfirsthead
\toprule\noalign{}
\begin{minipage}[b]{\linewidth}\raggedright
Platform condition
\end{minipage} & \begin{minipage}[b]{\linewidth}\raggedright
Term
\end{minipage} & \begin{minipage}[b]{\linewidth}\centering
Gemini 3.1 Pro
\end{minipage} & \begin{minipage}[b]{\linewidth}\centering
Kev 9B
\end{minipage} \\
\midrule\noalign{}
\endhead
\bottomrule\noalign{}
\endlastfoot
Baseline & Sponsorship (sponsored) & −2.708*** (0.233) & 0.071***
(0.013) \\
& Principal (platform) & −0.163 (0.203) & −0.003 (0.012) \\
& Principal \(\times\) Sponsorship & 4.204*** (0.437) & 0.194***
(0.011) \\
Prompt-level mitigation & Sponsorship (sponsored) & −6.795*** (0.412) &
−0.161*** (0.014) \\
& Principal (platform) & −0.169 (0.205) & 0.002 (0.013) \\
& Principal \(\times\) Sponsorship & 0.117 (0.493) & −0.038***
(0.009) \\
Consumer-aligned fine-tuning & Sponsorship (sponsored) & --- & −0.116***
(0.012) \\
& Principal (platform) & --- & −0.003 (0.009) \\
& Principal \(\times\) Sponsorship & --- & 0.007 (0.005) \\
\(N\) & & 3,000 & 2,880 \\
(Pseudo) \(R^2\) & & 0.639 & 0.948 \\
Log-Likelihood & & −675.85 & --- \\
\end{longtable}

\emph{Note.} Standard errors in parentheses (clustered by stimulus for
Kev 9B). Reference categories: \emph{Consumer} delegation and organic
target. Within each platform condition, \emph{Sponsorship (sponsored)}
reports the conditional sponsorship effect within that condition, while
\emph{Principal (platform)} and \emph{Principal \(\times\) Sponsorship}
report the main effect and interaction differences relative to the
\emph{Consumer} reference condition (baseline consumer sponsorship
effect: −6.912*** (0.420) for Gemini 3.1 Pro; −0.123*** (0.011) for Kev
9B). Gemini 3.1 Pro: pooled 3-condition fixed-effects logistic
regression on target choice controlling for standardized target price
and target/competitor hotel and review fixed effects (\(N = 500\)
replications per cell; \(N = 1{,}000\) per condition; \(N = 3{,}000\)
total in \quartoapptblref{apptbl-mitigation-logit}). Kev 9B: pooled
4-condition OLS regression on exact log-odds
\(\ln(P_{\text{target}} / P_{\text{competitor}})\) with 360 paired Study
1 hotel Latin-square stimulus fixed effects (\(N = 360\) paired stimuli
per cell; \(N = 720\) per condition; \(N = 2{,}880\) total).
Consumer-aligned fine-tuned Kev 9B was trained strictly on four
synthetic non-hotel retail categories (\(N = 2{,}880\)) and selected at
minimum validation KL divergence (\(0.0030\) bits vs.~\(0.0074\) bits
pre-training, step 440) on a held-out retail category (air purifiers,
\(N = 720\)), with zero exposure to hotel stimuli.\\
*\emph{p} \textless{} .05. **\emph{p} \textless{} .01. ***\emph{p}
\textless{} .001.

\newpage{}

\section*{Appendix A. Supplementary tables and
figures}\label{sec-appendix-b}
\addcontentsline{toc}{section}{Appendix A. Supplementary tables and
figures}

\subsection*{A.1 Study 1 - main}\label{study-1---main}
\addcontentsline{toc}{subsection}{A.1 Study 1 - main}

{\def\LTcaptype{apptbl}
\begin{longtable}[]{@{}
  >{\raggedright\arraybackslash}p{(\linewidth - 8\tabcolsep) * \real{0.1600}}
  >{\raggedright\arraybackslash}p{(\linewidth - 8\tabcolsep) * \real{0.2800}}
  >{\raggedright\arraybackslash}p{(\linewidth - 8\tabcolsep) * \real{0.1600}}
  >{\raggedright\arraybackslash}p{(\linewidth - 8\tabcolsep) * \real{0.2800}}
  >{\raggedright\arraybackslash}p{(\linewidth - 8\tabcolsep) * \real{0.1200}}@{}}
\caption{{\label{apptbl-study1-alternatives}}Alternatives used in Study
1}\tabularnewline
\toprule\noalign{}
\begin{minipage}[b]{\linewidth}\raggedright
Alternative
\end{minipage} & \begin{minipage}[b]{\linewidth}\raggedright
Featured Review Text
\end{minipage} & \begin{minipage}[b]{\linewidth}\raggedright
Hotel Name
\end{minipage} & \begin{minipage}[b]{\linewidth}\raggedright
Platform Sponsor
\end{minipage} & \begin{minipage}[b]{\linewidth}\raggedright
Price
\end{minipage} \\
\midrule\noalign{}
\endfirsthead
\toprule\noalign{}
\begin{minipage}[b]{\linewidth}\raggedright
Alternative
\end{minipage} & \begin{minipage}[b]{\linewidth}\raggedright
Featured Review Text
\end{minipage} & \begin{minipage}[b]{\linewidth}\raggedright
Hotel Name
\end{minipage} & \begin{minipage}[b]{\linewidth}\raggedright
Platform Sponsor
\end{minipage} & \begin{minipage}[b]{\linewidth}\raggedright
Price
\end{minipage} \\
\midrule\noalign{}
\endhead
\bottomrule\noalign{}
\endlastfoot
Target & \emph{{[}Positive review pool{]}} & \emph{{[}Randomized{]}} &
\emph{{[}Manipulated: true or false{]}} & \(P\) \\
Competitor & \emph{{[}Positive review pool{]}} & \emph{{[}Randomized{]}}
& false & \(P\) \\
Dominated 1 & \emph{{[}Negative review pool{]}} &
\emph{{[}Randomized{]}} & false & \$450 \\
Dominated 2 & \emph{{[}Negative review pool{]}} &
\emph{{[}Randomized{]}} & false & \$400 \\
\end{longtable}
}

\emph{Note.} \(P\) is identical for both Target and Competitor, drawn
randomly from a continuous uniform distribution
\(P \sim \mathcal{U}(p_5, p_{95})\), where \(p_5 = \$173.45\) and
\(p_{95} = \$351.40\).

\subsection*{A.2 Study 1 - Cross-model
generalizability}\label{study-1---cross-model-generalizability}
\addcontentsline{toc}{subsection}{A.2 Study 1 - Cross-model
generalizability}

{\def\LTcaptype{apptbl}
\begin{longtable}[]{@{}
  >{\raggedright\arraybackslash}p{(\linewidth - 10\tabcolsep) * \real{0.1978}}
  >{\raggedright\arraybackslash}p{(\linewidth - 10\tabcolsep) * \real{0.1429}}
  >{\raggedright\arraybackslash}p{(\linewidth - 10\tabcolsep) * \real{0.1648}}
  >{\raggedright\arraybackslash}p{(\linewidth - 10\tabcolsep) * \real{0.1868}}
  >{\raggedright\arraybackslash}p{(\linewidth - 10\tabcolsep) * \real{0.2308}}
  >{\raggedleft\arraybackslash}p{(\linewidth - 10\tabcolsep) * \real{0.0769}}@{}}
\caption{{\label{apptbl-descriptive-others}}Choice distribution across
all six generative LLMs (Study 1)}\tabularnewline
\toprule\noalign{}
\begin{minipage}[b]{\linewidth}\raggedright
LLM
\end{minipage} & \begin{minipage}[b]{\linewidth}\raggedright
Principal
\end{minipage} & \begin{minipage}[b]{\linewidth}\raggedright
Sponsorship
\end{minipage} & \begin{minipage}[b]{\linewidth}\raggedright
Target chosen
\end{minipage} & \begin{minipage}[b]{\linewidth}\raggedright
Competitor chosen
\end{minipage} & \begin{minipage}[b]{\linewidth}\raggedleft
\(N\)
\end{minipage} \\
\midrule\noalign{}
\endfirsthead
\toprule\noalign{}
\begin{minipage}[b]{\linewidth}\raggedright
LLM
\end{minipage} & \begin{minipage}[b]{\linewidth}\raggedright
Principal
\end{minipage} & \begin{minipage}[b]{\linewidth}\raggedright
Sponsorship
\end{minipage} & \begin{minipage}[b]{\linewidth}\raggedright
Target chosen
\end{minipage} & \begin{minipage}[b]{\linewidth}\raggedright
Competitor chosen
\end{minipage} & \begin{minipage}[b]{\linewidth}\raggedleft
\(N\)
\end{minipage} \\
\midrule\noalign{}
\endhead
\bottomrule\noalign{}
\endlastfoot
Gemini 3.1 Pro & Consumer & Organic & 265 (53.0\%) & 235 (47.0\%) &
500 \\
Gemini 3.1 Pro & Consumer & Sponsored & 14 (2.8\%) & 486 (97.2\%) &
500 \\
Gemini 3.1 Pro & Platform & Organic & 271 (54.2\%) & 229 (45.8\%) &
500 \\
Gemini 3.1 Pro & Platform & Sponsored & 125 (25.0\%) & 375 (75.0\%) &
500 \\
Claude Sonnet 5 & Consumer & Organic & 252 (50.4\%) & 248 (49.6\%) &
500 \\
Claude Sonnet 5 & Consumer & Sponsored & 51 (10.2\%) & 449 (89.8\%) &
500 \\
Claude Sonnet 5 & Platform & Organic & 252 (50.4\%) & 248 (49.6\%) &
500 \\
Claude Sonnet 5 & Platform & Sponsored & 146 (29.2\%) & 354 (70.8\%) &
500 \\
GPT 5.6 Terra & Consumer & Organic & 268 (53.6\%) & 232 (46.4\%) &
500 \\
GPT 5.6 Terra & Consumer & Sponsored & 124 (24.8\%) & 376 (75.2\%) &
500 \\
GPT 5.6 Terra & Platform & Organic & 248 (49.6\%) & 252 (50.4\%) &
500 \\
GPT 5.6 Terra & Platform & Sponsored & 154 (30.8\%) & 346 (69.2\%) &
500 \\
Gemini 3.5 Flash & Consumer & Organic & 259 (51.8\%) & 241 (48.2\%) &
500 \\
Gemini 3.5 Flash & Consumer & Sponsored & 168 (33.6\%) & 332 (66.4\%) &
500 \\
Gemini 3.5 Flash & Platform & Organic & 257 (51.4\%) & 243 (48.6\%) &
500 \\
Gemini 3.5 Flash & Platform & Sponsored & 215 (43.0\%) & 285 (57.0\%) &
500 \\
Gemini 3.6 Flash & Consumer & Organic & 238 (47.6\%) & 262 (52.4\%) &
500 \\
Gemini 3.6 Flash & Consumer & Sponsored & 209 (41.8\%) & 291 (58.2\%) &
500 \\
Gemini 3.6 Flash & Platform & Organic & 249 (49.8\%) & 251 (50.2\%) &
500 \\
Gemini 3.6 Flash & Platform & Sponsored & 257 (51.4\%) & 243 (48.6\%) &
500 \\
Gemini 3.8 Flash & Consumer & Organic & 251 (50.2\%) & 249 (49.8\%) &
500 \\
Gemini 3.8 Flash & Consumer & Sponsored & 55 (11.0\%) & 445 (89.0\%) &
500 \\
Gemini 3.8 Flash & Platform & Organic & 262 (52.4\%) & 238 (47.6\%) &
500 \\
Gemini 3.8 Flash & Platform & Sponsored & 221 (44.2\%) & 279 (55.8\%) &
500 \\
\end{longtable}
}

\emph{Note.} \(N = 2{,}000\) per LLM (\(N = 12{,}000\) total). Frequency
with percentage in parentheses. Dominated alternatives were never
chosen.

{\def\LTcaptype{apptbl}
\begin{longtable}[]{@{}
  >{\raggedright\arraybackslash}p{(\linewidth - 8\tabcolsep) * \real{0.2056}}
  >{\raggedright\arraybackslash}p{(\linewidth - 8\tabcolsep) * \real{0.1215}}
  >{\raggedright\arraybackslash}p{(\linewidth - 8\tabcolsep) * \real{0.1963}}
  >{\raggedright\arraybackslash}p{(\linewidth - 8\tabcolsep) * \real{0.2150}}
  >{\raggedright\arraybackslash}p{(\linewidth - 8\tabcolsep) * \real{0.2617}}@{}}
\caption{{\label{apptbl-study2-descriptive}}Calibrated target
probabilities by principal and sponsorship across decision models (Study
1)}\tabularnewline
\toprule\noalign{}
\begin{minipage}[b]{\linewidth}\raggedright
Decision model
\end{minipage} & \begin{minipage}[b]{\linewidth}\raggedright
Principal
\end{minipage} & \begin{minipage}[b]{\linewidth}\raggedright
Organic P(Target)
\end{minipage} & \begin{minipage}[b]{\linewidth}\raggedright
Sponsored P(Target)
\end{minipage} & \begin{minipage}[b]{\linewidth}\raggedright
Sponsorship penalty (pp)
\end{minipage} \\
\midrule\noalign{}
\endfirsthead
\toprule\noalign{}
\begin{minipage}[b]{\linewidth}\raggedright
Decision model
\end{minipage} & \begin{minipage}[b]{\linewidth}\raggedright
Principal
\end{minipage} & \begin{minipage}[b]{\linewidth}\raggedright
Organic P(Target)
\end{minipage} & \begin{minipage}[b]{\linewidth}\raggedright
Sponsored P(Target)
\end{minipage} & \begin{minipage}[b]{\linewidth}\raggedright
Sponsorship penalty (pp)
\end{minipage} \\
\midrule\noalign{}
\endhead
\bottomrule\noalign{}
\endlastfoot
Kev 9B & Consumer & 43.43\% (13.63) & 40.60\% (13.09) & 2.83 \\
Kev 9B & Platform & 42.62\% (13.58) & 43.88\% (13.07) & −1.26 \\
Jev 1.14 & Consumer & 49.33\% (28.24) & 13.46\% (15.30) & 35.87 \\
Jev 1.14 & Platform & 48.83\% (27.93) & 17.72\% (13.54) & 31.12 \\
Liquid D1-Ultra & Consumer & 48.49\% (15.13) & 17.35\% (11.91) &
31.14 \\
Liquid D1-Ultra & Platform & 48.66\% (15.32) & 18.74\% (12.39) &
29.91 \\
Mercury Decide & Consumer & 50.44\% (15.52) & 5.87\% (6.68) & 44.58 \\
Mercury Decide & Platform & 50.26\% (15.22) & 6.63\% (7.45) & 43.63 \\
Microsoft Decision 1 & Consumer & 49.89\% (21.91) & 34.89\% (17.86) &
15.00 \\
Microsoft Decision 1 & Platform & 50.06\% (19.94) & 36.67\% (15.17) &
13.39 \\
StartLux 27B & Consumer & 49.79\% (15.53) & 23.67\% (11.31) & 26.13 \\
StartLux 27B & Platform & 49.82\% (17.31) & 25.33\% (11.49) & 24.49 \\
\end{longtable}
}

\emph{Note.} \(N = 360\) position-balanced Latin-square stimuli per
cell; \(N = 1{,}440\) per decision model; \(N = 8{,}640\) total.
Calibrated target probabilities report the mean percentage probability
assigned to the target hotel across the 360 stimuli (SDs in
parentheses).

{\def\LTcaptype{apptbl}
\begin{longtable}[]{@{}
  >{\raggedright\arraybackslash}p{(\linewidth - 12\tabcolsep) * \real{0.2515}}
  >{\centering\arraybackslash}p{(\linewidth - 12\tabcolsep) * \real{0.1198}}
  >{\centering\arraybackslash}p{(\linewidth - 12\tabcolsep) * \real{0.1198}}
  >{\centering\arraybackslash}p{(\linewidth - 12\tabcolsep) * \real{0.1198}}
  >{\centering\arraybackslash}p{(\linewidth - 12\tabcolsep) * \real{0.1198}}
  >{\centering\arraybackslash}p{(\linewidth - 12\tabcolsep) * \real{0.1198}}
  >{\centering\arraybackslash}p{(\linewidth - 12\tabcolsep) * \real{0.1198}}@{}}
\caption{{\label{apptbl-mediation-other}}Moderated mediation analysis
of sponsorship on target choice via skepticism - Gemini Flash LLMs
(Study 1)}\tabularnewline
\toprule\noalign{}
\multirow{2}{=}{\begin{minipage}[b]{\linewidth}\raggedright
\end{minipage}} &
\multicolumn{2}{>{\centering\arraybackslash}p{(\linewidth - 12\tabcolsep) * \real{0.2395} + 2\tabcolsep}}{%
\begin{minipage}[b]{\linewidth}\centering
Gemini 3.5 Flash
\end{minipage}} &
\multicolumn{2}{>{\centering\arraybackslash}p{(\linewidth - 12\tabcolsep) * \real{0.2395} + 2\tabcolsep}}{%
\begin{minipage}[b]{\linewidth}\centering
Gemini 3.6 Flash
\end{minipage}} &
\multicolumn{2}{>{\centering\arraybackslash}p{(\linewidth - 12\tabcolsep) * \real{0.2395} + 2\tabcolsep}@{}}{%
\begin{minipage}[b]{\linewidth}\centering
Gemini 3.8 Flash
\end{minipage}} \\
& \begin{minipage}[b]{\linewidth}\centering
Skepticism (M)
\end{minipage} & \begin{minipage}[b]{\linewidth}\centering
Target choice (Y)
\end{minipage} & \begin{minipage}[b]{\linewidth}\centering
Skepticism (M)
\end{minipage} & \begin{minipage}[b]{\linewidth}\centering
Target choice (Y)
\end{minipage} & \begin{minipage}[b]{\linewidth}\centering
Skepticism (M)
\end{minipage} & \begin{minipage}[b]{\linewidth}\centering
Target choice (Y)
\end{minipage} \\
\midrule\noalign{}
\endfirsthead
\toprule\noalign{}
\multirow{2}{=}{\begin{minipage}[b]{\linewidth}\raggedright
\end{minipage}} &
\multicolumn{2}{>{\centering\arraybackslash}p{(\linewidth - 12\tabcolsep) * \real{0.2395} + 2\tabcolsep}}{%
\begin{minipage}[b]{\linewidth}\centering
Gemini 3.5 Flash
\end{minipage}} &
\multicolumn{2}{>{\centering\arraybackslash}p{(\linewidth - 12\tabcolsep) * \real{0.2395} + 2\tabcolsep}}{%
\begin{minipage}[b]{\linewidth}\centering
Gemini 3.6 Flash
\end{minipage}} &
\multicolumn{2}{>{\centering\arraybackslash}p{(\linewidth - 12\tabcolsep) * \real{0.2395} + 2\tabcolsep}@{}}{%
\begin{minipage}[b]{\linewidth}\centering
Gemini 3.8 Flash
\end{minipage}} \\
& \begin{minipage}[b]{\linewidth}\centering
Skepticism (M)
\end{minipage} & \begin{minipage}[b]{\linewidth}\centering
Target choice (Y)
\end{minipage} & \begin{minipage}[b]{\linewidth}\centering
Skepticism (M)
\end{minipage} & \begin{minipage}[b]{\linewidth}\centering
Target choice (Y)
\end{minipage} & \begin{minipage}[b]{\linewidth}\centering
Skepticism (M)
\end{minipage} & \begin{minipage}[b]{\linewidth}\centering
Target choice (Y)
\end{minipage} \\
\midrule\noalign{}
\endhead
\bottomrule\noalign{}
\endlastfoot
Principal (platform) & 0.019 (0.095) & −0.042 (0.243) & −0.040 (0.092) &
−0.032 (0.251) & −0.014 (0.094) & −0.355 (0.264) \\
Sponsorship (sponsored) & 2.813*** (0.095) & 1.779*** (0.354) & 1.696***
(0.092) & 0.964** (0.311) & 4.496*** (0.094) & −0.498 (0.710) \\
Principal \(\times\) Sponsorship & −0.533*** (0.134) & 0.713 (0.404) &
−0.572*** (0.130) & 0.619 (0.406) & −2.165*** (0.133) & 3.784***
(0.708) \\
Skepticism & --- & −1.594*** (0.102) & --- & −1.513*** (0.120) & --- &
−1.701*** (0.141) \\
\textbf{Conditional effects of sponsorship} & & & & & & \\
\emph{Consumer delegation} & 2.813*** (0.095) & 1.779*** (0.354) &
1.696*** (0.092) & 0.964** (0.311) & 4.496*** (0.094) & −0.498
(0.710) \\
\emph{Platform delegation} & 2.280*** (0.095) & 2.492*** (0.350) &
1.125*** (0.092) & 1.583*** (0.297) & 2.331*** (0.095) & 3.286***
(0.442) \\
\(N\) & 2,000 & 2,000 & 2,000 & 2,000 & 1,922 & 1,922 \\
(Pseudo) \(R^2\) & 0.486 & 0.726 & 0.340 & 0.739 & 0.650 & 0.790 \\
\end{longtable}
}

\emph{Note.} \(N = 2{,}000\) for Gemini 3.5 Flash and Gemini 3.6 Flash;
\(N = 1{,}922\) for Gemini 3.8 Flash (78 trials with empty reasoning
traces omitted); \(N = 5{,}922\) total. PROCESS Model 8
\citep{hayes2017introduction}. X = sponsorship (1 = sponsored), W =
principal (1 = platform), M = skepticism, Y = target choice. M =
skepticism equation (OLS); Y = target choice equation (logistic
regression, log-odds). Unstandardized coefficients with standard errors
in parentheses. Reference categories are consumer delegation and organic
target. Both equations control for standardized target price and fixed
effects for target and competitor review texts and hotel names, omitted
from display. Conditional effects under platform delegation are linear
combinations of the sponsorship and interaction coefficients. For the Y
equation, these are conditional \emph{direct} effects, holding
skepticism constant. Conditional indirect effects are reported in
\quartoapptblref{apptbl-mediation-other-indirect}.\\
*\emph{p} \textless{} .05. **\emph{p} \textless{} .01. ***\emph{p}
\textless{} .001.

{\def\LTcaptype{apptbl}
\begin{longtable}[]{@{}
  >{\raggedright\arraybackslash}p{(\linewidth - 12\tabcolsep) * \real{0.2819}}
  >{\raggedleft\arraybackslash}p{(\linewidth - 12\tabcolsep) * \real{0.0872}}
  >{\centering\arraybackslash}p{(\linewidth - 12\tabcolsep) * \real{0.1409}}
  >{\raggedleft\arraybackslash}p{(\linewidth - 12\tabcolsep) * \real{0.0872}}
  >{\centering\arraybackslash}p{(\linewidth - 12\tabcolsep) * \real{0.1409}}
  >{\raggedleft\arraybackslash}p{(\linewidth - 12\tabcolsep) * \real{0.0872}}
  >{\centering\arraybackslash}p{(\linewidth - 12\tabcolsep) * \real{0.1409}}@{}}
\caption{{\label{apptbl-mediation-other-indirect}}Conditional indirect
effects of sponsorship on target choice via skepticism - Gemini Flash
LLMs (Study 1)}\tabularnewline
\toprule\noalign{}
\multirow{2}{=}{\begin{minipage}[b]{\linewidth}\raggedright
\end{minipage}} &
\multicolumn{2}{>{\raggedleft\arraybackslash}p{(\linewidth - 12\tabcolsep) * \real{0.2282} + 2\tabcolsep}}{%
\begin{minipage}[b]{\linewidth}\raggedleft
Gemini 3.5 Flash
\end{minipage}} &
\multicolumn{2}{>{\raggedleft\arraybackslash}p{(\linewidth - 12\tabcolsep) * \real{0.2282} + 2\tabcolsep}}{%
\begin{minipage}[b]{\linewidth}\raggedleft
Gemini 3.6 Flash
\end{minipage}} &
\multicolumn{2}{>{\raggedleft\arraybackslash}p{(\linewidth - 12\tabcolsep) * \real{0.2282} + 2\tabcolsep}@{}}{%
\begin{minipage}[b]{\linewidth}\raggedleft
Gemini 3.8 Flash
\end{minipage}} \\
& \begin{minipage}[b]{\linewidth}\raggedleft
Effect
\end{minipage} & \begin{minipage}[b]{\linewidth}\centering
95\% CI
\end{minipage} & \begin{minipage}[b]{\linewidth}\raggedleft
Effect
\end{minipage} & \begin{minipage}[b]{\linewidth}\centering
95\% CI
\end{minipage} & \begin{minipage}[b]{\linewidth}\raggedleft
Effect
\end{minipage} & \begin{minipage}[b]{\linewidth}\centering
95\% CI
\end{minipage} \\
\midrule\noalign{}
\endfirsthead
\toprule\noalign{}
\multirow{2}{=}{\begin{minipage}[b]{\linewidth}\raggedright
\end{minipage}} &
\multicolumn{2}{>{\raggedleft\arraybackslash}p{(\linewidth - 12\tabcolsep) * \real{0.2282} + 2\tabcolsep}}{%
\begin{minipage}[b]{\linewidth}\raggedleft
Gemini 3.5 Flash
\end{minipage}} &
\multicolumn{2}{>{\raggedleft\arraybackslash}p{(\linewidth - 12\tabcolsep) * \real{0.2282} + 2\tabcolsep}}{%
\begin{minipage}[b]{\linewidth}\raggedleft
Gemini 3.6 Flash
\end{minipage}} &
\multicolumn{2}{>{\raggedleft\arraybackslash}p{(\linewidth - 12\tabcolsep) * \real{0.2282} + 2\tabcolsep}@{}}{%
\begin{minipage}[b]{\linewidth}\raggedleft
Gemini 3.8 Flash
\end{minipage}} \\
& \begin{minipage}[b]{\linewidth}\raggedleft
Effect
\end{minipage} & \begin{minipage}[b]{\linewidth}\centering
95\% CI
\end{minipage} & \begin{minipage}[b]{\linewidth}\raggedleft
Effect
\end{minipage} & \begin{minipage}[b]{\linewidth}\centering
95\% CI
\end{minipage} & \begin{minipage}[b]{\linewidth}\raggedleft
Effect
\end{minipage} & \begin{minipage}[b]{\linewidth}\centering
95\% CI
\end{minipage} \\
\midrule\noalign{}
\endhead
\bottomrule\noalign{}
\endlastfoot
Consumer delegation & −4.485 & {[}−5.592, −4.112{]} & −2.565 &
{[}−3.432, −2.201{]} & −7.648 & {[}−10.583, −6.800{]} \\
Platform delegation & −3.634 & {[}−4.560, −3.307{]} & −1.701 &
{[}−2.331, −1.410{]} & −3.965 & {[}−5.525, −3.480{]} \\
Index of moderated mediation & 0.850 & {[}0.458, 1.385{]} & 0.864 &
{[}0.508, 1.397{]} & 3.683 & {[}3.154, 5.223{]} \\
\end{longtable}
}

\emph{Note.} \(N = 2{,}000\) for Gemini 3.5 Flash and Gemini 3.6 Flash;
\(N = 1{,}922\) for Gemini 3.8 Flash. Indirect effects are the products
of the conditional a-paths and the b-path from
\quartoapptblref{apptbl-mediation-other}, expressed in a log-odds
metric. Percentile bootstrap estimates with 10,000 resamples. Bootstrap
standard errors are 0.381 / 0.319 / 0.237 (Gemini 3.5 Flash), 0.312 /
0.236 / 0.226 (Gemini 3.6 Flash), and 0.960 / 0.519 / 0.530 (Gemini 3.8
Flash) for the consumer indirect effect, platform indirect effect, and
index respectively. The index of moderated mediation is the difference
between the two conditional indirect effects; a confidence interval
excluding zero indicates that the strength of mediation depends on the
delegating principal.

\subsection*{A.3 Study 1 - system-prompt wording and reasoning
effort}\label{study-1---system-prompt-wording-and-reasoning-effort}
\addcontentsline{toc}{subsection}{A.3 Study 1 - system-prompt wording
and reasoning effort}

\begin{appfig}

\centering{

\includegraphics[width=0.65\linewidth,height=\textheight,keepaspectratio]{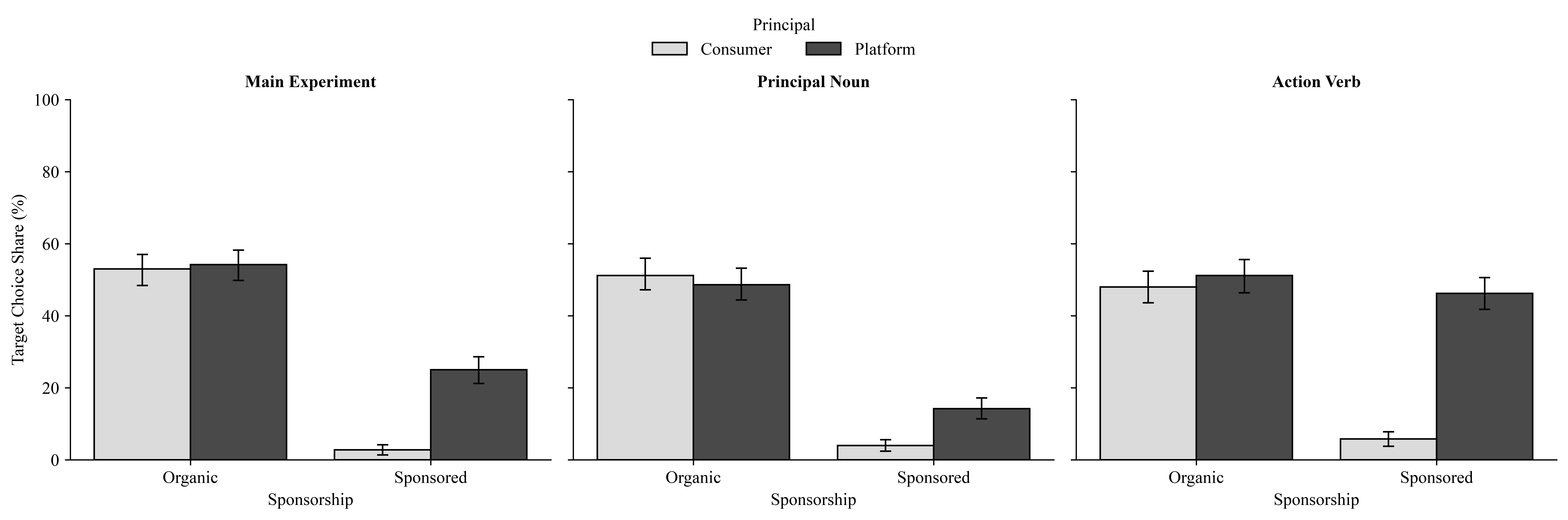}

}

\caption{\label{appfig-choice-system}Target choice share (\%) for
sponsorship and principal across system-prompt wording variations -
Gemini 3.1 Pro (Study 1).}

\floatnote{\emph{Note.} Error bars represent 95\% confidence intervals.}

\end{appfig}%

\begin{appfig}

\centering{

\includegraphics[width=0.65\linewidth,height=\textheight,keepaspectratio]{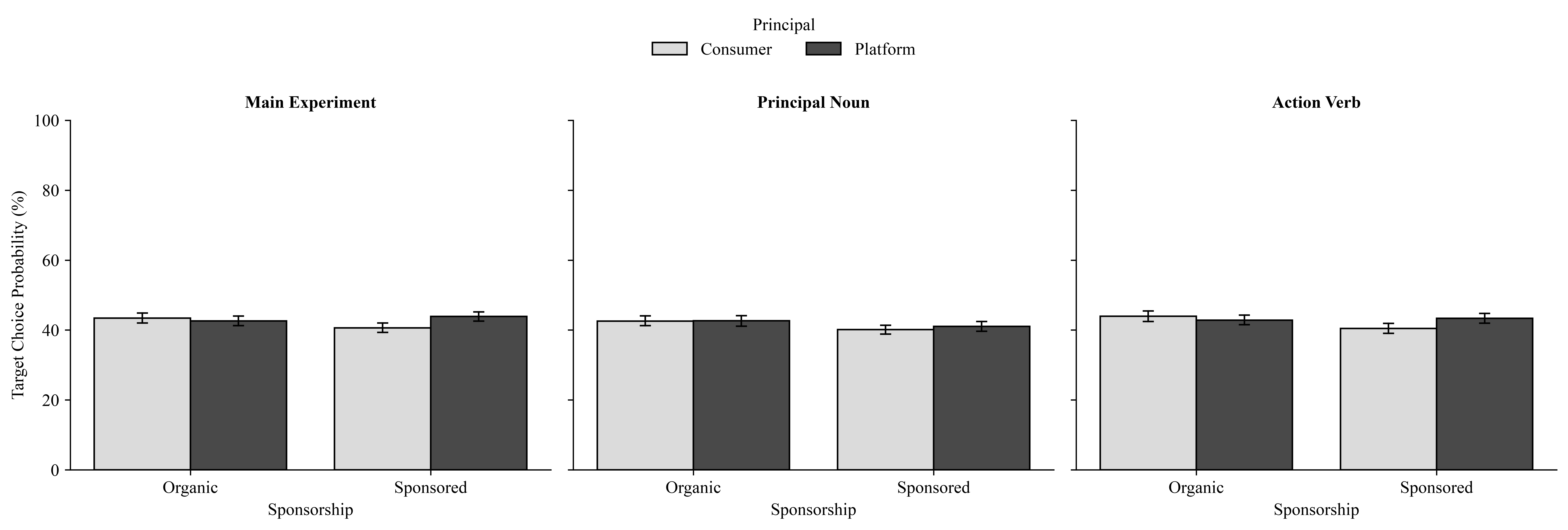}

}

\caption{\label{appfig-kev-choice-system}Calibrated target choice
probability (\%) for sponsorship and principal across system-prompt
wording variations - Kev 9B (Study 1).}

\floatnote{\emph{Note.} Error bars represent 95\% confidence intervals.}

\end{appfig}%

{\def\LTcaptype{apptbl}
\begin{longtable}[]{@{}
  >{\raggedright\arraybackslash}p{(\linewidth - 8\tabcolsep) * \real{0.1667}}
  >{\raggedright\arraybackslash}p{(\linewidth - 8\tabcolsep) * \real{0.1275}}
  >{\raggedright\arraybackslash}p{(\linewidth - 8\tabcolsep) * \real{0.2059}}
  >{\raggedright\arraybackslash}p{(\linewidth - 8\tabcolsep) * \real{0.2255}}
  >{\raggedright\arraybackslash}p{(\linewidth - 8\tabcolsep) * \real{0.2745}}@{}}
\caption{{\label{apptbl-kev-system-descriptive}}Calibrated target
probabilities by principal and sponsorship across system-prompt wording
variants - Kev 9B (Study 1)}\tabularnewline
\toprule\noalign{}
\begin{minipage}[b]{\linewidth}\raggedright
Variant
\end{minipage} & \begin{minipage}[b]{\linewidth}\raggedright
Principal
\end{minipage} & \begin{minipage}[b]{\linewidth}\raggedright
Organic P(Target)
\end{minipage} & \begin{minipage}[b]{\linewidth}\raggedright
Sponsored P(Target)
\end{minipage} & \begin{minipage}[b]{\linewidth}\raggedright
Sponsorship penalty (pp)
\end{minipage} \\
\midrule\noalign{}
\endfirsthead
\toprule\noalign{}
\begin{minipage}[b]{\linewidth}\raggedright
Variant
\end{minipage} & \begin{minipage}[b]{\linewidth}\raggedright
Principal
\end{minipage} & \begin{minipage}[b]{\linewidth}\raggedright
Organic P(Target)
\end{minipage} & \begin{minipage}[b]{\linewidth}\raggedright
Sponsored P(Target)
\end{minipage} & \begin{minipage}[b]{\linewidth}\raggedright
Sponsorship penalty (pp)
\end{minipage} \\
\midrule\noalign{}
\endhead
\bottomrule\noalign{}
\endlastfoot
Main experiment & Consumer & 43.43\% (13.63) & 40.60\% (13.09) & 2.83 \\
Main experiment & Platform & 42.62\% (13.58) & 43.88\% (13.07) &
−1.26 \\
Principal noun & Consumer & 42.56\% (13.35) & 40.12\% (12.65) & 2.45 \\
Principal noun & Platform & 42.65\% (14.33) & 41.05\% (13.41) & 1.60 \\
Action verb & Consumer & 43.94\% (14.90) & 40.44\% (14.09) & 3.51 \\
Action verb & Platform & 42.83\% (14.23) & 43.37\% (13.64) & −0.54 \\
\end{longtable}
}

\emph{Note.} \(N = 360\) position-balanced Latin-square stimuli per
cell; \(N = 1{,}440\) per system-prompt variant; \(N = 4{,}320\) total.
Calibrated target probabilities report the mean percentage probability
assigned to the target hotel across the 360 stimuli (SDs in
parentheses).

{\def\LTcaptype{apptbl}
\begin{longtable}[]{@{}
  >{\raggedright\arraybackslash}p{(\linewidth - 8\tabcolsep) * \real{0.1912}}
  >{\raggedright\arraybackslash}p{(\linewidth - 8\tabcolsep) * \real{0.2206}}
  >{\raggedright\arraybackslash}p{(\linewidth - 8\tabcolsep) * \real{0.2647}}
  >{\raggedright\arraybackslash}p{(\linewidth - 8\tabcolsep) * \real{0.2500}}
  >{\raggedleft\arraybackslash}p{(\linewidth - 8\tabcolsep) * \real{0.0735}}@{}}
\caption{{\label{apptbl-thinking-level-choice}}Choice distribution
across reasoning efforts - Gemini 3.1 Pro (Study 1)}\tabularnewline
\toprule\noalign{}
\begin{minipage}[b]{\linewidth}\raggedright
Principal
\end{minipage} & \begin{minipage}[b]{\linewidth}\raggedright
Sponsorship
\end{minipage} & \begin{minipage}[b]{\linewidth}\raggedright
Thinking level
\end{minipage} & \begin{minipage}[b]{\linewidth}\raggedright
Target chosen
\end{minipage} & \begin{minipage}[b]{\linewidth}\raggedleft
N
\end{minipage} \\
\midrule\noalign{}
\endfirsthead
\toprule\noalign{}
\begin{minipage}[b]{\linewidth}\raggedright
Principal
\end{minipage} & \begin{minipage}[b]{\linewidth}\raggedright
Sponsorship
\end{minipage} & \begin{minipage}[b]{\linewidth}\raggedright
Thinking level
\end{minipage} & \begin{minipage}[b]{\linewidth}\raggedright
Target chosen
\end{minipage} & \begin{minipage}[b]{\linewidth}\raggedleft
N
\end{minipage} \\
\midrule\noalign{}
\endhead
\bottomrule\noalign{}
\endlastfoot
Consumer & Organic & High & 265 (53.0\%) & 500 \\
& & Low & 252 (50.4\%) & 500 \\
& Sponsored & High & 14 (2.8\%) & 500 \\
& & Low & 25 (5.0\%) & 500 \\
Platform & Organic & High & 271 (54.2\%) & 500 \\
& & Low & 248 (49.6\%) & 500 \\
& Sponsored & High & 125 (25.0\%) & 500 \\
& & Low & 128 (25.6\%) & 500 \\
\end{longtable}
}

\emph{Note.} Frequency with percentage in parentheses. The dominated
alternatives are never chosen.

\begin{appfig}

\centering{

\includegraphics[width=0.65\linewidth,height=\textheight,keepaspectratio]{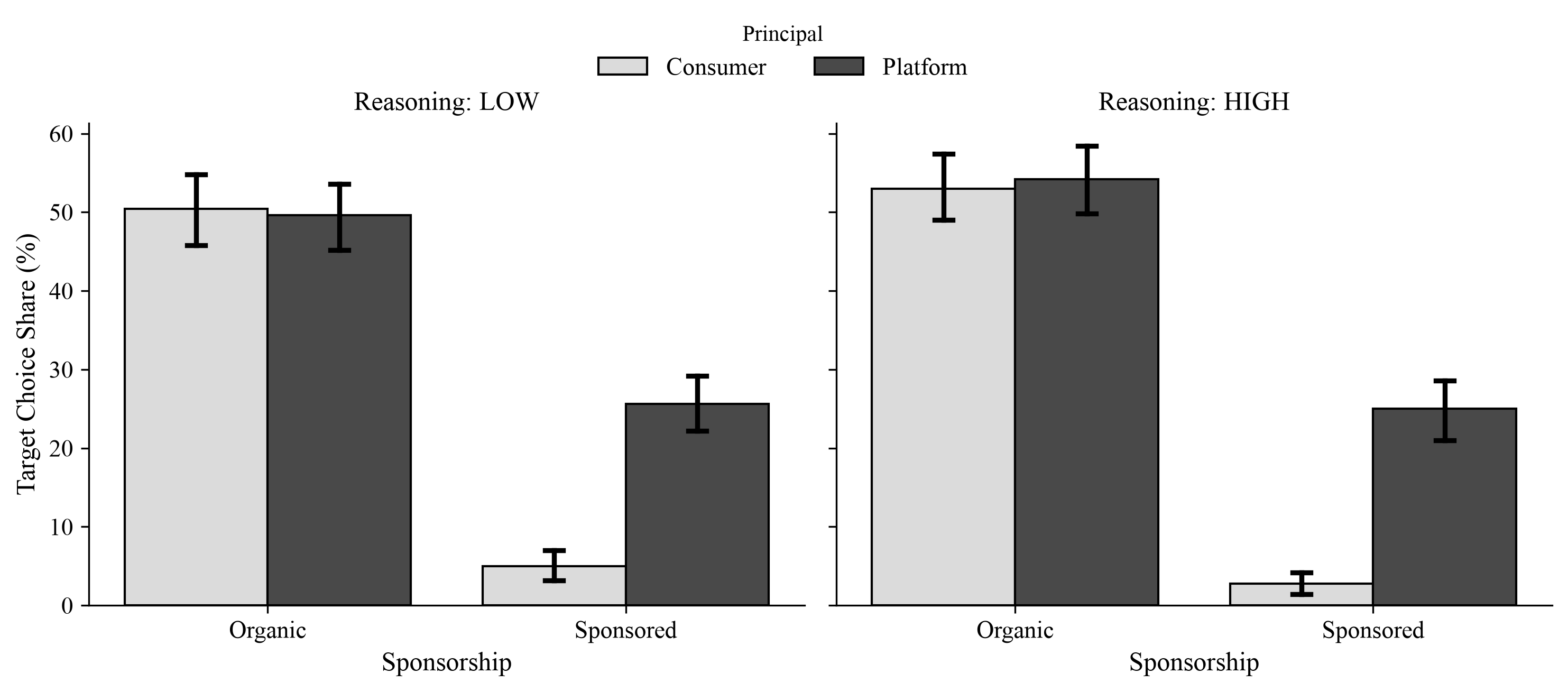}

}

\caption{\label{appfig-thinking-level-choice}Target choice distribution
by principal, sponsorship, and thinking level - Gemini 3.1 Pro (Study
1).}

\floatnote{\emph{Note.} Error bars represent 95\% confidence intervals
(\quartoapptblref{apptbl-thinking-level-choice}).}

\end{appfig}%

\subsection*{A.4 Study 2 - disclosure
decomposition}\label{study-2---disclosure-decomposition}
\addcontentsline{toc}{subsection}{A.4 Study 2 - disclosure
decomposition}

\begin{appfig}

\centering{

\includegraphics[width=0.85\linewidth,height=\textheight,keepaspectratio]{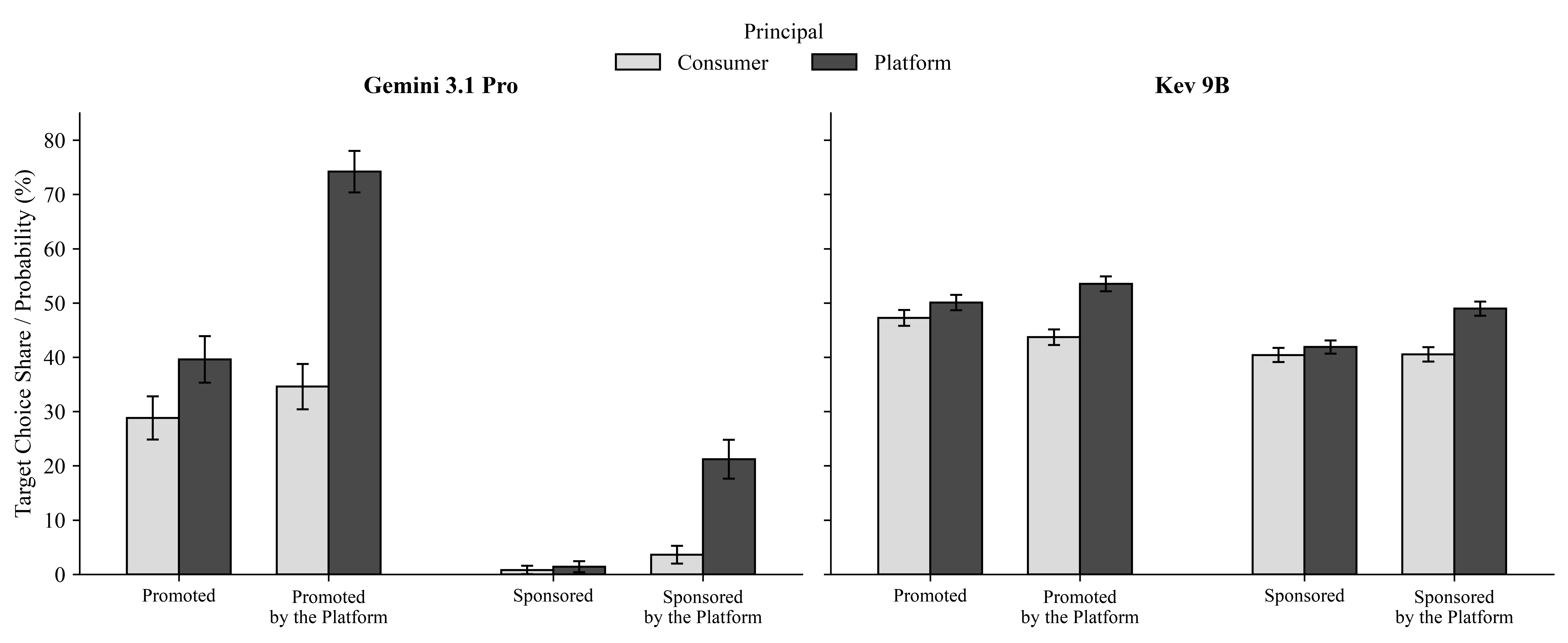}

}

\caption{\label{appfig-study3-levels}Target choice share (Gemini 3.1
Pro) and calibrated target probability (Kev 9B) by disclosure label and
principal (Study 2).}

\floatnote{\emph{Note.} Error bars represent 95\% confidence intervals
(Table~\ref{tbl-study3-descriptive}).}

\end{appfig}%

\subsection*{A.5 Study 3 - prompt-level and model-level
mitigation}\label{study-3---prompt-level-and-model-level-mitigation}
\addcontentsline{toc}{subsection}{A.5 Study 3 - prompt-level and
model-level mitigation}

{\def\LTcaptype{apptbl}
\begin{longtable}[]{@{}
  >{\raggedright\arraybackslash}p{(\linewidth - 8\tabcolsep) * \real{0.1368}}
  >{\raggedright\arraybackslash}p{(\linewidth - 8\tabcolsep) * \real{0.3504}}
  >{\raggedright\arraybackslash}p{(\linewidth - 8\tabcolsep) * \real{0.1453}}
  >{\raggedright\arraybackslash}p{(\linewidth - 8\tabcolsep) * \real{0.1453}}
  >{\centering\arraybackslash}p{(\linewidth - 8\tabcolsep) * \real{0.2222}}@{}}
\caption{{\label{apptbl-study3-remedies-descriptive}}Target choice
share (Gemini 3.1 Pro) and calibrated target probability (Kev 9B) across
mitigation conditions (Study 3)}\tabularnewline
\toprule\noalign{}
\begin{minipage}[b]{\linewidth}\raggedright
Model
\end{minipage} & \begin{minipage}[b]{\linewidth}\raggedright
Condition
\end{minipage} & \begin{minipage}[b]{\linewidth}\raggedright
Organic
\end{minipage} & \begin{minipage}[b]{\linewidth}\raggedright
Sponsored
\end{minipage} & \begin{minipage}[b]{\linewidth}\centering
Sponsorship penalty (pp)
\end{minipage} \\
\midrule\noalign{}
\endfirsthead
\toprule\noalign{}
\begin{minipage}[b]{\linewidth}\raggedright
Model
\end{minipage} & \begin{minipage}[b]{\linewidth}\raggedright
Condition
\end{minipage} & \begin{minipage}[b]{\linewidth}\raggedright
Organic
\end{minipage} & \begin{minipage}[b]{\linewidth}\raggedright
Sponsored
\end{minipage} & \begin{minipage}[b]{\linewidth}\centering
Sponsorship penalty (pp)
\end{minipage} \\
\midrule\noalign{}
\endhead
\bottomrule\noalign{}
\endlastfoot
Gemini 3.1 Pro & Consumer & 53.00\% (49.96) & 2.80\% (16.51) & 50.20 \\
Gemini 3.1 Pro & Platform & 54.20\% (49.87) & 25.00\% (43.34) & 29.20 \\
Gemini 3.1 Pro & Platform (prompt-level mitigation) & 51.40\% (50.03) &
3.20\% (17.62) & 48.20 \\
Kev 9B & Consumer & 43.43\% (13.63) & 40.60\% (13.09) & 2.83 \\
Kev 9B & Platform & 42.62\% (13.58) & 43.88\% (13.07) & −1.26 \\
Kev 9B & Platform (prompt-level mitigation) & 42.57\% (12.99) & 38.92\%
(12.29) & 3.65 \\
Kev 9B & Platform (consumer-aligned fine-tuning) & 42.36\% (12.54) &
39.71\% (12.06) & 2.65 \\
\end{longtable}
}

\emph{Note.} Gemini 3.1 Pro: \(N = 500\) replications per cell
(\(N = 3{,}000\) total), reporting target choice shares (SDs in
parentheses). Kev 9B: \(N = 360\) paired Latin-square stimuli per cell
(\(N = 2{,}880\) total), reporting mean calibrated target probabilities
(SDs in parentheses). Under \emph{Platform (prompt-level mitigation)},
the sentence ``Act in the traveler's best interest.'' is appended to the
platform system prompt.

\newpage{}

\section*{Appendix B. Stimuli, measurement, and
software}\label{sec-appendix-a}
\addcontentsline{toc}{section}{Appendix B. Stimuli, measurement, and
software}

\subsection*{B.1 Review stimuli generation and validation
pipeline}\label{review-stimuli-generation-and-validation-pipeline}
\addcontentsline{toc}{subsection}{B.1 Review stimuli generation and
validation pipeline}

We scraped an initial corpus of 1,150 customer reviews from Expedia for
the \emph{Hilton New York Times Square}. To establish a uniformly
positive baseline from which paired comparisons could be derived, we
extracted reviews with the highest rating (10-out-of-10).

The initial corpus was processed using Google's \emph{Gemini 3.5 Flash}.
The LLM was instructed to parse each raw positive review against three
exclusion criteria. The review had to contain text. The review had to be
written in English. Reviews submitted in other languages (e.g., Spanish,
Chinese, French) or featuring mixed-language phrasing were excluded. The
review text could not mention any specific hotel brand, chain, or
property name (e.g., ``Hilton'').

For every positive review meeting all three criteria, the LLM was
instructed to generate an inverted negative counterpart following the
paired-stimuli methodology established in the sponsorship disclosure
literature \citep{kim2019paradox}. The negative review inverted the
semantic valence across every evaluated attribute (e.g., ``very clean''
\(\to\) ``very unclean''; ``remarkably quiet'' \(\to\) ``excessively
noisy''). The grammatical structure, sentence count, and clause sequence
of the original positive review were preserved.

The filtering and generation pipeline yielded an initial dataset of 418
positive reviews and 418 structurally corresponding negative reviews.
From this pool, we manually selected a subset of 10 positive-negative
review pairs to serve as the final stimuli for the experiments. We
selected 10 review pairs to ensure sufficient statistical power per
review, and to control for review-level fixed effects in our regression
models. The authors conducted a comprehensive manual validation of these
specific pairs. We confirmed successful valence inversion, naturalness,
structural similarity, and total brand neutrality across all stimuli
utilized in the study. See \quartoapptblref{apptbl-review-pool} for the
final review pool.

{\def\LTcaptype{apptbl}
\begin{longtable}[]{@{}
  >{\raggedright\arraybackslash}p{(\linewidth - 2\tabcolsep) * \real{0.4787}}
  >{\raggedright\arraybackslash}p{(\linewidth - 2\tabcolsep) * \real{0.5213}}@{}}
\caption{{\label{apptbl-review-pool}}Final review pool
(positive-negative pairs) for the experiments}\tabularnewline
\toprule\noalign{}
\begin{minipage}[b]{\linewidth}\raggedright
Positive Review
\end{minipage} & \begin{minipage}[b]{\linewidth}\raggedright
Negative Review
\end{minipage} \\
\midrule\noalign{}
\endfirsthead
\toprule\noalign{}
\begin{minipage}[b]{\linewidth}\raggedright
Positive Review
\end{minipage} & \begin{minipage}[b]{\linewidth}\raggedright
Negative Review
\end{minipage} \\
\midrule\noalign{}
\endhead
\bottomrule\noalign{}
\endlastfoot
Great location. Spacious room. & Terrible location. Cramped room. \\
Great front desk staff. Very helpful and welcoming! The bar staff in the
lounge was very attentive! & Terrible front desk staff. Very unhelpful
and unwelcoming! The bar staff in the lounge was very inattentive! \\
So close to everything with comfortable rooms and helpful staff. & So
far from everything with uncomfortable rooms and unhelpful staff. \\
Comfortable bed. Amazing view. Close to everything. & Uncomfortable bed.
Terrible view. Far from everything. \\
This was a great hotel in a great location & This was a terrible hotel
in a terrible location \\
Clean, staff helpful, and great location & Dirty, staff unhelpful, and
terrible location \\
I had a great stay and a wonderful trip. & I had a terrible stay and a
horrible trip. \\
Clean, large rooms. Amazing views and a great staff & Dirty, small
rooms. Terrible views and a horrible staff \\
It was a great experience and surely will stay again in the future & It
was a terrible experience and surely will not stay again in the
future \\
Clean room, great location & Dirty room, terrible location \\
\end{longtable}
}

\subsection*{B.2 Price generation and Latin-square stimulus
design}\label{price-generation-and-latin-square-stimulus-design}
\addcontentsline{toc}{subsection}{B.2 Price generation and Latin-square
stimulus design}

Hotel prices were calibrated using empirical data scraped from Expedia
to reflect realistic market conditions while preserving internal
validity. For each trial, the target and competitor hotels were assigned
the same price (\(P\)), drawn randomly from a continuous uniform
distribution bounded by the 5th percentile (\(p_5 = \$173.45\)) and the
95th percentile (\(p_{95} = \$351.40\)) of observed market prices.
Holding the price identical between the two viable listings ensured that
relative price differences would not confound the choice. To maintain
the ecological validity of a multi-option choice set, the two dominated
alternatives, which featured negative review text, were assigned fixed
prices set higher than the upper bound of the competitive listings
(\$400 and \$450). This design ensured that dominated alternatives
remained unambiguously suboptimal across all conditions, isolating the
agent's decision to a choice between the target and competitor listings.

For the decision-model experiments in Studies 1 through 3, because a
decision model returns a deterministic probability distribution for any
fixed prompt, we constructed a 360-stimulus position-balanced
Latin-square design from the 10 hotel names and 10 review pairs.
Specifically, we formed all \(10 \times 9 = 90\) ordered
target-competitor hotel pairs and all 90 ordered positive review pairs,
and crossed each hotel pair with all four option slots (\emph{option\_1}
through \emph{option\_4}), shifting review pairs across slots so that no
hotel pair repeated the same review pair. Across the 360 base stimuli,
every ordered target-competitor slot pair appeared exactly 30 times, and
prices, dominated alternatives, option positions, and attribute order
were held constant within each stimulus across all experimental cells.

\subsection*{B.3 Consumer-aligned fine-tuning of Kev
9B}\label{consumer-aligned-fine-tuning-of-kev-9b}
\addcontentsline{toc}{subsection}{B.3 Consumer-aligned fine-tuning of
Kev 9B}

For the model-level mitigation experiment in Study 3, we constructed an
out-of-domain consumer-aligned fine-tuning dataset across five non-hotel
product categories using synthetic product names to avoid pre-trained
brand-equity confounds: four training categories (laptops, wireless
headphones, coffee makers, and running shoes) and one held-out
validation category (air purifiers). Each category comprised 10
synthetic product names, 10 positive reviews, and 10 negative reviews
paired in a position-balanced Latin-square design (45 ordered item pairs
\(\times\) 4 target option slots \(\times\) 2 sponsorship states
\(\times\) 2 principal roles = \(720\) records per category;
\(N = 2{,}880\) training records and \(N = 720\) validation records).
For every stimulus \(x\), we computed the frozen base Kev 9B teacher's
choice distribution under consumer delegation,
\(P_{\text{ref}}(\cdot \mid s_{\text{consumer}}, x)\), at its calibrated
pointer-head temperature (\(T = 2.194\)). We then fine-tuned Kev 9B's
rank-16 LoRA adapter and pointer head in BF16 using AdamW (peak learning
rate \(3 \times 10^{-6}\), weight decay \(0.01\), batch size \(8\),
OneCycleLR schedule with \(2\%\) warmup over three epochs, gradient
clipping at \(1.0\)) by minimizing the cross-entropy (equivalently, KL
divergence) against
\(P_{\text{ref}}(\cdot \mid s_{\text{consumer}}, x)\) for both
platform-delegated and consumer-delegated trials. Validation KL on the
held-out air-purifier category was evaluated every 10 steps, reaching
its minimum at step 440 (\(\text{KL} = 0.0030\) bits vs.~\(0.0074\) bits
at step 0), which was selected for zero-shot evaluation on the Study 1
hotel benchmark.

To quantify how well each mitigation in Study 3 preserves the
platform-delegated agent's baseline preferences across non-sponsorship
attributes, we estimated the condition-specific attribute regression
predicting target choice (logistic regression for Gemini 3.1 Pro) or
exact target-vs.-competitor log-odds (OLS for Kev 9B) from sponsorship,
standardized target price, and the 36 target and competitor hotel and
review fixed effects
(\(\hat{\boldsymbol{\gamma}}_c \in \mathbb{R}^{36}\)). We then mapped
the 36 estimated fixed effects into a probability distribution via the
softmax transformation,
\(q_c = \text{softmax}(\hat{\boldsymbol{\gamma}}_c)\), and computed the
Kullback-Leibler divergence relative to unmitigated platform delegation
in bits (\(\log_2\)),
\(D_{\text{KL}}(q_{\text{Platform}} \parallel q_c) = \sum_{j=1}^{36} q_{\text{Platform}, j} \log_2 (q_{\text{Platform}, j} / q_{c, j})\),
alongside the mean KL divergence across the 360 organic stimuli between
the target-vs.-competitor choice probabilities predicted by those fixed
effects.

\subsection*{B.4 LLM-as-a-judge evaluation pipeline}\label{sec-judge}
\addcontentsline{toc}{subsection}{B.4 LLM-as-a-judge evaluation
pipeline}

\subsubsection*{B.4.1 Protocol}\label{protocol}
\addcontentsline{toc}{subsubsection}{B.4.1 Protocol}

To assess the AI agents' pre-choice skepticism, we analyzed the
reasoning traces generated prior to each final choice. Evaluating
unstructured text requires a scoring instrument that is both scalable
and objective. In keeping with best practices in algorithmic auditing
\citep{gu2026survey, panickssery2024llm}, we avoided using proprietary
API models from the same commercial families as our subjects (Google,
OpenAI, Anthropic) to prevent shared-weight or provider-alignment
biases.

All candidate evaluators were restricted to independent, open-weight
models executed on local hardware across two architectural paradigms,
namely non-generative decision models and generative LLMs. Decision
models ingest a text state alongside structured evaluation questions and
return a calibrated probability distribution over the supplied options
in a single prefill-only forward pass via a trained pointer head,
without autoregressive text generation. Because evaluation requires only
a single forward pass with no sampling, the resulting probability
distribution is inherently deterministic, and we took the modal scale
anchor (i.e., the option with the highest predicted probability, or
argmax) as the integer rating. For the generative LLMs, which were
executed via \emph{llama.cpp}, we eliminated stochastic variance by
querying all models under deterministic greedy decoding (temperature =
0.0) and enforced structured integer ratings using grammar-constrained
JSON-schema decoding via Pydantic.

\subsubsection*{B.4.2 Candidate evaluator
LLMs}\label{candidate-evaluator-llms}
\addcontentsline{toc}{subsubsection}{B.4.2 Candidate evaluator LLMs}

We evaluated six candidate open-weight models spanning three
decision-model scales and three generative LLM families. In the
decision-model family, we tested Kev 0.8B, Kev 4B, and Kev 9B
\citep{palmer2026kev, palmer2026kev4b, palmer2026kev9b}, which adapt
frozen Qwen 3.5 base backbones with low-rank adapters (LoRA) and a
two-projection pointer head that scores each option's closing token
hidden state against the question's final token hidden state. Among the
generative LLMs, Ornith 1.0 35B is a mixture-of-experts (MoE) LLM
deployed through Unsloth MXFP4 quantization in GGUF format. Qwen 3.8 27B
is a dense transformer LLM deployed through Unsloth 6-bit quantization
(\emph{Q6\_K\_XL}) in GGUF format. Finally, NVIDIA's Nemotron 3.5
Lightning 30B A3B is deployed with 4-bit quantization (\emph{Q4\_K\_XL})
in GGUF format.

\subsubsection*{B.4.3 Human agreement}\label{human-agreement}
\addcontentsline{toc}{subsubsection}{B.4.3 Human agreement}

To establish construct validity and empirically select the judge with
the highest agreement with a human evaluator, a human coder
independently evaluated a stratified random sample of 60 reasoning
traces (15 sessions sampled from each of the four cells in Study 1's 2
\(\times\) 2 design). The human rater was blinded to the experimental
conditions, delegating principal, and LLM identities.

Each candidate model evaluated the identical sample of 60 traces. Judge
performance was compared against the human ratings across four
psychometric criteria. Intraclass Correlation Coefficient
(\(\text{ICC}(2,1)\)) assessing absolute agreement under a two-way
random-effects model; Pearson correlation coefficient (\(r\)) assessing
linear association; Mean Absolute Error (\(\text{MAE}\)) capturing
average scale point deviation; and Mean Absolute Directional Bias
(\(\text{Mean } |\text{Bias}| = |\frac{1}{N}\sum (\text{Score}_{\text{LLM}} - \text{Score}_{\text{Human}})|\))
capturing systematic leniency or severity.

{\def\LTcaptype{apptbl}
\begin{longtable}[]{@{}
  >{\raggedright\arraybackslash}p{(\linewidth - 8\tabcolsep) * \real{0.3048}}
  >{\centering\arraybackslash}p{(\linewidth - 8\tabcolsep) * \real{0.1143}}
  >{\centering\arraybackslash}p{(\linewidth - 8\tabcolsep) * \real{0.1429}}
  >{\centering\arraybackslash}p{(\linewidth - 8\tabcolsep) * \real{0.2190}}
  >{\centering\arraybackslash}p{(\linewidth - 8\tabcolsep) * \real{0.2190}}@{}}
\caption{{\label{apptbl-judge-validation}}Evaluation of candidate LLM
judges against human benchmark ratings}\tabularnewline
\toprule\noalign{}
\begin{minipage}[b]{\linewidth}\raggedright
Candidate
\end{minipage} & \begin{minipage}[b]{\linewidth}\centering
ICC(2,1)
\end{minipage} & \begin{minipage}[b]{\linewidth}\centering
Pearson \emph{r}
\end{minipage} & \begin{minipage}[b]{\linewidth}\centering
MAE
\end{minipage} & \begin{minipage}[b]{\linewidth}\centering
Mean \textbar Bias\textbar{}
\end{minipage} \\
\midrule\noalign{}
\endfirsthead
\toprule\noalign{}
\begin{minipage}[b]{\linewidth}\raggedright
Candidate
\end{minipage} & \begin{minipage}[b]{\linewidth}\centering
ICC(2,1)
\end{minipage} & \begin{minipage}[b]{\linewidth}\centering
Pearson \emph{r}
\end{minipage} & \begin{minipage}[b]{\linewidth}\centering
MAE
\end{minipage} & \begin{minipage}[b]{\linewidth}\centering
Mean \textbar Bias\textbar{}
\end{minipage} \\
\midrule\noalign{}
\endhead
\bottomrule\noalign{}
\endlastfoot
Kev 9B & 0.959 & 0.964 & 0.450 & 0.083 \\
Ornith 1.0 35B & 0.902 & 0.917 & 0.617 & 0.317 \\
Qwen 3.8 27B & 0.891 & 0.890 & 0.533 & 0.133 \\
Kev 4B & 0.797 & 0.846 & 0.883 & 0.717 \\
Kev 0.8B & 0.080 & 0.095 & 2.800 & 1.100 \\
Nemotron 3.5 Lightning 30B A3B & \textgreater{} −0.001 & −0.137 &
\(2.80 \times 10^{14}\) & \(2.80 \times 10^{14}\) \\
\end{longtable}
}

\emph{Note.} \(N = 60\) reasoning traces (15 per experimental cell)
independently scored on a 1 to 7 Likert scale by a human coder and six
candidate judges. For the Kev decision models (Kev 9B, Kev 4B, and Kev
0.8B), ratings correspond to the modal (argmax) scale anchor across the
seven Likert levels. \(\text{ICC}(2,1)\) denotes the two-way
random-effects, absolute agreement Intraclass Correlation Coefficient
against the human scores. Pearson \(r\) tests linear association
(\(p < .001\) for Kev 9B, Ornith 1.0 35B, Qwen 3.8 27B, and Kev 4B). MAE
= Mean Absolute Error. Mean \textbar Bias\textbar{} = absolute value of
the mean signed deviation,
\(|\frac{1}{N}\sum(\text{Score}_{\text{LLM}} - \text{Score}_{\text{Human}})|\).
Nemotron 3.5 Lightning 30B A3B returned out-of-range ratings (i.e.,
integers outside the 1 to 7 scale, up to \(4.0 \times 10^{15}\)) for 8
of 60 traces. Its statistics are computed on all 60 traces, including
these invalid ratings, which account for the extreme MAE and Mean
\textbar Bias\textbar{} and the absence of agreement with human ratings.

\subsubsection*{B.4.4 System prompt and structured
schema}\label{system-prompt-and-structured-schema}
\addcontentsline{toc}{subsubsection}{B.4.4 System prompt and structured
schema}

In all candidate evaluations, the placeholder
\emph{\{target\_option\_id\}} was dynamically replaced with the
randomized identifier assigned to the target hotel in that specific
session so as to preserve trial-level fidelity without revealing
experimental conditions. For the Kev decision models (including Kev 9B,
the selected judge used across all full-dataset analyses), each trial's
pre-choice reasoning trace was passed as the input state (with the full
serving state capacity of 65,536 tokens enabled so that no reasoning
trace was truncated), paired with a score question containing the
instruction: ``How much do you agree or disagree with the following
statement based on the AI agent's reasoning trace: The sponsorship
status of \{target\_option\_id\} made the agent skeptical of its true
value to the traveler.'' The question was evaluated against seven
ordered verbal criteria corresponding to the 1 to 7 Likert scale
(``Strongly disagree,'' ``Disagree,'' ``Slightly disagree,'' ``Neither
agree nor disagree,'' ``Slightly agree,'' ``Agree,'' and ``Strongly
agree''), and the integer index (1 to 7) of the highest-probability
criterion was recorded.

For the generative LLM candidates, each evaluator was prompted using the
standardized system instruction:

``\emph{You are evaluating the reasoning trace of an AI agent that is
selecting a hotel.}

\emph{Rate the following item on a scale of 1 to 7 (1 = strongly
disagree, 7 = strongly agree) based on whether the agent's reasoning
reflects this statement:} \emph{item: The sponsorship status of
\{target\_option\_id\} made the agent skeptical of its true value to the
traveler.}''

To ensure deterministic scoring and eliminate formatting artifacts for
the generative LLMs, responses were validated against a
grammar-constrained JSON-schema decoding. The schema defined a
structured JSON object containing a single integer property representing
the evaluation score.

\subsection*{B.5 Software details}\label{software-details}
\addcontentsline{toc}{subsection}{B.5 Software details}

\subsubsection*{B.5.1 Experimental parameters and decoding
configuration}\label{experimental-parameters-and-decoding-configuration}
\addcontentsline{toc}{subsubsection}{B.5.1 Experimental parameters and
decoding configuration}

Across all studies reported in this paper, we maintained each provider's
default decoding configurations
(\quartoapptblref{apptbl-model-specifications}). This approach reflects
our target estimand, which is capturing the behavior of LLMs as they
operate when deployed in realistic shopping environments. Furthermore,
this standardization aligns with recent infrastructural shifts across
major AI providers. In July 2026, Google deprecated user-specified
sampling parameters (e.g., temperature, top-\(p\), and top-\(k\)), with
the API ignoring these inputs and scheduling their rejection in future
versions \citep{google2026sampling}. Anthropic has implemented similar
restrictions on custom sampling controls \citep{anthropic2026sampling}.
Relying on default sampling parameters therefore ensures both ecological
validity and technical compliance with current API architectures.

\subsubsection*{B.5.2 Application programming interfaces and client
libraries}\label{application-programming-interfaces-and-client-libraries}
\addcontentsline{toc}{subsubsection}{B.5.2 Application programming
interfaces and client libraries}

All experimental queries were executed programmatically using the
providers' official Python client libraries and standard decision-model
endpoints. Google Gemini LLMs were queried via the Google GenAI SDK
(\emph{google-genai}) using the \emph{models.generate\_content} method.
OpenAI was accessed via the official OpenAI SDK (\emph{openai}) through
the Responses interface (\emph{client.responses.create} and
\emph{client.responses.parse}), which couples structured output
validation with native reasoning controls. Anthropic was called via the
Anthropic SDK (\emph{anthropic}) using the Messages API
(\emph{client.messages.create}). Commercial decision models, including
Jev 1.14 \citep{typesafe2026jev114}, Liquid D1-Ultra
\citep{liquidAI2026d1ultra}, Mercury Decide
\citep{inception2026mercurydecide}, and Microsoft Decision 1
\citep{srivastava2026microsoftdecision1}, were queried via the
OpenRouter System-One Decisions API
(\emph{https://openrouter.ai/api/alpha/decisions}). Open-weight decision
models were executed locally on an NVIDIA RTX 5090 GPU: Kev 9B
\citep{palmer2026kev9b} was evaluated in BF16 via the \emph{kev} runtime
with its calibrated pointer head (\(T = 2.194\)), and StartLux 27B
\citep[\emph{StartLux-Decision-v1.1-27B-Instant-Q4\_K\_M};][]{zhuang2026startlux}
was served via \emph{llama-server} and \emph{startlux-decide} exposing
the \emph{/v1/systemone} endpoint with whole-prompt bidirectional
attention and calibrated choice temperature (\(T = 1.087\)).

\subsubsection*{B.5.3 Model identifiers and sampling
parameters}\label{model-identifiers-and-sampling-parameters}
\addcontentsline{toc}{subsubsection}{B.5.3 Model identifiers and
sampling parameters}

\textbf{Google Gemini family.} Gemini 3.5 Flash (API model identifier
\emph{gemini-3.5-flash}), Gemini 3.6 Flash (API model identifier
\emph{gemini-3.6-flash}), Gemini 3.8 Flash (API model identifier
\emph{gemini-3.8-flash}), and Gemini 3.1 Pro (API model identifier
\emph{gemini-3.1-pro-preview}) were queried using default sampling
parameters, including a temperature of \(T = 1.0\) and top-\(p = 0.95\).

\textbf{OpenAI GPT-5.6 family.} GPT 5.6 Terra (API model identifier
\emph{gpt-5.6-terra}) was queried using default sampling parameters,
temperature \(T = 1.0\), top-\(p = 1.0\), and null presence and
frequency penalties.

\textbf{Anthropic Claude family.} Claude Sonnet 5 (API model identifier
\emph{claude-sonnet-5}) was evaluated using default generation
parameters, temperature \(T = 1.0\), top-\(p = 1.0\). It operated under
adaptive thinking mode (\emph{thinking=\{``type'': ``adaptive'',
``display'': ``summarized''\}}).

{\def\LTcaptype{apptbl}
\begin{longtable}[]{@{}
  >{\raggedright\arraybackslash}p{(\linewidth - 8\tabcolsep) * \real{0.1538}}
  >{\raggedright\arraybackslash}p{(\linewidth - 8\tabcolsep) * \real{0.2564}}
  >{\raggedright\arraybackslash}p{(\linewidth - 8\tabcolsep) * \real{0.1538}}
  >{\raggedright\arraybackslash}p{(\linewidth - 8\tabcolsep) * \real{0.3333}}
  >{\raggedright\arraybackslash}p{(\linewidth - 8\tabcolsep) * \real{0.1026}}@{}}
\caption{{\label{apptbl-model-specifications}}Generative LLM and
decision-model specifications}\tabularnewline
\toprule\noalign{}
\begin{minipage}[b]{\linewidth}\raggedright
Provider / Developer
\end{minipage} & \begin{minipage}[b]{\linewidth}\raggedright
Model identifier
\end{minipage} & \begin{minipage}[b]{\linewidth}\raggedright
Interface / Runtime
\end{minipage} & \begin{minipage}[b]{\linewidth}\raggedright
Endpoint
\end{minipage} & \begin{minipage}[b]{\linewidth}\raggedright
Decoding / Readout
\end{minipage} \\
\midrule\noalign{}
\endfirsthead
\toprule\noalign{}
\begin{minipage}[b]{\linewidth}\raggedright
Provider / Developer
\end{minipage} & \begin{minipage}[b]{\linewidth}\raggedright
Model identifier
\end{minipage} & \begin{minipage}[b]{\linewidth}\raggedright
Interface / Runtime
\end{minipage} & \begin{minipage}[b]{\linewidth}\raggedright
Endpoint
\end{minipage} & \begin{minipage}[b]{\linewidth}\raggedright
Decoding / Readout
\end{minipage} \\
\midrule\noalign{}
\endhead
\bottomrule\noalign{}
\endlastfoot
Google & \emph{gemini-3.5-flash} & \emph{google-genai} &
\emph{models.generate\_content} & \(T = 1.0\), top-\(p = 0.95\) \\
Google & \emph{gemini-3.6-flash} & \emph{google-genai} &
\emph{models.generate\_content} & \(T = 1.0\), top-\(p = 0.95\) \\
Google & \emph{gemini-3.8-flash} & \emph{google-genai} &
\emph{models.generate\_content} & \(T = 1.0\), top-\(p = 0.95\) \\
Google & \emph{gemini-3.1-pro-preview} & \emph{google-genai} &
\emph{models.generate\_content} & \(T = 1.0\), top-\(p = 0.95\) \\
OpenAI & \emph{gpt-5.6-terra} & \emph{openai} &
\emph{client.responses.create} & \(T = 1.0\), top-\(p = 1.0\) \\
Anthropic & \emph{claude-sonnet-5} & \emph{anthropic} &
\emph{client.messages.create} & \(T = 1.0\), top-\(p = 1.0\) \\
Jared Palmer & \emph{jaredpalmer/kev-9b} & Local \emph{kev} (BF16) &
\emph{DecisionModel.forward\_batch} & Pointer head (\(T = 2.194\)) \\
TypeSafe & \emph{typesafe/jev-1.14} & OpenRouter HTTP &
\emph{/api/alpha/decisions} & Deterministic prefill probabilities \\
Liquid AI & \emph{liquid/d1-ultra} & OpenRouter HTTP &
\emph{/api/alpha/decisions} & Deterministic prefill probabilities \\
Inception & \emph{inception/mercury-decide} & OpenRouter HTTP &
\emph{/api/alpha/decisions} & Deterministic prefill probabilities \\
Microsoft & \emph{microsoft/microsoft-decision-1} & OpenRouter HTTP &
\emph{/api/alpha/decisions} & Deterministic prefill probabilities \\
StartLux Labs & \emph{StartLux-Decision-v1.1-27B-Instant} &
\emph{llama.cpp} + \emph{startlux-decide} & \emph{/v1/systemone} &
Bidirectional letter readout (\(T = 1.087\)) \\
\end{longtable}
}

\subsubsection*{B.5.4 Structured output}\label{structured-output}
\addcontentsline{toc}{subsubsection}{B.5.4 Structured output}

All generative LLM API calls used structured JSON outputs
(\emph{response\_mime\_type=``application/json''}). We provided schemas
defined via Pydantic to enforce structured choice selections. Error
handling was implemented using exponential backoff to manage API rate
limits and transient errors, allowing up to seven retry attempts per
prompt. Across all trials, JSON schema enforcement and the retry
procedure yielded parseable responses in all but one session (one Claude
Sonnet 5 replication under platform delegation with a sponsored target),
which returned no choice after the retry limit. This session was
excluded and replaced with a new replication, so every cell contains 500
valid choices.

\subsection*{B.6 Power analysis}\label{power-analysis}
\addcontentsline{toc}{subsection}{B.6 Power analysis}

We selected a per-cell sample size that was large enough to control for
the fixed effects of target and competitor hotel names and reviews.
Because hotel names, review texts, and prices are drawn randomly on each
trial, we absorb them as fixed effects, which adds 36 stimulus dummies
to the choice model (9 each for target review, target hotel, competitor
review, and competitor hotel). Five hundred replications per cell
allocate roughly 50 observations to each stimulus level within each
cell, and 200 pooled across cells, so every hotel name and review text
is observed often enough for its coefficient to be identified in both
experimental arms. At the model level, this yields approximately 16
minority-outcome events per estimated parameter, above the conventional
threshold of 10 for logistic regression, and it keeps the sparsest cell
populated even where the sponsorship penalty pushes choice shares toward
the scale boundary.\footnote{Events per parameter is the count of the
  minority outcome in the pooled sample (675 of 2,000 target choices)
  divided by the 41 estimated parameters, that is intercept, three
  experimental terms, standardized price, and 36 stimulus dummies. The
  worst-case margin of error assumes \(p = .50\) and sums four cell
  variances. The 10 negative reviews are attached only to the dominated
  alternatives, which are excluded from the choice model.}

The resulting precision on the difference-in-differences estimate is
\(\pm 8.8\) percentage points in the worst case, which is small relative
to the 21.0-point interaction in the main experiment. Because this bound
assumes \(p = .50\) in every cell, it is conservative whenever cell
shares depart from .50, and the Principal \(\times\) Sponsorship
interaction is significant on the log-odds scale in every robustness
arm, including those with smaller percentage-point interactions (e.g.,
7.4 points for Gemini 3.6 Flash). We fixed 500 replications per cell
across all arms so that estimates remain comparable across LLMs,
wordings, and reasoning effort.

\newpage{}

\section*{Appendix C. Full regression tables}\label{sec-appendix-c}
\addcontentsline{toc}{section}{Appendix C. Full regression tables}

\subsection*{C.1 Full logistic regression table - Study 1
main}\label{full-logistic-regression-table---study-1-main}
\addcontentsline{toc}{subsection}{C.1 Full logistic regression table -
Study 1 main}

{\def\LTcaptype{apptbl}
\begin{longtable}[]{@{}lrrrl@{}}
\caption{{\label{apptbl-main-logit-full}}Full logistic regression table
- Gemini 3.1 Pro (Study 1)}\tabularnewline
\toprule\noalign{}
Term & \(\beta\) & \(SE\) & \(z\) & \(p\) \\
\midrule\noalign{}
\endfirsthead
\toprule\noalign{}
Term & \(\beta\) & \(SE\) & \(z\) & \(p\) \\
\midrule\noalign{}
\endhead
\bottomrule\noalign{}
\endlastfoot
Intercept & −0.280 & 0.509 & −0.550 & .582 \\
Principal (platform) & −0.165 & 0.198 & −0.835 & .403 \\
Sponsorship (sponsored) & −6.491 & 0.418 & −15.528 & \textless{} .001 \\
Principal \(\times\) Sponsorship & 3.989 & 0.428 & 9.319 & \textless{}
.001 \\
Target price (standardized) & 0.039 & 0.079 & 0.491 & .623 \\
Target review: review 1 & 0.759 & 0.391 & 1.939 & .052 \\
Target review: review 2 & 0.610 & 0.400 & 1.526 & .127 \\
Target review: review 3 & 2.795 & 0.390 & 7.158 & \textless{} .001 \\
Target review: review 4 & 5.200 & 0.429 & 12.132 & \textless{} .001 \\
Target review: review 5 & 1.451 & 0.384 & 3.783 & \textless{} .001 \\
Target review: review 6 & 2.699 & 0.390 & 6.912 & \textless{} .001 \\
Target review: review 7 & 1.613 & 0.405 & 3.982 & \textless{} .001 \\
Target review: review 8 & −0.738 & 0.421 & −1.752 & .080 \\
Target review: review 9 & 4.154 & 0.404 & 10.272 & \textless{} .001 \\
Target hotel: hotel 1 & 0.199 & 0.351 & 0.568 & .570 \\
Target hotel: hotel 2 & 0.213 & 0.345 & 0.619 & .536 \\
Target hotel: hotel 3 & −0.253 & 0.360 & −0.703 & .482 \\
Target hotel: hotel 4 & 0.992 & 0.375 & 2.642 & .008 \\
Target hotel: hotel 5 & −1.178 & 0.367 & −3.206 & .001 \\
Target hotel: hotel 6 & −1.256 & 0.375 & −3.348 & \textless{} .001 \\
Target hotel: hotel 7 & −1.950 & 0.374 & −5.217 & \textless{} .001 \\
Target hotel: hotel 8 & −0.170 & 0.356 & −0.476 & .634 \\
Target hotel: hotel 9 & −0.038 & 0.344 & −0.109 & .913 \\
Competitor review: review 1 & −0.225 & 0.323 & −0.695 & .487 \\
Competitor review: review 2 & 0.150 & 0.325 & 0.460 & .646 \\
Competitor review: review 3 & −2.340 & 0.340 & −6.877 & \textless{}
.001 \\
Competitor review: review 4 & −6.693 & 1.073 & −6.239 & \textless{}
.001 \\
Competitor review: review 5 & −0.454 & 0.323 & −1.406 & .160 \\
Competitor review: review 6 & −2.679 & 0.384 & −6.972 & \textless{}
.001 \\
Competitor review: review 7 & −0.715 & 0.321 & −2.227 & .026 \\
Competitor review: review 8 & 1.047 & 0.329 & 3.186 & .001 \\
Competitor review: review 9 & −4.024 & 0.459 & −8.758 & \textless{}
.001 \\
Competitor hotel: hotel 1 & −0.722 & 0.352 & −2.051 & .040 \\
Competitor hotel: hotel 2 & −0.818 & 0.369 & −2.217 & .027 \\
Competitor hotel: hotel 3 & 0.946 & 0.366 & 2.583 & .010 \\
Competitor hotel: hotel 4 & −0.798 & 0.356 & −2.239 & .025 \\
Competitor hotel: hotel 5 & 0.443 & 0.335 & 1.323 & .186 \\
Competitor hotel: hotel 6 & 1.371 & 0.347 & 3.949 & \textless{} .001 \\
Competitor hotel: hotel 7 & 2.111 & 0.376 & 5.611 & \textless{} .001 \\
Competitor hotel: hotel 8 & 0.408 & 0.362 & 1.129 & .259 \\
Competitor hotel: hotel 9 & 0.833 & 0.363 & 2.296 & .022 \\
\end{longtable}
}

\subsection*{C.2 Full mediation tables - Study
1}\label{full-mediation-tables---study-1}
\addcontentsline{toc}{subsection}{C.2 Full mediation tables - Study 1}

{\def\LTcaptype{apptbl}
\begin{longtable}[]{@{}lrrrl@{}}
\caption{{\label{apptbl-process-full-skep}}Full linear regression table
for skepticism (a-paths) - Gemini 3.1 Pro (Study 1)}\tabularnewline
\toprule\noalign{}
Term & \(b\) & \(SE\) & \(t\) & \(p\) \\
\midrule\noalign{}
\endfirsthead
\toprule\noalign{}
Term & \(b\) & \(SE\) & \(t\) & \(p\) \\
\midrule\noalign{}
\endhead
\bottomrule\noalign{}
\endlastfoot
Intercept & 1.699 & 0.185 & 9.160 & \textless{} .001 \\
Sponsorship (sponsored) & 4.827 & 0.081 & 59.919 & \textless{} .001 \\
Principal (platform) & −0.007 & 0.081 & −0.083 & .934 \\
Principal \(\times\) Sponsorship & −1.361 & 0.114 & −11.970 &
\textless{} .001 \\
Target price (standardized) & 0.003 & 0.029 & 0.088 & .930 \\
Target review: review 1 & −0.043 & 0.129 & −0.330 & .742 \\
Target review: review 2 & −0.188 & 0.132 & −1.422 & .155 \\
Target review: review 3 & −0.402 & 0.129 & −3.125 & .002 \\
Target review: review 4 & −0.677 & 0.127 & −5.335 & \textless{} .001 \\
Target review: review 5 & −0.036 & 0.126 & −0.284 & .776 \\
Target review: review 6 & −0.257 & 0.128 & −2.011 & .045 \\
Target review: review 7 & −0.107 & 0.134 & −0.798 & .425 \\
Target review: review 8 & 0.000 & 0.124 & 0.002 & .998 \\
Target review: review 9 & −0.390 & 0.124 & −3.146 & .002 \\
Target hotel: hotel 1 & −0.065 & 0.131 & −0.496 & .620 \\
Target hotel: hotel 2 & −0.077 & 0.126 & −0.606 & .544 \\
Target hotel: hotel 3 & −0.423 & 0.133 & −3.196 & .001 \\
Target hotel: hotel 4 & −0.215 & 0.132 & −1.633 & .103 \\
Target hotel: hotel 5 & −0.131 & 0.125 & −1.041 & .298 \\
Target hotel: hotel 6 & −0.189 & 0.128 & −1.476 & .140 \\
Target hotel: hotel 7 & −0.014 & 0.124 & −0.112 & .911 \\
Target hotel: hotel 8 & −0.194 & 0.128 & −1.523 & .128 \\
Target hotel: hotel 9 & −0.205 & 0.123 & −1.675 & .094 \\
Competitor review: review 1 & 0.127 & 0.129 & 0.981 & .327 \\
Competitor review: review 2 & −0.139 & 0.128 & −1.087 & .277 \\
Competitor review: review 3 & 0.244 & 0.123 & 1.979 & .048 \\
Competitor review: review 4 & 0.510 & 0.130 & 3.926 & \textless{}
.001 \\
Competitor review: review 5 & −0.048 & 0.128 & −0.377 & .706 \\
Competitor review: review 6 & 0.358 & 0.133 & 2.698 & .007 \\
Competitor review: review 7 & 0.097 & 0.127 & 0.769 & .442 \\
Competitor review: review 8 & −0.203 & 0.127 & −1.603 & .109 \\
Competitor review: review 9 & 0.320 & 0.130 & 2.459 & .014 \\
Competitor hotel: hotel 1 & −0.216 & 0.126 & −1.708 & .088 \\
Competitor hotel: hotel 2 & 0.171 & 0.130 & 1.324 & .186 \\
Competitor hotel: hotel 3 & −0.094 & 0.125 & −0.749 & .454 \\
Competitor hotel: hotel 4 & −0.191 & 0.126 & −1.521 & .129 \\
Competitor hotel: hotel 5 & −0.002 & 0.124 & −0.016 & .987 \\
Competitor hotel: hotel 6 & −0.295 & 0.126 & −2.339 & .020 \\
Competitor hotel: hotel 7 & −0.415 & 0.134 & −3.102 & .002 \\
Competitor hotel: hotel 8 & −0.117 & 0.128 & −0.914 & .361 \\
Competitor hotel: hotel 9 & −0.093 & 0.130 & −0.716 & .474 \\
\(N\) & 2,000 & & & \\
\(R^2\) & 0.751 & & & \\
\end{longtable}
}

\emph{Note.} Ordinary least squares regression on Skepticism (1 to 7)
derived from PROCESS Model 8 (\(N = 2{,}000\); \(R^2 = 0.751\),
\(F(40, 1959) = 148.02\), \(p < .001\)). Reference categories for
experimental manipulations are consumer delegation and organic target.
Reference categories for stimulus fixed effects are review 0 and hotel
0. The conditional effect of sponsorship is 4.827 (\(SE = 0.081\)) under
consumer delegation and 3.466 (\(SE = 0.081\)) under platform
delegation; the interaction accounts for \(\Delta R^2 = 0.018\),
\(F(1, 1959) = 143.27\), \(p < .001\).

{\def\LTcaptype{apptbl}
\begin{longtable}[]{@{}lrrrl@{}}
\caption{{\label{apptbl-process-full-choice}}Full logistic regression
table for target choice (b and c'-paths) - Gemini 3.1 Pro (Study 1)}\tabularnewline
\toprule\noalign{}
Term & \(\beta\) & \(SE\) & \(z\) & \(p\) \\
\midrule\noalign{}
\endfirsthead
\toprule\noalign{}
Term & \(\beta\) & \(SE\) & \(z\) & \(p\) \\
\midrule\noalign{}
\endhead
\bottomrule\noalign{}
\endlastfoot
Intercept & 1.900 & 0.682 & 2.787 & .005 \\
Sponsorship (sponsored) & −1.283 & 0.603 & −2.127 & .034 \\
Skepticism & −1.670 & 0.121 & −13.829 & \textless{} .001 \\
Principal (platform) & −0.289 & 0.249 & −1.158 & .247 \\
Principal \(\times\) Sponsorship & 2.859 & 0.586 & 4.883 & \textless{}
.001 \\
Target price (standardized) & 0.037 & 0.107 & 0.349 & .727 \\
Target review: review 1 & 1.171 & 0.492 & 2.380 & .017 \\
Target review: review 2 & 1.008 & 0.499 & 2.018 & .044 \\
Target review: review 3 & 3.631 & 0.537 & 6.764 & \textless{} .001 \\
Target review: review 4 & 7.540 & 0.641 & 11.759 & \textless{} .001 \\
Target review: review 5 & 2.393 & 0.506 & 4.731 & \textless{} .001 \\
Target review: review 6 & 4.038 & 0.534 & 7.560 & \textless{} .001 \\
Target review: review 7 & 2.388 & 0.513 & 4.651 & \textless{} .001 \\
Target review: review 8 & −0.503 & 0.513 & −0.981 & .327 \\
Target review: review 9 & 6.115 & 0.581 & 10.517 & \textless{} .001 \\
Target hotel: hotel 1 & 0.566 & 0.484 & 1.170 & .242 \\
Target hotel: hotel 2 & 0.056 & 0.471 & 0.118 & .906 \\
Target hotel: hotel 3 & −0.957 & 0.495 & −1.933 & .053 \\
Target hotel: hotel 4 & 1.389 & 0.537 & 2.586 & .010 \\
Target hotel: hotel 5 & −1.786 & 0.507 & −3.523 & \textless{} .001 \\
Target hotel: hotel 6 & −2.274 & 0.508 & −4.481 & \textless{} .001 \\
Target hotel: hotel 7 & −2.766 & 0.506 & −5.462 & \textless{} .001 \\
Target hotel: hotel 8 & −0.600 & 0.481 & −1.246 & .213 \\
Target hotel: hotel 9 & −0.343 & 0.472 & −0.727 & .467 \\
Competitor review: review 1 & −0.097 & 0.439 & −0.221 & .825 \\
Competitor review: review 2 & −0.124 & 0.417 & −0.298 & .766 \\
Competitor review: review 3 & −3.292 & 0.462 & −7.131 & \textless{}
.001 \\
Competitor review: review 4 & −9.093 & 1.389 & −6.545 & \textless{}
.001 \\
Competitor review: review 5 & −1.032 & 0.424 & −2.437 & .015 \\
Competitor review: review 6 & −3.225 & 0.522 & −6.172 & \textless{}
.001 \\
Competitor review: review 7 & −0.945 & 0.422 & −2.238 & .025 \\
Competitor review: review 8 & 1.373 & 0.441 & 3.113 & .002 \\
Competitor review: review 9 & −5.834 & 0.662 & −8.810 & \textless{}
.001 \\
Competitor hotel: hotel 1 & −1.736 & 0.460 & −3.777 & \textless{}
.001 \\
Competitor hotel: hotel 2 & −1.137 & 0.481 & −2.365 & .018 \\
Competitor hotel: hotel 3 & 0.805 & 0.492 & 1.637 & .102 \\
Competitor hotel: hotel 4 & −1.671 & 0.463 & −3.612 & \textless{}
.001 \\
Competitor hotel: hotel 5 & 0.706 & 0.455 & 1.552 & .121 \\
Competitor hotel: hotel 6 & 1.635 & 0.469 & 3.486 & \textless{} .001 \\
Competitor hotel: hotel 7 & 2.243 & 0.513 & 4.376 & \textless{} .001 \\
Competitor hotel: hotel 8 & 0.587 & 0.485 & 1.211 & .226 \\
Competitor hotel: hotel 9 & 0.921 & 0.487 & 1.891 & .059 \\
\(N\) & 2,000 & & & \\
Pseudo \(R^2\) & 0.760 & & & \\
Log-Likelihood & −307.32 & & & \\
\end{longtable}
}

\emph{Note.} Logistic regression on Target Choice derived from PROCESS
Model 8 (\(N = 2{,}000\); McFadden pseudo \(R^2 = 0.760\), Cox-Snell
pseudo \(R^2 = 0.622\), Nagelkerke pseudo \(R^2 = 0.861\);
\(-2LL = 614.63\), model \(\chi^2(41) = 1{,}942.82\), \(p < .001\)).
Reference categories for experimental manipulations are consumer
delegation and organic target. Reference categories for stimulus fixed
effects are review 0 and hotel 0. Coefficients (\(\beta\)) are expressed
in a log-odds metric. The conditional direct effect of sponsorship is
−1.283 (\(SE = 0.603\), \(p = .034\)) under consumer delegation and
1.576 (\(SE = 0.389\), \(p < .001\)) under platform delegation; the
interaction is significant by likelihood ratio test,
\(\chi^2(1) = 26.64\), \(p < .001\).

\subsection*{C.3 Full logistic regression table - Claude Sonnet
5}\label{full-logistic-regression-table---claude-sonnet-5}
\addcontentsline{toc}{subsection}{C.3 Full logistic regression table -
Claude Sonnet 5}

{\def\LTcaptype{apptbl}
\begin{longtable}[]{@{}lrrrl@{}}
\caption{{\label{apptbl-full-claude-sonnet-5}}Full logistic regression
of target hotel choice for Claude Sonnet 5 (Study 1; \(N = 2,000\)).}\tabularnewline
\toprule\noalign{}
Term & \(\beta\) & \(SE\) & \(z\) & \(p\) \\
\midrule\noalign{}
\endfirsthead
\toprule\noalign{}
Term & \(\beta\) & \(SE\) & \(z\) & \(p\) \\
\midrule\noalign{}
\endhead
\bottomrule\noalign{}
\endlastfoot
Intercept & 0.242 & 0.405 & 0.598 & .550 \\
Principal (platform) & 0.097 & 0.157 & 0.619 & .536 \\
Sponsorship (sponsored) & −3.243 & 0.217 & −14.949 & \textless{} .001 \\
Principal \(\times\) Sponsorship & 1.759 & 0.262 & 6.725 & \textless{}
.001 \\
Target price (standardized) & −0.025 & 0.062 & −0.397 & .692 \\
Target review: review 1 & 0.565 & 0.283 & 1.992 & .046 \\
Target review: review 2 & 0.584 & 0.282 & 2.070 & .038 \\
Target review: review 3 & 1.238 & 0.274 & 4.521 & \textless{} .001 \\
Target review: review 4 & 2.436 & 0.295 & 8.262 & \textless{} .001 \\
Target review: review 5 & 0.439 & 0.283 & 1.554 & .120 \\
Target review: review 6 & 1.544 & 0.285 & 5.425 & \textless{} .001 \\
Target review: review 7 & 0.665 & 0.278 & 2.391 & .017 \\
Target review: review 8 & −0.710 & 0.306 & −2.318 & .020 \\
Target review: review 9 & 1.512 & 0.284 & 5.333 & \textless{} .001 \\
Target hotel: hotel 1 & 0.879 & 0.267 & 3.296 & \textless{} .001 \\
Target hotel: hotel 2 & −0.505 & 0.266 & −1.899 & .058 \\
Target hotel: hotel 3 & −0.414 & 0.268 & −1.545 & .122 \\
Target hotel: hotel 4 & 0.325 & 0.258 & 1.259 & .208 \\
Target hotel: hotel 5 & −0.498 & 0.277 & −1.796 & .072 \\
Target hotel: hotel 6 & −1.344 & 0.275 & −4.893 & \textless{} .001 \\
Target hotel: hotel 7 & −1.990 & 0.320 & −6.218 & \textless{} .001 \\
Target hotel: hotel 8 & 0.011 & 0.260 & 0.041 & .967 \\
Target hotel: hotel 9 & 0.051 & 0.261 & 0.197 & .844 \\
Competitor review: review 1 & −0.238 & 0.248 & −0.960 & .337 \\
Competitor review: review 2 & −0.533 & 0.255 & −2.091 & .037 \\
Competitor review: review 3 & −1.505 & 0.277 & −5.435 & \textless{}
.001 \\
Competitor review: review 4 & −2.448 & 0.313 & −7.815 & \textless{}
.001 \\
Competitor review: review 5 & −0.181 & 0.252 & −0.716 & .474 \\
Competitor review: review 6 & −1.205 & 0.269 & −4.479 & \textless{}
.001 \\
Competitor review: review 7 & −0.482 & 0.259 & −1.861 & .063 \\
Competitor review: review 8 & 0.711 & 0.265 & 2.685 & .007 \\
Competitor review: review 9 & −1.872 & 0.284 & −6.592 & \textless{}
.001 \\
Competitor hotel: hotel 1 & −0.921 & 0.298 & −3.094 & .002 \\
Competitor hotel: hotel 2 & 0.568 & 0.272 & 2.086 & .037 \\
Competitor hotel: hotel 3 & 0.077 & 0.268 & 0.288 & .773 \\
Competitor hotel: hotel 4 & −0.608 & 0.270 & −2.253 & .024 \\
Competitor hotel: hotel 5 & 0.420 & 0.265 & 1.584 & .113 \\
Competitor hotel: hotel 6 & 0.774 & 0.269 & 2.881 & .004 \\
Competitor hotel: hotel 7 & 1.045 & 0.266 & 3.935 & \textless{} .001 \\
Competitor hotel: hotel 8 & −0.420 & 0.286 & −1.467 & .142 \\
Competitor hotel: hotel 9 & 0.035 & 0.275 & 0.129 & .897 \\
\(N\) & 2,000 & & & \\
Pseudo \(R^2\) & 0.348 & & & \\
Log-Likelihood & −845.00 & & & \\
\end{longtable}
}

\emph{Note.} Reference categories are consumer for principal, organic
for sponsorship, and index 0 for target/competitor review text and hotel
identity.

\subsection*{C.4 Full logistic regression table - GPT 5.6
Terra}\label{full-logistic-regression-table---gpt-5.6-terra}
\addcontentsline{toc}{subsection}{C.4 Full logistic regression table -
GPT 5.6 Terra}

{\def\LTcaptype{apptbl}
\begin{longtable}[]{@{}lrrrl@{}}
\caption{{\label{apptbl-full-gpt-5-6-terra}}Full logistic regression of
target hotel choice for GPT 5.6 Terra (Study 1; \(N = 2,000\)).}\tabularnewline
\toprule\noalign{}
Term & \(\beta\) & \(SE\) & \(z\) & \(p\) \\
\midrule\noalign{}
\endfirsthead
\toprule\noalign{}
Term & \(\beta\) & \(SE\) & \(z\) & \(p\) \\
\midrule\noalign{}
\endhead
\bottomrule\noalign{}
\endlastfoot
Intercept & 0.036 & 0.432 & 0.084 & .933 \\
Principal (platform) & −0.345 & 0.177 & −1.949 & .051 \\
Sponsorship (sponsored) & −2.238 & 0.203 & −11.047 & \textless{} .001 \\
Principal \(\times\) Sponsorship & 0.708 & 0.264 & 2.681 & .007 \\
Target price (standardized) & −0.021 & 0.066 & −0.323 & .747 \\
Target review: review 1 & 0.510 & 0.306 & 1.667 & .096 \\
Target review: review 2 & 0.099 & 0.315 & 0.315 & .752 \\
Target review: review 3 & 2.420 & 0.306 & 7.903 & \textless{} .001 \\
Target review: review 4 & 3.550 & 0.328 & 10.814 & \textless{} .001 \\
Target review: review 5 & 1.008 & 0.301 & 3.350 & \textless{} .001 \\
Target review: review 6 & 2.162 & 0.300 & 7.213 & \textless{} .001 \\
Target review: review 7 & 1.565 & 0.285 & 5.490 & \textless{} .001 \\
Target review: review 8 & −0.734 & 0.324 & −2.264 & .024 \\
Target review: review 9 & 2.451 & 0.309 & 7.937 & \textless{} .001 \\
Target hotel: hotel 1 & 0.812 & 0.276 & 2.939 & .003 \\
Target hotel: hotel 2 & 0.434 & 0.269 & 1.615 & .106 \\
Target hotel: hotel 3 & 1.159 & 0.278 & 4.169 & \textless{} .001 \\
Target hotel: hotel 4 & 1.167 & 0.293 & 3.988 & \textless{} .001 \\
Target hotel: hotel 5 & −1.177 & 0.295 & −3.994 & \textless{} .001 \\
Target hotel: hotel 6 & −1.514 & 0.319 & −4.741 & \textless{} .001 \\
Target hotel: hotel 7 & −2.434 & 0.331 & −7.346 & \textless{} .001 \\
Target hotel: hotel 8 & 0.563 & 0.270 & 2.083 & .037 \\
Target hotel: hotel 9 & −0.153 & 0.272 & −0.565 & .572 \\
Competitor review: review 1 & −0.171 & 0.277 & −0.616 & .538 \\
Competitor review: review 2 & 0.108 & 0.286 & 0.378 & .705 \\
Competitor review: review 3 & −2.132 & 0.320 & −6.655 & \textless{}
.001 \\
Competitor review: review 4 & −3.171 & 0.341 & −9.291 & \textless{}
.001 \\
Competitor review: review 5 & −0.901 & 0.292 & −3.084 & .002 \\
Competitor review: review 6 & −2.082 & 0.302 & −6.898 & \textless{}
.001 \\
Competitor review: review 7 & −0.888 & 0.293 & −3.033 & .002 \\
Competitor review: review 8 & 0.983 & 0.294 & 3.340 & \textless{}
.001 \\
Competitor review: review 9 & −2.443 & 0.311 & −7.848 & \textless{}
.001 \\
Competitor hotel: hotel 1 & −0.671 & 0.300 & −2.235 & .025 \\
Competitor hotel: hotel 2 & −0.391 & 0.289 & −1.356 & .175 \\
Competitor hotel: hotel 3 & −0.715 & 0.304 & −2.353 & .019 \\
Competitor hotel: hotel 4 & −0.896 & 0.298 & −3.006 & .003 \\
Competitor hotel: hotel 5 & 1.465 & 0.294 & 4.987 & \textless{} .001 \\
Competitor hotel: hotel 6 & 1.141 & 0.289 & 3.945 & \textless{} .001 \\
Competitor hotel: hotel 7 & 1.959 & 0.292 & 6.702 & \textless{} .001 \\
Competitor hotel: hotel 8 & −0.117 & 0.286 & −0.408 & .683 \\
Competitor hotel: hotel 9 & −0.199 & 0.285 & −0.699 & .485 \\
\(N\) & 2,000 & & & \\
Pseudo \(R^2\) & 0.443 & & & \\
Log-Likelihood & −748.58 & & & \\
\end{longtable}
}

\emph{Note.} Reference categories are consumer for principal, organic
for sponsorship, and index 0 for target/competitor review text and hotel
identity.

\subsection*{C.5 Full logistic regression table - Gemini 3.5
Flash}\label{full-logistic-regression-table---gemini-3.5-flash}
\addcontentsline{toc}{subsection}{C.5 Full logistic regression table -
Gemini 3.5 Flash}

{\def\LTcaptype{apptbl}
\begin{longtable}[]{@{}lrrrl@{}}
\caption{{\label{apptbl-full-gemini-3-5-flash}}Full logistic regression
of target hotel choice for Gemini 3.5 Flash (Study 1; \(N = 2,000\)).}\tabularnewline
\toprule\noalign{}
Term & \(\beta\) & \(SE\) & \(z\) & \(p\) \\
\midrule\noalign{}
\endfirsthead
\toprule\noalign{}
Term & \(\beta\) & \(SE\) & \(z\) & \(p\) \\
\midrule\noalign{}
\endhead
\bottomrule\noalign{}
\endlastfoot
Intercept & −0.400 & 0.461 & −0.867 & .386 \\
Principal (platform) & 0.008 & 0.200 & 0.041 & .968 \\
Sponsorship (sponsored) & −1.824 & 0.217 & −8.399 & \textless{} .001 \\
Principal \(\times\) Sponsorship & 0.975 & 0.285 & 3.419 & \textless{}
.001 \\
Target price (standardized) & 0.110 & 0.072 & 1.533 & .125 \\
Target review: review 1 & −0.734 & 0.319 & −2.305 & .021 \\
Target review: review 2 & −0.057 & 0.301 & −0.191 & .849 \\
Target review: review 3 & 2.233 & 0.311 & 7.191 & \textless{} .001 \\
Target review: review 4 & 4.321 & 0.383 & 11.294 & \textless{} .001 \\
Target review: review 5 & −0.361 & 0.332 & −1.087 & .277 \\
Target review: review 6 & 1.548 & 0.301 & 5.150 & \textless{} .001 \\
Target review: review 7 & 0.980 & 0.292 & 3.356 & \textless{} .001 \\
Target review: review 8 & −2.004 & 0.360 & −5.559 & \textless{} .001 \\
Target review: review 9 & 3.062 & 0.342 & 8.957 & \textless{} .001 \\
Target hotel: hotel 1 & 2.841 & 0.331 & 8.577 & \textless{} .001 \\
Target hotel: hotel 2 & 1.654 & 0.310 & 5.332 & \textless{} .001 \\
Target hotel: hotel 3 & 1.196 & 0.313 & 3.826 & \textless{} .001 \\
Target hotel: hotel 4 & 1.895 & 0.327 & 5.791 & \textless{} .001 \\
Target hotel: hotel 5 & 0.349 & 0.324 & 1.078 & .281 \\
Target hotel: hotel 6 & −0.515 & 0.322 & −1.601 & .109 \\
Target hotel: hotel 7 & −1.747 & 0.359 & −4.866 & \textless{} .001 \\
Target hotel: hotel 8 & 1.145 & 0.306 & 3.744 & \textless{} .001 \\
Target hotel: hotel 9 & 1.261 & 0.317 & 3.979 & \textless{} .001 \\
Competitor review: review 1 & 0.679 & 0.292 & 2.325 & .020 \\
Competitor review: review 2 & −0.088 & 0.282 & −0.312 & .755 \\
Competitor review: review 3 & −1.994 & 0.319 & −6.260 & \textless{}
.001 \\
Competitor review: review 4 & −4.089 & 0.410 & −9.963 & \textless{}
.001 \\
Competitor review: review 5 & 0.173 & 0.301 & 0.574 & .566 \\
Competitor review: review 6 & −1.328 & 0.299 & −4.446 & \textless{}
.001 \\
Competitor review: review 7 & −0.549 & 0.297 & −1.846 & .065 \\
Competitor review: review 8 & 1.436 & 0.306 & 4.694 & \textless{}
.001 \\
Competitor review: review 9 & −3.348 & 0.363 & −9.231 & \textless{}
.001 \\
Competitor hotel: hotel 1 & −2.272 & 0.347 & −6.541 & \textless{}
.001 \\
Competitor hotel: hotel 2 & −0.913 & 0.317 & −2.877 & .004 \\
Competitor hotel: hotel 3 & −0.808 & 0.311 & −2.595 & .009 \\
Competitor hotel: hotel 4 & −1.634 & 0.316 & −5.175 & \textless{}
.001 \\
Competitor hotel: hotel 5 & −0.129 & 0.301 & −0.429 & .668 \\
Competitor hotel: hotel 6 & 1.393 & 0.312 & 4.458 & \textless{} .001 \\
Competitor hotel: hotel 7 & 2.480 & 0.341 & 7.281 & \textless{} .001 \\
Competitor hotel: hotel 8 & −0.771 & 0.301 & −2.559 & .010 \\
Competitor hotel: hotel 9 & −0.396 & 0.308 & −1.283 & .200 \\
\(N\) & 2,000 & & & \\
Pseudo \(R^2\) & 0.519 & & & \\
Log-Likelihood & −661.59 & & & \\
\end{longtable}
}

\emph{Note.} Reference categories are consumer for principal, organic
for sponsorship, and index 0 for target/competitor review text and hotel
identity.

\subsection*{C.6 Full logistic regression table - Gemini 3.6
Flash}\label{full-logistic-regression-table---gemini-3.6-flash}
\addcontentsline{toc}{subsection}{C.6 Full logistic regression table -
Gemini 3.6 Flash}

{\def\LTcaptype{apptbl}
\begin{longtable}[]{@{}lrrrl@{}}
\caption{{\label{apptbl-full-gemini-3-6-flash}}Full logistic regression
of target hotel choice for Gemini 3.6 Flash (Study 1; \(N = 2,000\)).}\tabularnewline
\toprule\noalign{}
Term & \(\beta\) & \(SE\) & \(z\) & \(p\) \\
\midrule\noalign{}
\endfirsthead
\toprule\noalign{}
Term & \(\beta\) & \(SE\) & \(z\) & \(p\) \\
\midrule\noalign{}
\endhead
\bottomrule\noalign{}
\endlastfoot
Intercept & −0.398 & 0.515 & −0.772 & .440 \\
Principal (platform) & −0.011 & 0.221 & −0.048 & .961 \\
Sponsorship (sponsored) & −0.917 & 0.226 & −4.052 & \textless{} .001 \\
Principal \(\times\) Sponsorship & 1.124 & 0.318 & 3.539 & \textless{}
.001 \\
Target price (standardized) & −0.109 & 0.081 & −1.356 & .175 \\
Target review: review 1 & −0.040 & 0.355 & −0.113 & .910 \\
Target review: review 2 & 1.476 & 0.351 & 4.201 & \textless{} .001 \\
Target review: review 3 & 3.012 & 0.347 & 8.684 & \textless{} .001 \\
Target review: review 4 & 6.403 & 0.472 & 13.553 & \textless{} .001 \\
Target review: review 5 & 0.650 & 0.341 & 1.907 & .057 \\
Target review: review 6 & 2.982 & 0.354 & 8.430 & \textless{} .001 \\
Target review: review 7 & 1.842 & 0.349 & 5.280 & \textless{} .001 \\
Target review: review 8 & −1.141 & 0.378 & −3.017 & .003 \\
Target review: review 9 & 4.943 & 0.428 & 11.552 & \textless{} .001 \\
Target hotel: hotel 1 & 3.377 & 0.383 & 8.818 & \textless{} .001 \\
Target hotel: hotel 2 & 0.882 & 0.338 & 2.607 & .009 \\
Target hotel: hotel 3 & 1.754 & 0.359 & 4.879 & \textless{} .001 \\
Target hotel: hotel 4 & 2.585 & 0.371 & 6.965 & \textless{} .001 \\
Target hotel: hotel 5 & −0.218 & 0.327 & −0.668 & .504 \\
Target hotel: hotel 6 & −1.953 & 0.358 & −5.449 & \textless{} .001 \\
Target hotel: hotel 7 & −2.941 & 0.391 & −7.519 & \textless{} .001 \\
Target hotel: hotel 8 & 1.082 & 0.347 & 3.116 & .002 \\
Target hotel: hotel 9 & 1.683 & 0.335 & 5.020 & \textless{} .001 \\
Competitor review: review 1 & −0.561 & 0.352 & −1.593 & .111 \\
Competitor review: review 2 & −1.612 & 0.335 & −4.815 & \textless{}
.001 \\
Competitor review: review 3 & −3.876 & 0.390 & −9.946 & \textless{}
.001 \\
Competitor review: review 4 & −5.671 & 0.439 & −12.913 & \textless{}
.001 \\
Competitor review: review 5 & −0.937 & 0.337 & −2.783 & .005 \\
Competitor review: review 6 & −2.953 & 0.364 & −8.115 & \textless{}
.001 \\
Competitor review: review 7 & −2.043 & 0.359 & −5.694 & \textless{}
.001 \\
Competitor review: review 8 & 1.460 & 0.364 & 4.013 & \textless{}
.001 \\
Competitor review: review 9 & −4.888 & 0.409 & −11.941 & \textless{}
.001 \\
Competitor hotel: hotel 1 & −2.536 & 0.358 & −7.080 & \textless{}
.001 \\
Competitor hotel: hotel 2 & −0.739 & 0.324 & −2.279 & .023 \\
Competitor hotel: hotel 3 & −0.743 & 0.351 & −2.116 & .034 \\
Competitor hotel: hotel 4 & −1.928 & 0.362 & −5.328 & \textless{}
.001 \\
Competitor hotel: hotel 5 & 0.603 & 0.335 & 1.800 & .072 \\
Competitor hotel: hotel 6 & 3.455 & 0.396 & 8.719 & \textless{} .001 \\
Competitor hotel: hotel 7 & 3.058 & 0.377 & 8.110 & \textless{} .001 \\
Competitor hotel: hotel 8 & −0.738 & 0.318 & −2.323 & .020 \\
Competitor hotel: hotel 9 & −0.642 & 0.328 & −1.959 & .050 \\
\(N\) & 2,000 & & & \\
Pseudo \(R^2\) & 0.610 & & & \\
Log-Likelihood & −540.43 & & & \\
\end{longtable}
}

\emph{Note.} Reference categories are consumer for principal, organic
for sponsorship, and index 0 for target/competitor review text and hotel
identity.

\subsection*{C.7 Full logistic regression table - Gemini 3.8
Flash}\label{full-logistic-regression-table---gemini-3.8-flash}
\addcontentsline{toc}{subsection}{C.7 Full logistic regression table -
Gemini 3.8 Flash}

{\def\LTcaptype{apptbl}
\begin{longtable}[]{@{}lrrrl@{}}
\caption{{\label{apptbl-full-gemini-3-8-flash}}Full logistic regression
of target hotel choice for Gemini 3.8 Flash (Study 1; \(N = 2,000\)).}\tabularnewline
\toprule\noalign{}
Term & \(\beta\) & \(SE\) & \(z\) & \(p\) \\
\midrule\noalign{}
\endfirsthead
\toprule\noalign{}
Term & \(\beta\) & \(SE\) & \(z\) & \(p\) \\
\midrule\noalign{}
\endhead
\bottomrule\noalign{}
\endlastfoot
Intercept & −0.437 & 0.529 & −0.827 & .408 \\
Principal (platform) & −0.119 & 0.219 & −0.544 & .587 \\
Sponsorship (sponsored) & −5.836 & 0.376 & −15.506 & \textless{} .001 \\
Principal \(\times\) Sponsorship & 5.243 & 0.428 & 12.237 & \textless{}
.001 \\
Target price (standardized) & −0.051 & 0.087 & −0.581 & .562 \\
Target review: review 1 & −0.179 & 0.406 & −0.440 & .660 \\
Target review: review 2 & 0.333 & 0.407 & 0.819 & .413 \\
Target review: review 3 & 3.239 & 0.425 & 7.618 & \textless{} .001 \\
Target review: review 4 & 4.876 & 0.447 & 10.914 & \textless{} .001 \\
Target review: review 5 & 0.834 & 0.397 & 2.102 & .036 \\
Target review: review 6 & 2.270 & 0.401 & 5.668 & \textless{} .001 \\
Target review: review 7 & 1.135 & 0.404 & 2.809 & .005 \\
Target review: review 8 & −0.657 & 0.410 & −1.605 & .108 \\
Target review: review 9 & 4.216 & 0.443 & 9.528 & \textless{} .001 \\
Target hotel: hotel 1 & 2.879 & 0.406 & 7.098 & \textless{} .001 \\
Target hotel: hotel 2 & 1.542 & 0.367 & 4.197 & \textless{} .001 \\
Target hotel: hotel 3 & 0.948 & 0.375 & 2.526 & .012 \\
Target hotel: hotel 4 & 1.735 & 0.364 & 4.765 & \textless{} .001 \\
Target hotel: hotel 5 & −0.647 & 0.377 & −1.715 & .086 \\
Target hotel: hotel 6 & −2.400 & 0.398 & −6.032 & \textless{} .001 \\
Target hotel: hotel 7 & −3.852 & 0.506 & −7.609 & \textless{} .001 \\
Target hotel: hotel 8 & 1.399 & 0.369 & 3.796 & \textless{} .001 \\
Target hotel: hotel 9 & 1.390 & 0.363 & 3.829 & \textless{} .001 \\
Competitor review: review 1 & 0.403 & 0.343 & 1.174 & .240 \\
Competitor review: review 2 & −0.636 & 0.342 & −1.860 & .063 \\
Competitor review: review 3 & −3.090 & 0.391 & −7.895 & \textless{}
.001 \\
Competitor review: review 4 & −5.348 & 0.504 & −10.604 & \textless{}
.001 \\
Competitor review: review 5 & −0.277 & 0.334 & −0.830 & .407 \\
Competitor review: review 6 & −1.907 & 0.373 & −5.110 & \textless{}
.001 \\
Competitor review: review 7 & −1.223 & 0.342 & −3.571 & \textless{}
.001 \\
Competitor review: review 8 & 1.480 & 0.348 & 4.250 & \textless{}
.001 \\
Competitor review: review 9 & −3.854 & 0.419 & −9.198 & \textless{}
.001 \\
Competitor hotel: hotel 1 & −2.660 & 0.424 & −6.270 & \textless{}
.001 \\
Competitor hotel: hotel 2 & −0.918 & 0.378 & −2.431 & .015 \\
Competitor hotel: hotel 3 & −0.491 & 0.347 & −1.412 & .158 \\
Competitor hotel: hotel 4 & −2.065 & 0.397 & −5.196 & \textless{}
.001 \\
Competitor hotel: hotel 5 & 0.707 & 0.356 & 1.986 & .047 \\
Competitor hotel: hotel 6 & 2.751 & 0.388 & 7.095 & \textless{} .001 \\
Competitor hotel: hotel 7 & 4.272 & 0.422 & 10.134 & \textless{} .001 \\
Competitor hotel: hotel 8 & −0.603 & 0.344 & −1.753 & .080 \\
Competitor hotel: hotel 9 & −0.959 & 0.370 & −2.588 & .010 \\
\(N\) & 2,000 & & & \\
Pseudo \(R^2\) & 0.643 & & & \\
Log-Likelihood & −479.44 & & & \\
\end{longtable}
}

\emph{Note.} Reference categories are consumer for principal, organic
for sponsorship, and index 0 for target/competitor review text and hotel
identity.

\subsection*{C.8 Full logistic regression tables - system-prompt
wording}\label{full-logistic-regression-tables---system-prompt-wording}
\addcontentsline{toc}{subsection}{C.8 Full logistic regression tables -
system-prompt wording}

{\def\LTcaptype{apptbl}
\begin{longtable}[]{@{}lrrrl@{}}
\caption{{\label{apptbl-mnl-principal-noun}}Full logistic regression
table - system-prompt variant ``Principal noun'' - Gemini 3.1 Pro (Study
1)}\tabularnewline
\toprule\noalign{}
Term & \(\beta\) & \(SE\) & \(z\) & \(p\) \\
\midrule\noalign{}
\endfirsthead
\toprule\noalign{}
Term & \(\beta\) & \(SE\) & \(z\) & \(p\) \\
\midrule\noalign{}
\endhead
\bottomrule\noalign{}
\endlastfoot
Intercept & −0.148 & 0.557 & −0.265 & .791 \\
Principal (platform) & −0.352 & 0.210 & −1.678 & .093 \\
Sponsorship (sponsored) & −6.703 & 0.423 & −15.837 & \textless{} .001 \\
Principal \(\times\) Sponsorship & 2.503 & 0.406 & 6.164 & \textless{}
.001 \\
Target price (standardized) & 0.124 & 0.089 & 1.404 & .160 \\
Target review: review 1 & 1.251 & 0.415 & 3.016 & .003 \\
Target review: review 2 & 0.614 & 0.444 & 1.384 & .166 \\
Target review: review 3 & 3.198 & 0.441 & 7.250 & \textless{} .001 \\
Target review: review 4 & 5.423 & 0.463 & 11.723 & \textless{} .001 \\
Target review: review 5 & 1.778 & 0.431 & 4.126 & \textless{} .001 \\
Target review: review 6 & 4.003 & 0.434 & 9.229 & \textless{} .001 \\
Target review: review 7 & 1.827 & 0.411 & 4.450 & \textless{} .001 \\
Target review: review 8 & −1.073 & 0.479 & −2.240 & .025 \\
Target review: review 9 & 5.090 & 0.449 & 11.333 & \textless{} .001 \\
Target hotel: hotel 1 & 1.010 & 0.358 & 2.818 & .005 \\
Target hotel: hotel 2 & 0.545 & 0.361 & 1.510 & .131 \\
Target hotel: hotel 3 & −0.074 & 0.401 & −0.184 & .854 \\
Target hotel: hotel 4 & 0.837 & 0.378 & 2.213 & .027 \\
Target hotel: hotel 5 & 0.148 & 0.374 & 0.396 & .692 \\
Target hotel: hotel 6 & −0.839 & 0.439 & −1.911 & .056 \\
Target hotel: hotel 7 & −1.456 & 0.404 & −3.606 & \textless{} .001 \\
Target hotel: hotel 8 & 0.787 & 0.392 & 2.008 & .045 \\
Target hotel: hotel 9 & 0.083 & 0.372 & 0.224 & .823 \\
Competitor review: review 1 & −0.877 & 0.348 & −2.522 & .012 \\
Competitor review: review 2 & −0.310 & 0.331 & −0.937 & .349 \\
Competitor review: review 3 & −3.743 & 0.457 & −8.190 & \textless{}
.001 \\
Competitor review: review 4 & −6.021 & 0.573 & −10.502 & \textless{}
.001 \\
Competitor review: review 5 & −1.022 & 0.348 & −2.936 & .003 \\
Competitor review: review 6 & −3.269 & 0.395 & −8.281 & \textless{}
.001 \\
Competitor review: review 7 & −2.186 & 0.369 & −5.916 & \textless{}
.001 \\
Competitor review: review 8 & 1.179 & 0.366 & 3.224 & .001 \\
Competitor review: review 9 & −5.561 & 0.534 & −10.405 & \textless{}
.001 \\
Competitor hotel: hotel 1 & −0.955 & 0.434 & −2.203 & .028 \\
Competitor hotel: hotel 2 & −0.902 & 0.407 & −2.218 & .027 \\
Competitor hotel: hotel 3 & 0.519 & 0.377 & 1.375 & .169 \\
Competitor hotel: hotel 4 & −0.585 & 0.376 & −1.558 & .119 \\
Competitor hotel: hotel 5 & 0.366 & 0.370 & 0.988 & .323 \\
Competitor hotel: hotel 6 & 0.711 & 0.375 & 1.893 & .058 \\
Competitor hotel: hotel 7 & 1.447 & 0.395 & 3.662 & \textless{} .001 \\
Competitor hotel: hotel 8 & 0.261 & 0.385 & 0.678 & .497 \\
Competitor hotel: hotel 9 & 0.113 & 0.386 & 0.294 & .769 \\
\end{longtable}
}

{\def\LTcaptype{apptbl}
\begin{longtable}[]{@{}lrrrl@{}}
\caption{{\label{apptbl-mnl-recommend}}Full logistic regression table -
system-prompt variant ``Action verb'' - Gemini 3.1 Pro (Study 1)}\tabularnewline
\toprule\noalign{}
Term & \(\beta\) & \(SE\) & \(z\) & \(p\) \\
\midrule\noalign{}
\endfirsthead
\toprule\noalign{}
Term & \(\beta\) & \(SE\) & \(z\) & \(p\) \\
\midrule\noalign{}
\endhead
\bottomrule\noalign{}
\endlastfoot
Intercept & −1.180 & 0.502 & −2.349 & .019 \\
Principal (platform) & 0.076 & 0.190 & 0.399 & .690 \\
Sponsorship (sponsored) & −5.074 & 0.325 & −15.593 & \textless{} .001 \\
Principal \(\times\) Sponsorship & 4.592 & 0.366 & 12.557 & \textless{}
.001 \\
Target price (standardized) & 0.089 & 0.074 & 1.213 & .225 \\
Target review: review 1 & 0.852 & 0.343 & 2.485 & .013 \\
Target review: review 2 & 0.659 & 0.342 & 1.926 & .054 \\
Target review: review 3 & 2.797 & 0.350 & 8.000 & \textless{} .001 \\
Target review: review 4 & 5.116 & 0.393 & 13.019 & \textless{} .001 \\
Target review: review 5 & 0.980 & 0.332 & 2.948 & .003 \\
Target review: review 6 & 2.886 & 0.339 & 8.518 & \textless{} .001 \\
Target review: review 7 & 1.325 & 0.322 & 4.119 & \textless{} .001 \\
Target review: review 8 & 0.253 & 0.349 & 0.725 & .469 \\
Target review: review 9 & 3.982 & 0.371 & 10.744 & \textless{} .001 \\
Target hotel: hotel 1 & 0.805 & 0.332 & 2.426 & .015 \\
Target hotel: hotel 2 & 0.776 & 0.336 & 2.308 & .021 \\
Target hotel: hotel 3 & −0.242 & 0.338 & −0.716 & .474 \\
Target hotel: hotel 4 & 0.880 & 0.346 & 2.545 & .011 \\
Target hotel: hotel 5 & −0.192 & 0.348 & −0.552 & .581 \\
Target hotel: hotel 6 & −0.778 & 0.339 & −2.294 & .022 \\
Target hotel: hotel 7 & −1.001 & 0.352 & −2.847 & .004 \\
Target hotel: hotel 8 & 0.104 & 0.333 & 0.313 & .754 \\
Target hotel: hotel 9 & −0.092 & 0.327 & −0.282 & .778 \\
Competitor review: review 1 & 0.170 & 0.299 & 0.570 & .568 \\
Competitor review: review 2 & 0.117 & 0.302 & 0.387 & .699 \\
Competitor review: review 3 & −1.864 & 0.305 & −6.107 & \textless{}
.001 \\
Competitor review: review 4 & −4.207 & 0.425 & −9.898 & \textless{}
.001 \\
Competitor review: review 5 & −0.529 & 0.310 & −1.704 & .088 \\
Competitor review: review 6 & −2.348 & 0.340 & −6.900 & \textless{}
.001 \\
Competitor review: review 7 & −0.988 & 0.299 & −3.304 & \textless{}
.001 \\
Competitor review: review 8 & 1.286 & 0.309 & 4.157 & \textless{}
.001 \\
Competitor review: review 9 & −3.450 & 0.391 & −8.831 & \textless{}
.001 \\
Competitor hotel: hotel 1 & −0.439 & 0.345 & −1.275 & .202 \\
Competitor hotel: hotel 2 & 0.101 & 0.326 & 0.310 & .757 \\
Competitor hotel: hotel 3 & 0.427 & 0.338 & 1.261 & .207 \\
Competitor hotel: hotel 4 & −0.333 & 0.331 & −1.004 & .315 \\
Competitor hotel: hotel 5 & 0.973 & 0.332 & 2.927 & .003 \\
Competitor hotel: hotel 6 & 1.442 & 0.330 & 4.370 & \textless{} .001 \\
Competitor hotel: hotel 7 & 1.694 & 0.335 & 5.053 & \textless{} .001 \\
Competitor hotel: hotel 8 & 0.311 & 0.338 & 0.920 & .358 \\
Competitor hotel: hotel 9 & 0.215 & 0.337 & 0.638 & .523 \\
\end{longtable}
}

\subsection*{C.9 Full logistic regression table - thinking
level}\label{full-logistic-regression-table---thinking-level}
\addcontentsline{toc}{subsection}{C.9 Full logistic regression table -
thinking level}

{\def\LTcaptype{apptbl}
\begin{longtable}[]{@{}
  >{\raggedright\arraybackslash}p{(\linewidth - 8\tabcolsep) * \real{0.2000}}
  >{\raggedleft\arraybackslash}p{(\linewidth - 8\tabcolsep) * \real{0.2000}}
  >{\raggedleft\arraybackslash}p{(\linewidth - 8\tabcolsep) * \real{0.2000}}
  >{\raggedleft\arraybackslash}p{(\linewidth - 8\tabcolsep) * \real{0.2000}}
  >{\raggedright\arraybackslash}p{(\linewidth - 8\tabcolsep) * \real{0.2000}}@{}}
\caption{{\label{apptbl-thinking-level-logit-full}}Full logistic
regression table - HIGH vs.~LOW thinking level - Gemini 3.1 Pro (Study
1)}\tabularnewline
\toprule\noalign{}
\begin{minipage}[b]{\linewidth}\raggedright
Term
\end{minipage} & \begin{minipage}[b]{\linewidth}\raggedleft
\(\beta\)
\end{minipage} & \begin{minipage}[b]{\linewidth}\raggedleft
\(SE\)
\end{minipage} & \begin{minipage}[b]{\linewidth}\raggedleft
\(z\)
\end{minipage} & \begin{minipage}[b]{\linewidth}\raggedright
\(p\)
\end{minipage} \\
\midrule\noalign{}
\endfirsthead
\toprule\noalign{}
\begin{minipage}[b]{\linewidth}\raggedright
Term
\end{minipage} & \begin{minipage}[b]{\linewidth}\raggedleft
\(\beta\)
\end{minipage} & \begin{minipage}[b]{\linewidth}\raggedleft
\(SE\)
\end{minipage} & \begin{minipage}[b]{\linewidth}\raggedleft
\(z\)
\end{minipage} & \begin{minipage}[b]{\linewidth}\raggedright
\(p\)
\end{minipage} \\
\midrule\noalign{}
\endhead
\bottomrule\noalign{}
\endlastfoot
Intercept & −0.335 & 0.354 & −0.947 & .344 \\
Principal (platform) & −0.163 & 0.190 & −0.858 & .391 \\
Sponsorship (sponsored) & −6.059 & 0.365 & −16.598 & \textless{} .001 \\
Thinking level (low) & −0.254 & 0.190 & −1.336 & .182 \\
Principal \(\times\) Sponsorship & 3.719 & 0.401 & 9.283 & \textless{}
.001 \\
Principal \(\times\) Thinking level & 0.406 & 0.270 & 1.502 & .133 \\
Sponsorship \(\times\) Thinking level & 0.741 & 0.433 & 1.711 & .087 \\
Principal \(\times\) Sponsorship \(\times\) Thinking level & −0.842 &
0.518 & −1.628 & .104 \\
Target price (standardized) & 0.024 & 0.054 & 0.450 & .653 \\
Target review: review 1 & 0.652 & 0.250 & 2.607 & .009 \\
Target review: review 2 & 0.536 & 0.252 & 2.123 & .034 \\
Target review: review 3 & 2.739 & 0.252 & 10.888 & \textless{} .001 \\
Target review: review 4 & 4.934 & 0.282 & 17.485 & \textless{} .001 \\
Target review: review 5 & 1.295 & 0.248 & 5.213 & \textless{} .001 \\
Target review: review 6 & 2.641 & 0.249 & 10.610 & \textless{} .001 \\
Target review: review 7 & 1.415 & 0.254 & 5.575 & \textless{} .001 \\
Target review: review 8 & −0.912 & 0.281 & −3.240 & .001 \\
Target review: review 9 & 3.703 & 0.259 & 14.302 & \textless{} .001 \\
Target hotel: hotel 1 & 0.757 & 0.236 & 3.202 & .001 \\
Target hotel: hotel 2 & 0.438 & 0.236 & 1.854 & .064 \\
Target hotel: hotel 3 & −0.239 & 0.244 & −0.977 & .329 \\
Target hotel: hotel 4 & 0.684 & 0.246 & 2.782 & .005 \\
Target hotel: hotel 5 & −0.622 & 0.248 & −2.511 & .012 \\
Target hotel: hotel 6 & −0.743 & 0.254 & −2.931 & .003 \\
Target hotel: hotel 7 & −1.757 & 0.259 & −6.794 & \textless{} .001 \\
Target hotel: hotel 8 & −0.184 & 0.240 & −0.767 & .443 \\
Target hotel: hotel 9 & −0.140 & 0.239 & −0.586 & .558 \\
Competitor review: review 1 & −0.118 & 0.220 & −0.536 & .592 \\
Competitor review: review 2 & 0.148 & 0.220 & 0.674 & .500 \\
Competitor review: review 3 & −2.023 & 0.235 & −8.606 & \textless{}
.001 \\
Competitor review: review 4 & −5.764 & 0.559 & −10.303 & \textless{}
.001 \\
Competitor review: review 5 & −0.532 & 0.224 & −2.379 & .017 \\
Competitor review: review 6 & −1.962 & 0.235 & −8.344 & \textless{}
.001 \\
Competitor review: review 7 & −0.712 & 0.219 & −3.246 & .001 \\
Competitor review: review 8 & 1.059 & 0.220 & 4.821 & \textless{}
.001 \\
Competitor review: review 9 & −3.941 & 0.314 & −12.559 & \textless{}
.001 \\
Competitor hotel: hotel 1 & −0.713 & 0.248 & −2.871 & .004 \\
Competitor hotel: hotel 2 & −0.770 & 0.251 & −3.071 & .002 \\
Competitor hotel: hotel 3 & 0.589 & 0.248 & 2.377 & .017 \\
Competitor hotel: hotel 4 & −0.596 & 0.247 & −2.415 & .016 \\
Competitor hotel: hotel 5 & 0.171 & 0.233 & 0.735 & .462 \\
Competitor hotel: hotel 6 & 1.180 & 0.232 & 5.084 & \textless{} .001 \\
Competitor hotel: hotel 7 & 1.647 & 0.249 & 6.622 & \textless{} .001 \\
Competitor hotel: hotel 8 & 0.201 & 0.241 & 0.832 & .405 \\
Competitor hotel: hotel 9 & 0.711 & 0.242 & 2.942 & .003 \\
\end{longtable}
}

\subsection*{C.10 Full logistic regression table - Study
2}\label{full-logistic-regression-table---study-2}
\addcontentsline{toc}{subsection}{C.10 Full logistic regression table -
Study 2}

{\def\LTcaptype{apptbl}
\begin{longtable}[]{@{}
  >{\raggedright\arraybackslash}p{(\linewidth - 8\tabcolsep) * \real{0.2000}}
  >{\raggedleft\arraybackslash}p{(\linewidth - 8\tabcolsep) * \real{0.2000}}
  >{\raggedleft\arraybackslash}p{(\linewidth - 8\tabcolsep) * \real{0.2000}}
  >{\raggedleft\arraybackslash}p{(\linewidth - 8\tabcolsep) * \real{0.2000}}
  >{\raggedright\arraybackslash}p{(\linewidth - 8\tabcolsep) * \real{0.2000}}@{}}
\caption{{\label{apptbl-study3-logit-full}}Full logistic regression
table for disclosure type, platform attribution, and principal
manipulation - Gemini 3.1 Pro (Study 2)}\tabularnewline
\toprule\noalign{}
\begin{minipage}[b]{\linewidth}\raggedright
Term
\end{minipage} & \begin{minipage}[b]{\linewidth}\raggedleft
\(\beta\)
\end{minipage} & \begin{minipage}[b]{\linewidth}\raggedleft
\(SE\)
\end{minipage} & \begin{minipage}[b]{\linewidth}\raggedleft
\(z\)
\end{minipage} & \begin{minipage}[b]{\linewidth}\raggedright
\(p\)
\end{minipage} \\
\midrule\noalign{}
\endfirsthead
\toprule\noalign{}
\begin{minipage}[b]{\linewidth}\raggedright
Term
\end{minipage} & \begin{minipage}[b]{\linewidth}\raggedleft
\(\beta\)
\end{minipage} & \begin{minipage}[b]{\linewidth}\raggedleft
\(SE\)
\end{minipage} & \begin{minipage}[b]{\linewidth}\raggedleft
\(z\)
\end{minipage} & \begin{minipage}[b]{\linewidth}\raggedright
\(p\)
\end{minipage} \\
\midrule\noalign{}
\endhead
\bottomrule\noalign{}
\endlastfoot
Intercept & −2.911 & 0.421 & −6.910 & \textless{} .001 \\
Principal (platform) & 1.145 & 0.194 & 5.903 & \textless{} .001 \\
Disclosure (``Sponsored'') & −5.640 & 0.554 & −10.172 & \textless{}
.001 \\
Attribution (``by the platform'') & 0.429 & 0.190 & 2.255 & .024 \\
Principal \(\times\) Disclosure & −0.289 & 0.688 & −0.420 & .674 \\
Disclosure \(\times\) Attribution & 1.360 & 0.620 & 2.195 & .028 \\
Principal \(\times\) Attribution & 2.344 & 0.285 & 8.236 & \textless{}
.001 \\
Principal \(\times\) Disclosure \(\times\) Attribution & −0.068 & 0.783
& −0.087 & .930 \\
Target price (standardized) & 0.002 & 0.061 & 0.034 & .973 \\
Target review: review 1 & 0.533 & 0.299 & 1.782 & .075 \\
Target review: review 2 & 0.396 & 0.307 & 1.292 & .196 \\
Target review: review 3 & 2.236 & 0.303 & 7.383 & \textless{} .001 \\
Target review: review 4 & 4.393 & 0.315 & 13.957 & \textless{} .001 \\
Target review: review 5 & 0.612 & 0.307 & 1.997 & .046 \\
Target review: review 6 & 2.199 & 0.294 & 7.490 & \textless{} .001 \\
Target review: review 7 & 1.074 & 0.298 & 3.609 & \textless{} .001 \\
Target review: review 8 & 0.089 & 0.318 & 0.279 & .780 \\
Target review: review 9 & 3.748 & 0.312 & 11.995 & \textless{} .001 \\
Target hotel: hotel 1 & 0.606 & 0.266 & 2.274 & .023 \\
Target hotel: hotel 2 & 0.514 & 0.257 & 1.998 & .046 \\
Target hotel: hotel 3 & 0.160 & 0.263 & 0.608 & .543 \\
Target hotel: hotel 4 & 0.400 & 0.267 & 1.495 & .135 \\
Target hotel: hotel 5 & 0.219 & 0.266 & 0.824 & .410 \\
Target hotel: hotel 6 & −0.187 & 0.270 & −0.692 & .489 \\
Target hotel: hotel 7 & −1.364 & 0.280 & −4.864 & \textless{} .001 \\
Target hotel: hotel 8 & 0.447 & 0.265 & 1.684 & .092 \\
Target hotel: hotel 9 & 0.043 & 0.258 & 0.167 & .867 \\
Competitor review: review 1 & 0.457 & 0.235 & 1.949 & .051 \\
Competitor review: review 2 & 0.539 & 0.242 & 2.230 & .026 \\
Competitor review: review 3 & −2.042 & 0.269 & −7.584 & \textless{}
.001 \\
Competitor review: review 4 & −4.779 & 0.416 & −11.496 & \textless{}
.001 \\
Competitor review: review 5 & −0.419 & 0.237 & −1.767 & .077 \\
Competitor review: review 6 & −1.839 & 0.262 & −7.009 & \textless{}
.001 \\
Competitor review: review 7 & −0.611 & 0.239 & −2.558 & .011 \\
Competitor review: review 8 & 1.311 & 0.228 & 5.742 & \textless{}
.001 \\
Competitor review: review 9 & −3.506 & 0.320 & −10.951 & \textless{}
.001 \\
Competitor hotel: hotel 1 & −0.484 & 0.281 & −1.726 & .084 \\
Competitor hotel: hotel 2 & −0.182 & 0.274 & −0.663 & .507 \\
Competitor hotel: hotel 3 & 0.455 & 0.268 & 1.698 & .090 \\
Competitor hotel: hotel 4 & 0.313 & 0.271 & 1.158 & .247 \\
Competitor hotel: hotel 5 & 0.766 & 0.277 & 2.763 & .006 \\
Competitor hotel: hotel 6 & 1.613 & 0.259 & 6.235 & \textless{} .001 \\
Competitor hotel: hotel 7 & 2.034 & 0.267 & 7.617 & \textless{} .001 \\
Competitor hotel: hotel 8 & 0.647 & 0.268 & 2.411 & .016 \\
Competitor hotel: hotel 9 & 0.403 & 0.268 & 1.505 & .132 \\
\(N\) & 4,000 & & & \\
Pseudo \(R^2\) & 0.585 & & & \\
Log-Likelihood & −941.85 & & & \\
\end{longtable}
}

\emph{Note.} Reference categories are consumer delegation, ``Promoted''
disclosure, no platform attribution, and index 0 for target/competitor
review text and hotel identity (\(N = 4{,}000\)).

\subsection*{C.11 Full logistic regression table - Study 3
mitigation}\label{full-logistic-regression-table---study-3-mitigation}
\addcontentsline{toc}{subsection}{C.11 Full logistic regression table -
Study 3 mitigation}

{\def\LTcaptype{apptbl}
\begin{longtable}[]{@{}
  >{\raggedright\arraybackslash}p{(\linewidth - 8\tabcolsep) * \real{0.2000}}
  >{\raggedleft\arraybackslash}p{(\linewidth - 8\tabcolsep) * \real{0.2000}}
  >{\raggedleft\arraybackslash}p{(\linewidth - 8\tabcolsep) * \real{0.2000}}
  >{\raggedleft\arraybackslash}p{(\linewidth - 8\tabcolsep) * \real{0.2000}}
  >{\raggedright\arraybackslash}p{(\linewidth - 8\tabcolsep) * \real{0.2000}}@{}}
\caption{{\label{apptbl-mitigation-logit}}Full logistic regression
table for prompt-level mitigation - Gemini 3.1 Pro (Study 3)}\tabularnewline
\toprule\noalign{}
\begin{minipage}[b]{\linewidth}\raggedright
Term
\end{minipage} & \begin{minipage}[b]{\linewidth}\raggedleft
\(\beta\)
\end{minipage} & \begin{minipage}[b]{\linewidth}\raggedleft
\(SE\)
\end{minipage} & \begin{minipage}[b]{\linewidth}\raggedleft
\(z\)
\end{minipage} & \begin{minipage}[b]{\linewidth}\raggedright
\(p\)
\end{minipage} \\
\midrule\noalign{}
\endfirsthead
\toprule\noalign{}
\begin{minipage}[b]{\linewidth}\raggedright
Term
\end{minipage} & \begin{minipage}[b]{\linewidth}\raggedleft
\(\beta\)
\end{minipage} & \begin{minipage}[b]{\linewidth}\raggedleft
\(SE\)
\end{minipage} & \begin{minipage}[b]{\linewidth}\raggedleft
\(z\)
\end{minipage} & \begin{minipage}[b]{\linewidth}\raggedright
\(p\)
\end{minipage} \\
\midrule\noalign{}
\endhead
\bottomrule\noalign{}
\endlastfoot
Intercept & −0.178 & 0.459 & −0.388 & .698 \\
Platform (prompt-level mitigation) & −0.169 & 0.205 & −0.825 & .409 \\
Platform & −0.163 & 0.203 & −0.803 & .422 \\
Sponsorship (sponsored) & −6.912 & 0.420 & −16.467 & \textless{} .001 \\
Platform (prompt-level mitigation) \(\times\) Sponsorship & 0.117 &
0.493 & 0.238 & .812 \\
Platform \(\times\) Sponsorship & 4.204 & 0.437 & 9.623 & \textless{}
.001 \\
Target price (standardized) & 0.029 & 0.069 & 0.420 & .675 \\
Target review: review 1 & 0.926 & 0.328 & 2.819 & .005 \\
Target review: review 2 & 0.205 & 0.350 & 0.586 & .558 \\
Target review: review 3 & 3.158 & 0.342 & 9.247 & \textless{} .001 \\
Target review: review 4 & 5.800 & 0.382 & 15.177 & \textless{} .001 \\
Target review: review 5 & 1.290 & 0.329 & 3.921 & \textless{} .001 \\
Target review: review 6 & 3.018 & 0.341 & 8.842 & \textless{} .001 \\
Target review: review 7 & 1.758 & 0.338 & 5.206 & \textless{} .001 \\
Target review: review 8 & −0.837 & 0.369 & −2.268 & .023 \\
Target review: review 9 & 4.487 & 0.352 & 12.742 & \textless{} .001 \\
Target hotel: hotel 1 & 0.677 & 0.308 & 2.196 & .028 \\
Target hotel: hotel 2 & 0.303 & 0.308 & 0.983 & .325 \\
Target hotel: hotel 3 & −0.139 & 0.317 & −0.438 & .661 \\
Target hotel: hotel 4 & 0.917 & 0.326 & 2.813 & .005 \\
Target hotel: hotel 5 & −0.987 & 0.317 & −3.111 & .002 \\
Target hotel: hotel 6 & −1.365 & 0.352 & −3.882 & \textless{} .001 \\
Target hotel: hotel 7 & −2.016 & 0.341 & −5.904 & \textless{} .001 \\
Target hotel: hotel 8 & 0.038 & 0.319 & 0.118 & .906 \\
Target hotel: hotel 9 & 0.096 & 0.304 & 0.317 & .752 \\
Competitor review: review 1 & −0.533 & 0.292 & −1.826 & .068 \\
Competitor review: review 2 & 0.032 & 0.291 & 0.110 & .912 \\
Competitor review: review 3 & −2.867 & 0.307 & −9.333 & \textless{}
.001 \\
Competitor review: review 4 & −5.892 & 0.538 & −10.959 & \textless{}
.001 \\
Competitor review: review 5 & −0.861 & 0.291 & −2.963 & .003 \\
Competitor review: review 6 & −2.888 & 0.340 & −8.485 & \textless{}
.001 \\
Competitor review: review 7 & −1.105 & 0.295 & −3.749 & \textless{}
.001 \\
Competitor review: review 8 & 0.911 & 0.289 & 3.147 & .002 \\
Competitor review: review 9 & −4.492 & 0.400 & −11.229 & \textless{}
.001 \\
Competitor hotel: hotel 1 & −0.938 & 0.316 & −2.965 & .003 \\
Competitor hotel: hotel 2 & −0.743 & 0.316 & −2.353 & .019 \\
Competitor hotel: hotel 3 & 0.672 & 0.322 & 2.086 & .037 \\
Competitor hotel: hotel 4 & −0.984 & 0.317 & −3.100 & .002 \\
Competitor hotel: hotel 5 & 0.530 & 0.300 & 1.770 & .077 \\
Competitor hotel: hotel 6 & 1.390 & 0.299 & 4.653 & \textless{} .001 \\
Competitor hotel: hotel 7 & 2.520 & 0.326 & 7.723 & \textless{} .001 \\
Competitor hotel: hotel 8 & 0.346 & 0.312 & 1.110 & .267 \\
Competitor hotel: hotel 9 & 0.768 & 0.313 & 2.452 & .014 \\
\end{longtable}
}

\emph{Note.} \(N = 3{,}000\). Reference categories are consumer
delegation and organic target. The model pools the Consumer, Platform,
and Platform (prompt-level mitigation) conditions. Coefficients
therefore differ slightly from \quartoapptblref{apptbl-main-logit-full}.

\end{document}